\documentclass[aps,10pt,pre,floatfix,twocolumn,superscriptaddress]{revtex4-2}
\usepackage{latexsym,amssymb,amsmath,nccmath}
\usepackage{graphicx}
\usepackage{array}
\usepackage[caption=false]{subfig}
\usepackage{color}
\usepackage{csquotes}

\newcommand{\be}{\begin{equation}}
\newcommand{\ee}{\end{equation}}
\newcommand{\bea}{\begin{eqnarray}}
\newcommand{\eea}{\end{eqnarray}}

\begin{document}

\title{Role of spatial patterns in fracture of disordered multiphase materials}
\author{Rajat Pratap Singh Parihar}
\email{rajatsingh@live.com}
\affiliation{Department of Applied Mechanics, Indian Institute of Technology Madras, Chennai-600036, India}
\author{Dhiwakar V. Mani}
\affiliation{Department of Applied Mechanics, Indian Institute of Technology Madras, Chennai-600036, India}
\author{Anuradha Banerjee}
\email{anuban@iitm.ac.in}
\affiliation{Department of Applied Mechanics, Indian Institute of Technology Madras, Chennai-600036, India}
\author{R.Rajesh} 
\email{rrajesh@imsc.res.in}
\affiliation{The Institute of Mathematical Sciences, C.I.T. Campus, Taramani, Chennai-600113, India} 
\affiliation{Homi Bhabha National Institute, Training School Complex, Anushakti Nagar, Mumbai-400094, India}
\date{\today}

\begin{abstract}
Multi-phase materials, such as composite materials, exhibit multiple competing failure mechanisms during the growth of a macroscopic defect. For the simulation of the overall fracture process in such materials, we develop a two-phase spring network model that accounts for the architecture between the different components as well as the respective disorders in their failure characteristics. In the specific case of a plain weave architecture, we show that any offset between the layers reduces the delocalization of the stresses at the crack tip and thereby substantially lowers the strength and fracture toughness of the overall laminate. The avalanche statistics of the broken springs do not show a distinguishable dependence on the offsets between layers. The power law exponents are found to be much smaller than that of disordered spring network models in the absence of a crack. A discussion is developed on the possibility of the avalanche statistics being those near breakdown.

\end{abstract}

\maketitle

\section{Introduction}

In biological materials, nature's design exploits the beneficial aspects of combining different materials and the spatial patterns between them to develop lightweight material systems with significantly improved resistance to fracture~\cite{gao-2003, gupta-2009, sen-2011}. Synthetic or engineered composite materials, similarly, have aimed to expand the strength to weight as well as toughness to weight performances of available materials. Different spatial patterns, such as short stiff fiber reinforcements randomly arranged in a compliant matrix, long fibers unidirectionally arranged, or in a two- or three-dimensional woven architecture, etc., have been developed successfully~\cite{piggott-1994, puck-2002, mishnaevsky-2009, daniel-2009, wicks-2014, das-2018}. The more intricate the architecture, the higher the associated complexity of the deformation and in particular, the fracture process, due to several competing failure mechanisms, making predictive modeling significantly more challenging. For better design and life assessment of heterogeneous composite materials, in the present study, we develop a statistical approach to examine some of the factors that influence the defect tolerance of spatially patterned heterogeneous composite materials.

The effect of heterogeneity on the macroscopic mechanical performance of composite materials, measured in terms of stiffness, strength, and toughness, has been examined by several researchers in both statistical physics as well as engineering. Depending on the relevant length scales of a given system, many different theoretical frameworks have been adopted for analysis, ranging from atomistic simulations for design of materials to continuum models for design of components and structures~\cite{tan-2000, carol-2001a, carol-2001b, luccioni-2003, ansar-2011, zhang-2012}. Continuum description of composite material typically involves micro-mechanical modeling of a representative volume to arrive at the effective anisotropic elastic and plastic behavior. The homogenized anisotropic constitutive relations are then combined with a damage initiation and evolution criterion for simulation of progressive damage~\cite{camanho-2013, volger-2013, tan-2016}. While these models have been shown to accurately describe uniaxial responses in particular directions, the material behavior under multiaxial loading is, however, not well reproduced.~\cite{camanho-2013, volger-2013, tan-2016}.

Brittle heterogeneous materials, when subject to mechanical loads, are prone to micro-cracking at multiple sites. On the application of higher stresses, these micro-cracks tend to interact and grow, resulting in multiple events prior to final failure, as evident in the acoustic emission activity~\cite{rosti-2009, baro-2013}. Such complex fracture processes, which typically occur over large process zones, are stochastic in nature, and have been investigated comprehensively using lattice-based models for single-phase systems with disorder. In these lattice-based models, the continuum is discretized into a collection of interconnected particles, and a simple inter particle interaction is defined to account for the mechanical and fracture properties of the material~\cite{curtin-1990, alava-2006, pan-2018}. The scaling properties observed in acoustic emission experimental data during fracture of heterogeneous materials have been of significant interest in the field of statistical physics. Attempts have been made to interpret the scaling laws of energy release and fluctuations in temporal statistics via concepts of criticality and phase transitions~\cite{ray-2006, alava-2006, rosti-2009,pradhan-2006}. Qualitative features of the role of disorder in crack paths, and macroscopic response have also been investigated in detail because arbitrarily oriented, multiple crack paths, typically observed in materials like concrete, particulate composites, ceramics, cortical bones, etc., are easy to simulate using lattice-based models~\cite{karihaloo-2003, berton-2006, grassl-2009, ashwij-2016, ashwij-2017, ashwij-2018}.

Brittle multiphase heterogeneous systems are comparatively less well studied. Investigating the role of elastic heterogeneity in a two-phase system, studied using a network of hard and soft springs, Urabe \textit{et al.}~\cite{urabe-2010} showed the possibility of designing material systems tougher and stronger than individual components. The two-phase network was also utilized in the translaminar fracture simulation, under mode I and mixed mode, of woven composites~\cite{dhatreyi-2015}. In these studies, while the relative proportion of hard and soft bonds and their orientations was maintained to ensure equivalence with the macroscopic elastic modulus of the composite material, the details of the meso-structural patterns of the woven reinforcements were ignored, as the heterogeneity was randomly distributed in the domain.

The role of spatial patterns in fracture of heterogeneous systems was analyzed in a series of investigations by Buehler and co-workers~\cite{dimas-2014a,dimas-2015a, dimas-2015b}. Spatial patterning in a two-dimensional heterogeneous spring bead network was represented by introducing a correlation length in the spatial distribution of the Young's moduli modeled as a Gaussian process, such that any significant variation in the modulus occurs only over the correlation length. Toughening in the heterogeneous solid with disordered elastic fields was attributed to the \enquote{distribution-of-weakness} mechanism, resulting from the observed crack arrest and stress delocalization~\cite{dimas-2014a}. Further, some of these features could be reproduced in solvable models of rectangular blocks, rods, and plates~\cite{dimas-2015a, dimas-2015b}. 

How does the fracture process differ between laminates that are elastically equivalent but differ in the spatial patterns between their hard and soft phases? Under what relative fracture properties of the phases does the elastic heterogeneity result in enhancement of toughness? Do avalanche statistics depend on spatial patterning? In this paper, we explore the role of meso-structural patterns by considering a representative system of woven fiber-reinforced laminate. To incorporate the composite's mesoscopic spatial pattern, we develop a multiphase two-dimensional computational model using spring networks, the elastic properties of each phase being estimated using standard homogenization technique and the rule of mixture. Within the model, we study in detail the dependence of the elastic stress distribution and fracture properties such as strength, toughness, crack paths, and avalanche statistics on the spatial patterning of the laminate. In particular, we show that toughness is enhanced with increased elastic heterogeneity.

\section{\label{sec:model}Modeling and Simulation}

In this section, we develop a two-dimensional model for the analysis of plain weave laminates. Four steps are involved. First, the three-dimensional plain weave laminate is mapped onto a two-dimensional geometric pattern based on the local volume fraction of the yarns (Sec.~\ref{sec:geometric}). Second, from the two-dimensional geometric pattern, a discrete element model in the form of a two-dimensional spring network model made up of bonds of multiple types is constructed (Sec.~\ref{subsec:lsnmmod}). Third, a methodology of assigning local elastic parameters to springs is developed so that the experimentally observed macroscopic elastic behavior is reproduced (Sec.~\ref{subsec:elasticmodel}). Fourth, the algorithm for simulating the resultant spring network model using molecular dynamics is discussed (Sec.~\ref{subsec:md}).

\subsection{\label{sec:geometric} Geometric Modeling}

\begin{figure}
	\includegraphics[width=\columnwidth]{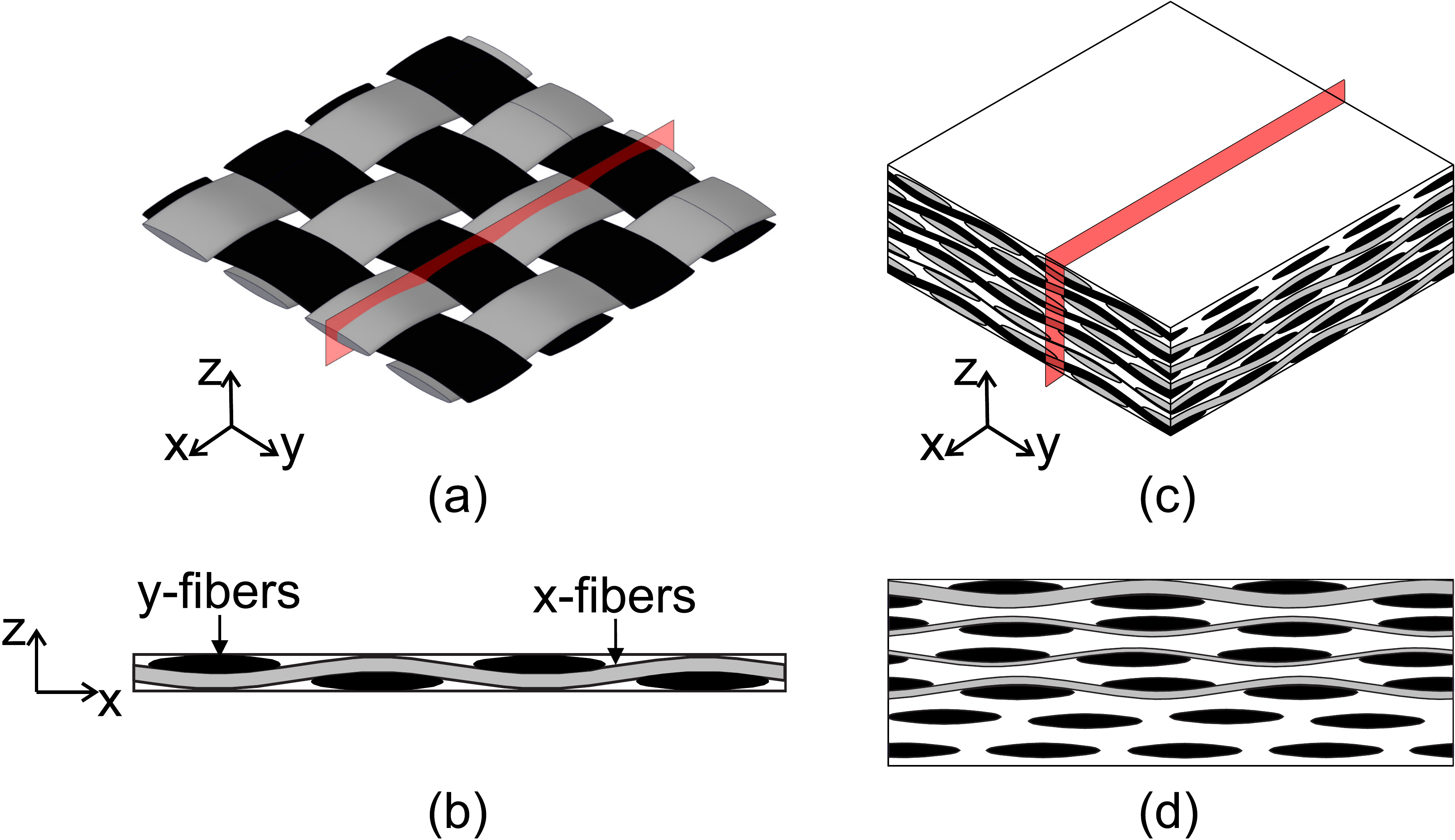}
	\caption{\label{img:plain_weave}Schematic diagrams of (a) a single lamina or layer with fibers woven in perpendicular directions, (b) cross-sectional view of the lamina, at the sectioning plane (shown in red [translucent cutting plane] in (a)), (c) a laminate in which the constituent laminae are stacked randomly, and (d) cross-sectional view of the laminate at the sectioning plane.}
\end{figure}

The representative material system being modeled is a soft matrix embedded with a harder reinforcement with a plain weave architecture. We focus on the most common reinforcement patterns in which yarns of fibers are interlaced in a textile-like geometric pattern. A typical pattern of a single layer or lamina is shown in Fig.~\ref{img:plain_weave}a. The pattern consists of two types of fibers: those oriented in the x-direction denoted as $ x $-fibers and those oriented in the orthogonal $ y $-direction denoted as $ y $-fibers. When sectioned in the thickness direction, a well defined geometric pattern emerges where, as shown in Fig.~\ref{img:plain_weave}b, the cross-section consists of $ y $-fibers appearing as isolated elliptic regions while the $ x $-fibers are continuous and have an oscillatory pattern.

Typical composite laminates consist of 6 to 8 such layers in the thickness direction ($z$-direction), with each layer having possible offsets in the $ xy $ plane (Fig.~\ref{img:plain_weave}c). Cross-sections of these laminates have 
periodically patterned rows that are phase shifted depending on the relative positioning of the layers with respect to the sectioning plane, as shown in Fig.~\ref{img:plain_weave}d. Note that depending on the relative positioning of any lamina with respect to the sectioning plane, the cross-sectional thickness of the $x$-fiber may vary from its maximum thickness (first layer in Fig.~\ref{img:plain_weave}d) to zero thickness (last layer in Fig.~\ref{img:plain_weave}d, where the grey tube is missing).

\begin{figure} 
\subfloat[\label{img:lamina_at_cs}]{
\includegraphics[width=0.96\columnwidth]{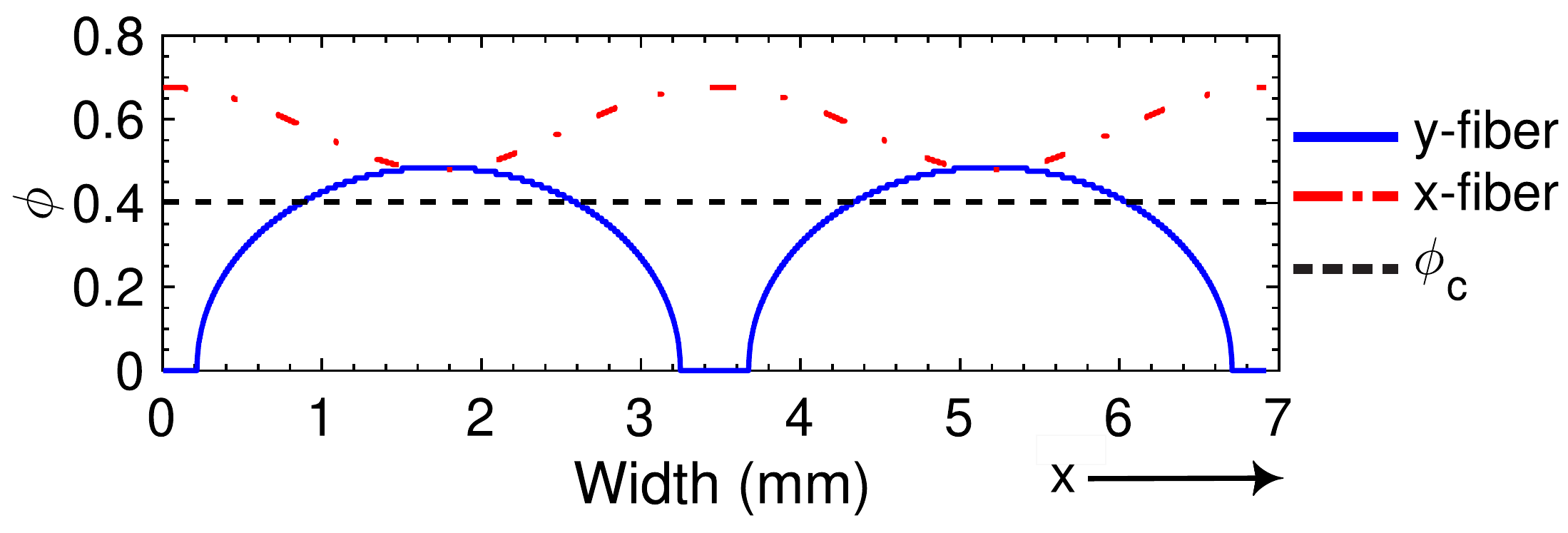}}
\vspace{0cm}
\subfloat[\label{img:laminate_at_cs}]{
\includegraphics[width=0.96\columnwidth]{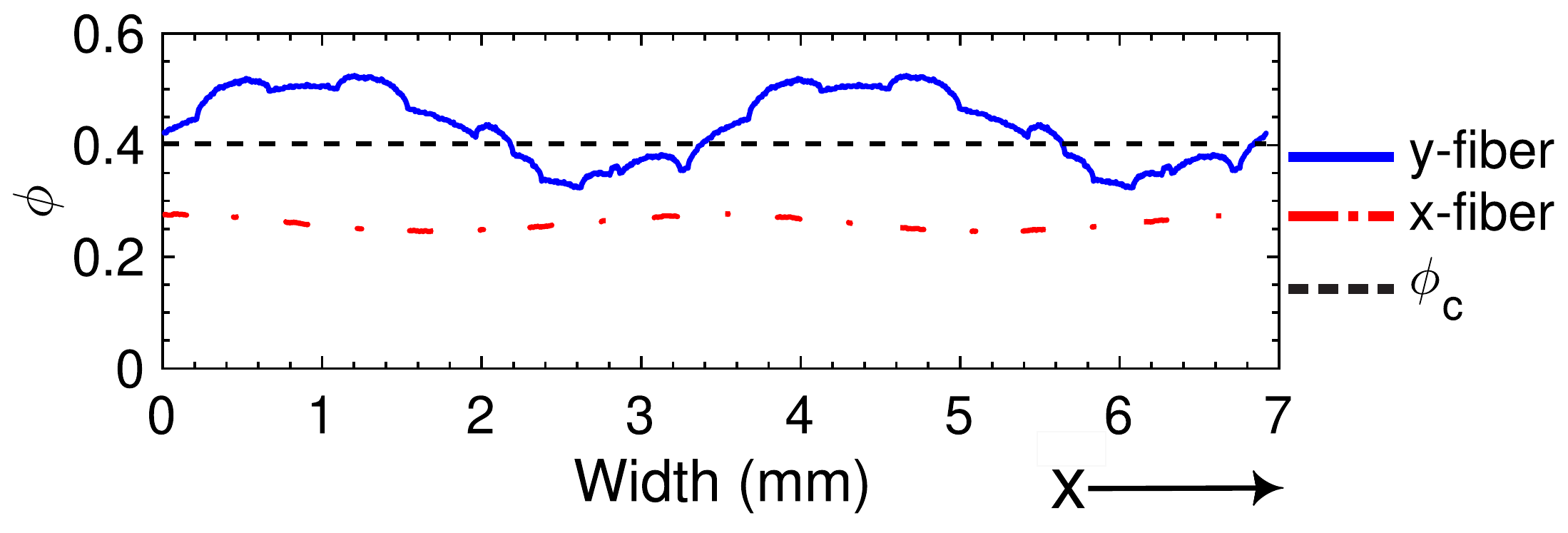}}
\caption{Variation of the volume fractions of x-fibers and y-fibers at a cross-section of (a) a single lamina and (b) a representative laminate of six layers. The horizontal dotted line $\phi_c$ is a threshold as defined in Table~\ref{table:rgn-cut}. \label{img:vf_variation}}
\end{figure}

For a typical single lamina, as well as the laminate, as shown in Fig.~\ref{img:plain_weave}, based on the pattern observed in the given cross-section, the volume fractions of $x$- and $y$-fibers along the section's length are shown in Fig.~\ref{img:lamina_at_cs}-\ref{img:laminate_at_cs}. In the cross-section of a single lamina, the volume fraction of the $y$-fibers varies between its maximum at the center of the elliptic cross-section and zero in the region between the ellipses as seen in Fig.~\ref{img:lamina_at_cs}. Correspondingly, the $ x $-fibers have a well-defined pattern as well. In a laminate, as shown in Fig.~\ref{img:laminate_at_cs}, the volume fraction of both $ y $-fibers as well as $ x $-fibers when averaged over the layers has comparatively less variation between the maximum, $ \phi_{\rm max} $, and minimum, $ \phi_{\rm min} $.

\begin{figure}
	\includegraphics[width=\columnwidth]{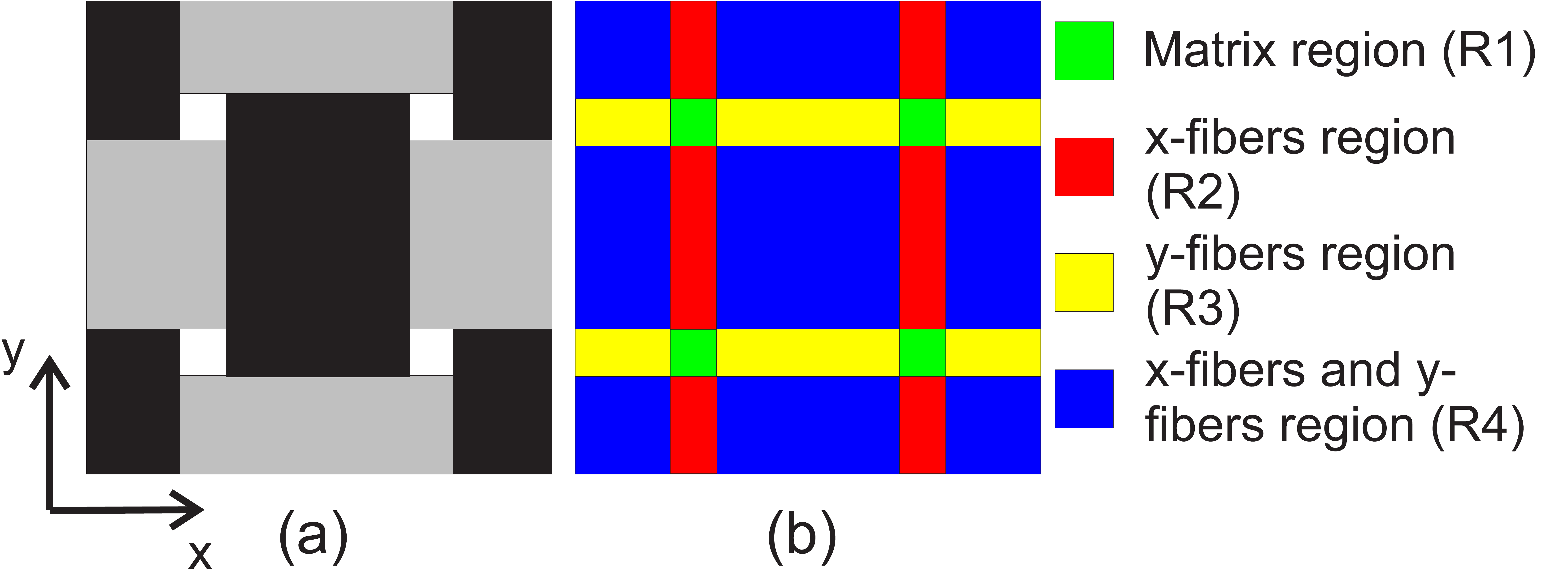}
	\centering
	\caption{(a) A representative area element of a single lamina. (b) The corresponding allocation of regions based on the configuration of $x$-fibers, $y$-fibers and matrix. \label{img:yarn_region}}
\end{figure}

A representative area element of a single lamina is shown in Fig.~\ref{img:yarn_region}a. It is composed of 4 interconnected regions: regions which have no fibers (referred as R1 and represented by the color {green [grey]} in Fig.~\ref{img:yarn_region}b), regions which have fibers only in the $ x $-direction (R2, {red [dark grey]}), regions which have fibers only in the $ y $-direction (R3, {yellow [light grey]}) and regions which have fibers in both directions (R4, {blue [black]}). The distinction between the fiber-rich regions and the matrix-rich (no fibers) regions becomes less apparent when several layers are stacked together in a laminate. For a simplified 2-dimensional representation of a laminate, the spatial distribution of volume fraction of fibers can be discretized by introducing a threshold cut-off $\phi_c$, such that it is approximated to be composed of fiber-rich and matrix-rich regions. For the present study, the criteria for allocating different regions based on the threshold and the volume fractions of $x$-fibers and $y$-fibers, was taken to be as per the rule in Table~\ref{table:rgn-cut}.
\begin{table}
	\centering
	\caption{Criteria for allocating different regions based on the threshold, $\phi_c$ and the volume fractions of $x$-fibers and $y$-fibers \label{table:rgn-cut}}
	\begin{ruledtabular}
		\begin{tabular} {m{5cm} m{1.5cm} m{1.5cm}}
			Regions & $ \phi$ (x-fiber) & $ \phi$ (y-fiber) \\
			Region rich in $ y $-fiber and $ x $-fiber (R4) & $ \geq \phi_c $ & $ \geq \phi_c $ \\
			$ y $-fiber rich region (R3) & $ < \phi_c $ & $ \geq \phi_c $ \\
			$ x $-fiber rich region (R2) & $ \geq \phi_c $ & $ < \phi_c $ \\
			Matrix rich region (R1) & $ < \phi_c $ & $ < \phi_c $ \\
		\end{tabular}
	\end{ruledtabular}
\end{table}

\begin{figure} \centering
\subfloat[\label{img:70_lam_random}]{
\includegraphics[width=0.31\columnwidth]{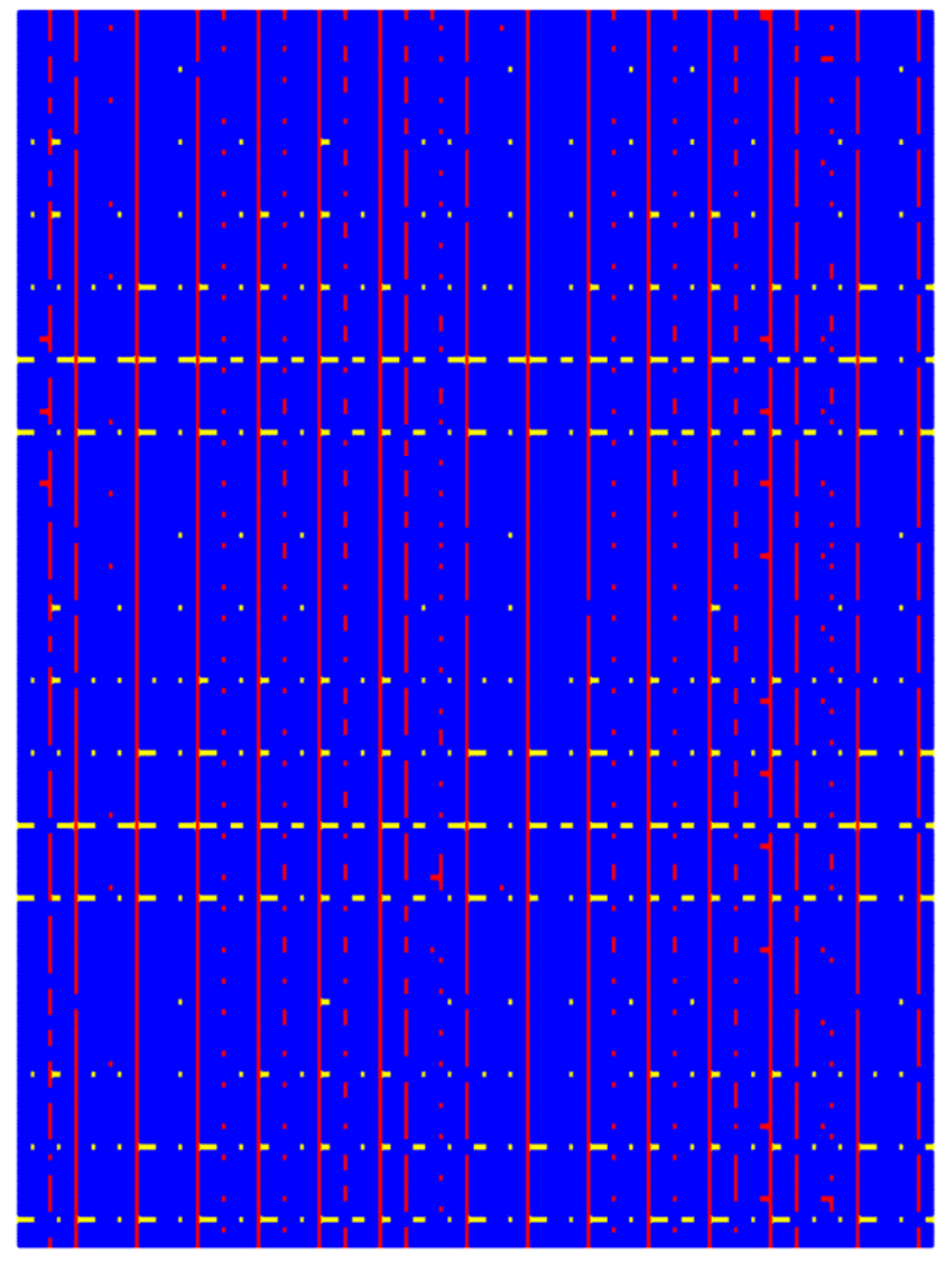}}
\hspace{0cm}
\subfloat[\label{img:100_lam_random_1}]{
\includegraphics[width=0.31\columnwidth]{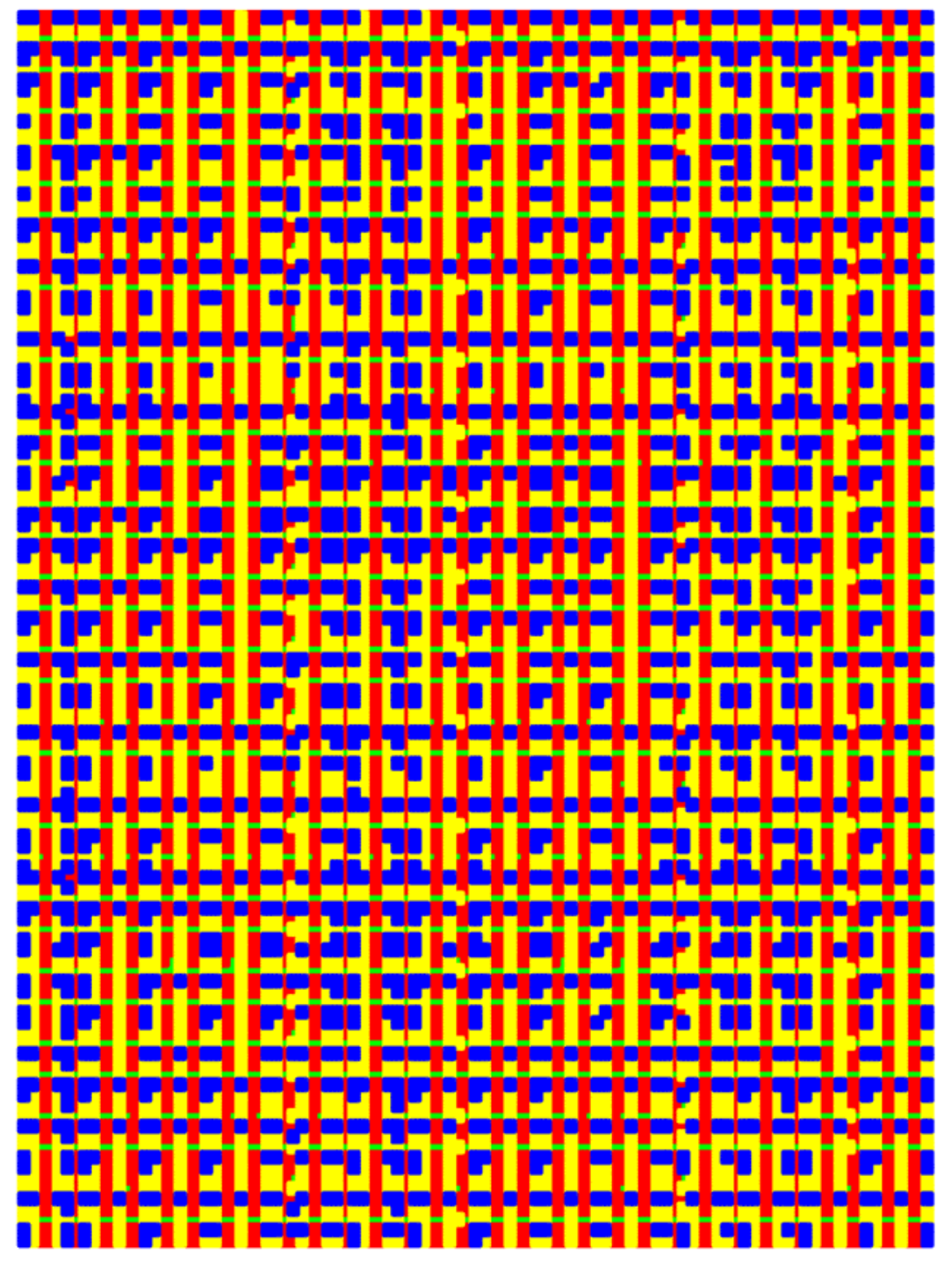}}
\hspace{0cm}
\subfloat[\label{img:136_lam_random}]{
\includegraphics[width=0.31\columnwidth]{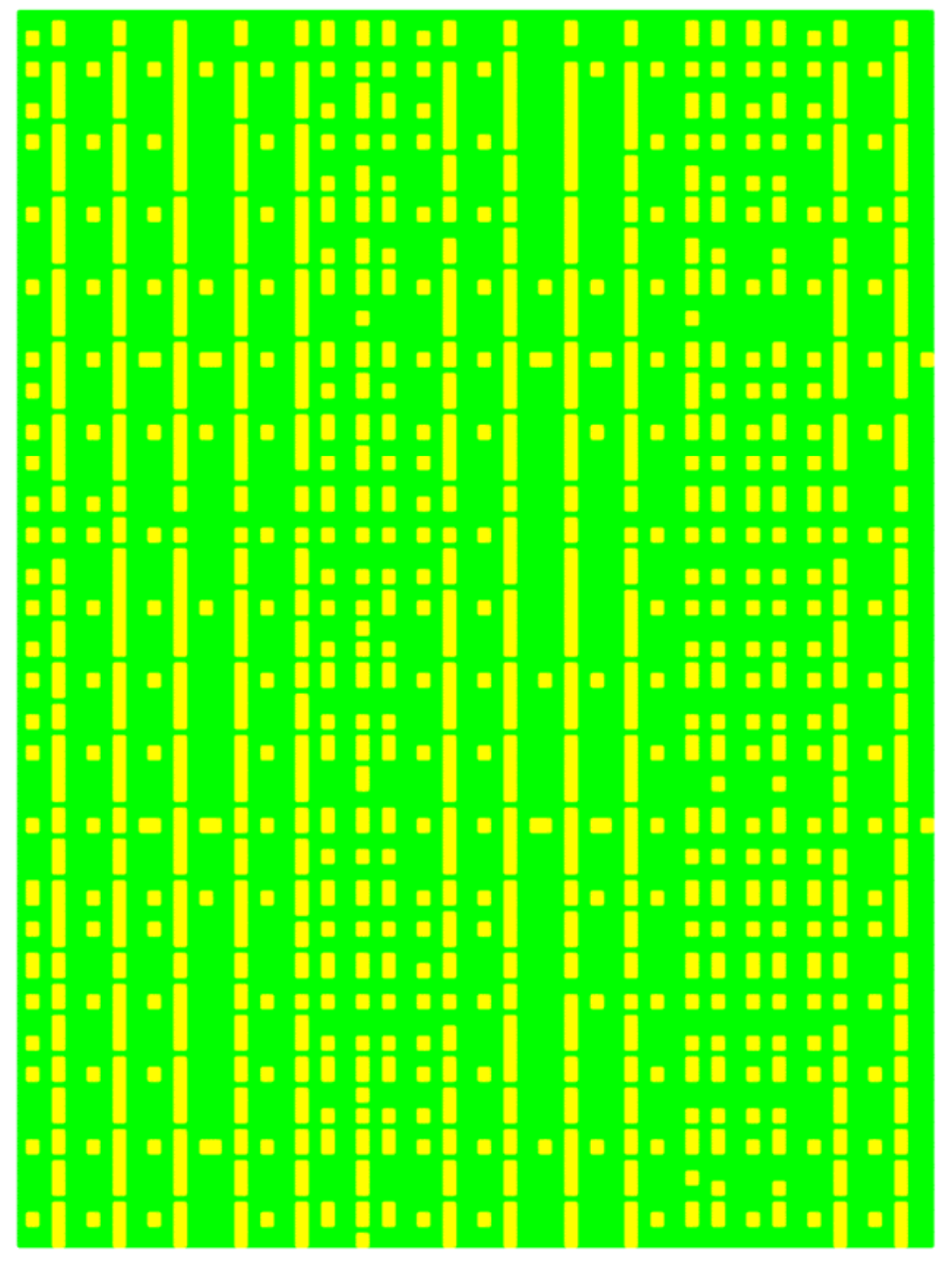}}
\caption{The different regions (R1-R4) when the ratio of the threshold $\phi_c$ to the average $\phi$ (y-fiber) is (a) 0.70, (b) 1.0 and (c) 1.36. (The color mapping is as in Fig.~\ref{img:yarn_region}) \label{img:laminates_random}}
\end{figure}

Thus, the choice of the threshold volume fraction would decide whether the laminate behavior is dominated by the matrix-rich (softer) regions or the fiber-rich (harder) regions. If $ \phi_c $ is chosen to be close to $ \phi_{\rm min} $, then most of the laminate would be of type R4 (fiber-rich hard material), while as we increase $ \phi_c $, the percentage of softer region would increase. Finally, when $ \phi_c $ is close to $ \phi_{\rm max} $, then most of the material would be matrix-rich. Examples of the decomposition for different choices of $\phi_c$ are shown in Fig.~\ref{img:laminates_random}.

{We use a threshold $\phi_c$, primarily to simplify the heterogeneity in elastic as well as fracture properties within the composite laminate, such that role of the hard and soft phases in the growth of damage can be differentiated easily. The simplification to a two-phase system also brings out the comparative differences between the behavior of spatially patterned two phase networks and the randomly distributed two phase networks reported in the existing literature~\cite{urabe-2010, dhatreyi-2015}. In simulation of actual experimental data, however, a more continuous variation in stiffness and fracture properties would probably be desirable. Predictions of such a model would have a stronger dependence on the lattice size.}

\subsection{Random spring network model (RSNM)} \label{subsec:lsnmmod}

\begin{figure}
	\includegraphics[width=\columnwidth]{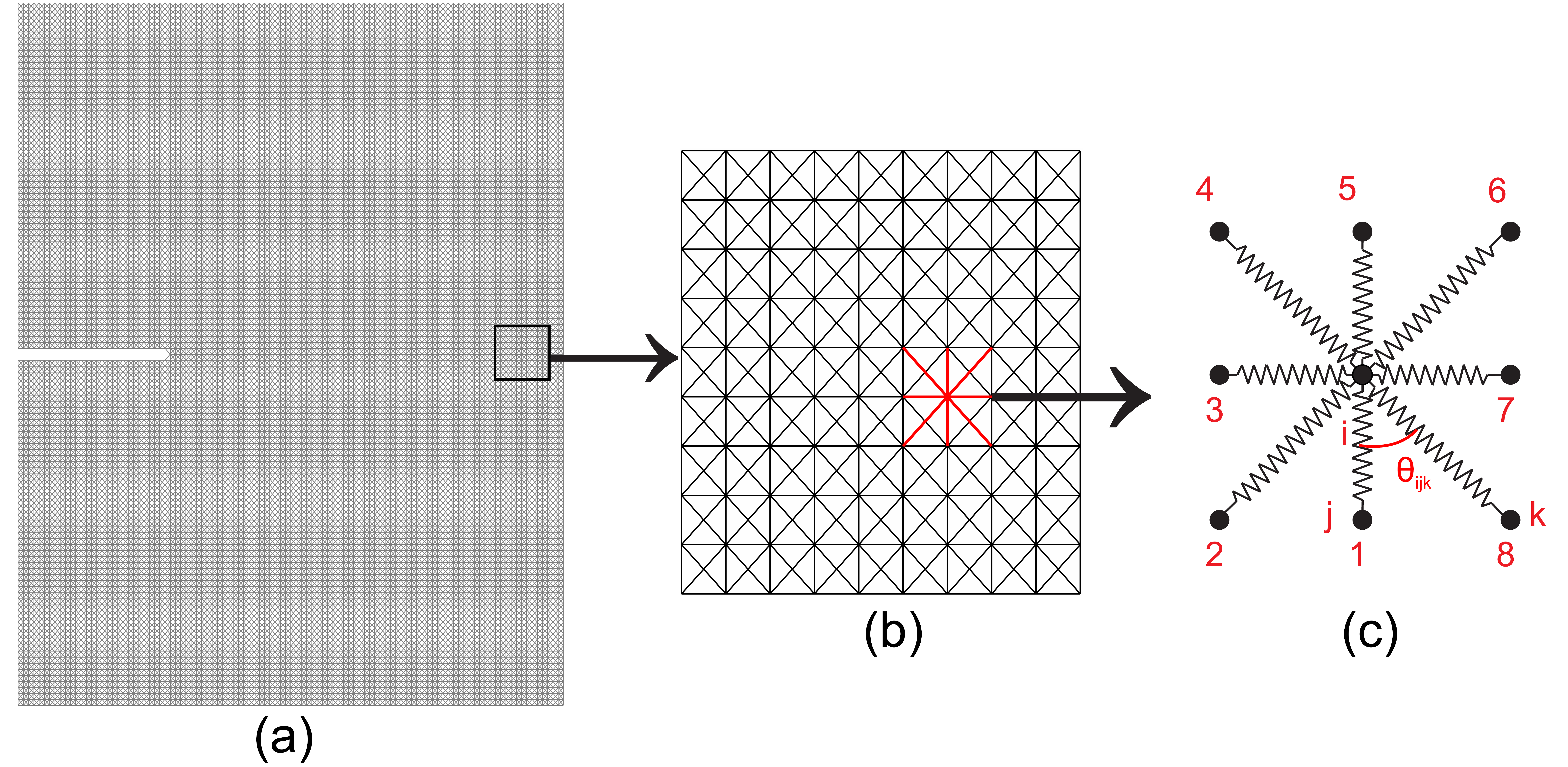}
	\centering
	\caption{(a) RSNM with a notch or crack. (b) Magnified view of spring network showing the spring connections. (c) 2-dimensional square lattice-springs connection with neighboring and next neighboring lattice points. \label{img:LSNM_exp}}
\end{figure}

The geometric pattern consisting of four regions, obtained from the laminate using the procedure in Sec.~\ref{sec:geometric} is now further discretized in space to convert it into a discrete element model. We discretize a rectangular domain of the laminate using a square lattice of size $105\times119$ with lattice spacing $a=0.5 \rm mm$, as shown in Fig.~\ref{img:LSNM_exp}a. Each lattice point, representative of an elemental area, is connected to its nearest and next-nearest neighbors with linear elastic extensional springs, as shown in Fig.~\ref{img:LSNM_exp}b. The resistance to shear deformation of the spring network is provided by torsional springs that resist changes to the angle, $ \theta_{ijk} $, subtended between the lattice point and its two adjacent neighbors, as shown in Fig.~\ref{img:LSNM_exp}c. For simulating fracture, a macroscopic crack is modeled by deleting the spring connections between the lattice points at the top and bottom surface of the crack which is located at mid-height whose length is approximately $0.28$ times the width of the system.

When the spring network is deformed from its initial configuration, the total potential energy stored, $ \Phi $, has both extensional component, $ \Phi_{\rm elongation} $ and rotational component, $ \Phi_{\rm rotation} $:
\begin{equation}
\Phi=\Phi_{\rm elongation}+\Phi_{\rm rotation},
\label{eqn:energy_total}
\end{equation}
For a spring network with $ N $ lattice points, the extensional energy is
\begin{equation}
\Phi_{\rm elongation}=\sum_{i=1}^{N}\sum_{j=1}^{4}\frac{1}{2}k_{ij}\{|\overrightarrow{r_{j}}-\overrightarrow{r_{i}}|-a_{ij}\}^2,
\label{eqn:energy_elong}
\end{equation}
{where $ \overrightarrow{r_{i}} $, $ \overrightarrow{r_{j}} $ are the position vectors of lattice sites $i$ and $j$, $ a_{ij} $ is the initial equilibrium distance between them, and $ k_{ij} $ is the elastic stiffness of spring joining $ i $ and $ j $. The rotational component is given by
\begin{equation}
\Phi_{\rm rotation} = \sum_{i=1}^{N}\sum_{\langle i j k\rangle}\frac{1}{2}c_{ijk}\left(\theta_{ijk}-\frac{\pi}{4}\right)^2,
\label{eqn:energy_bend}
\end{equation}
where the second sum is over the $8$ angles $\theta_{ijk}$ as shown in Fig.~\ref{img:LSNM_exp}c and $ c_{ijk} $ is the torsional stiffness of the spring resisting the change of the angle $\theta_{ijk}$ from $\pi/4$.} Using the equivalence of strain energy density of the spring network and that of a linear elastic isotropic continuum~\cite{monette-1994}, the elastic constants of the continuum, Young's modulus $E$ and Poisson's ratio $\nu$, are related to the local spring constants as~\cite{monette-1994}:
\bea
E&=&\dfrac{8k(k+c/a^2_0)}{3k+c/a^2_0},
\label{eqn:e_relation} \\
\nu &=& \dfrac{k-c/a^2_0}{3k+c/a^2_0}.
\label{eqn:nu_relation}
\eea
In the above RSNM, Poisson's ratio that can be simulated lie in the range $ (-1,1/3) $.

\begin{figure}
	\includegraphics[width=\columnwidth]{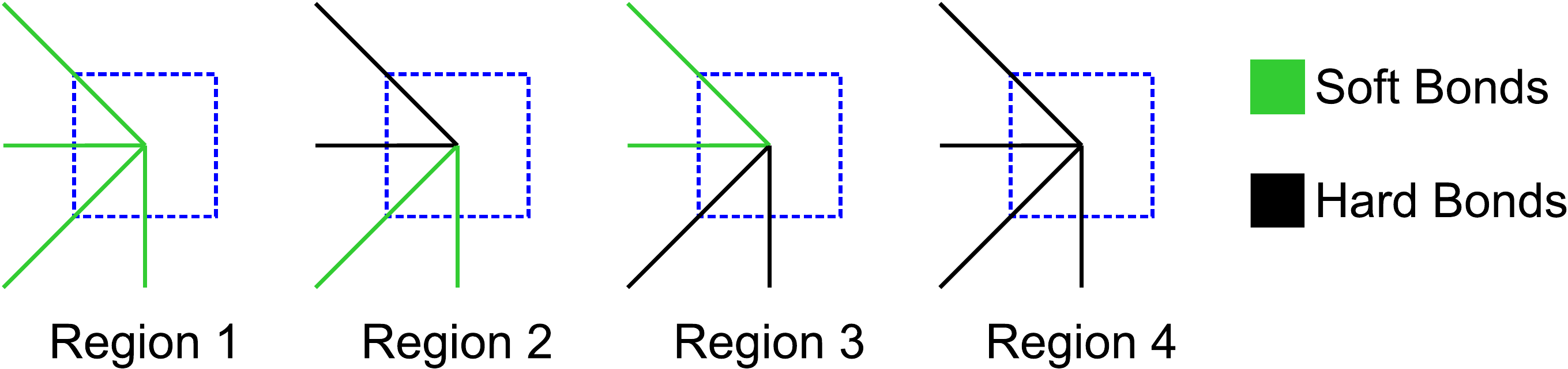}
	\centering
	\caption{Different types of springs based on the different regions of the laminate. \label{img:bond_direction_color}}
\end{figure}

In the above discussion, the relation between spring constants and the elastic constants are for a homogeneous material. To extend the RSNM to multi-phase heterogeneous material, {in particular a} plain weave laminate, springs connected to neighbors 1,2,3 and 4 of a lattice point have been assigned as hard or soft springs based on the region of that location (Fig.~\ref{img:bond_direction_color}). The regions which are rich in $y$-fibers (R3) have hard vertical and hard left adjacent diagonal springs, while other two springs are soft, and all others are soft springs. Similarly, regions which rich in $x$-fibers (R2) have hard horizontal and left adjacent diagonal springs while other springs are soft, matrix rich regions have all soft springs (R1) and regions rich in $x$- and $y$-fibers have all hard springs (R4) respectively.

\subsection{\label{subsec:elasticmodel}Elastic Modeling}

\begin{figure}
	\includegraphics[width=\columnwidth]{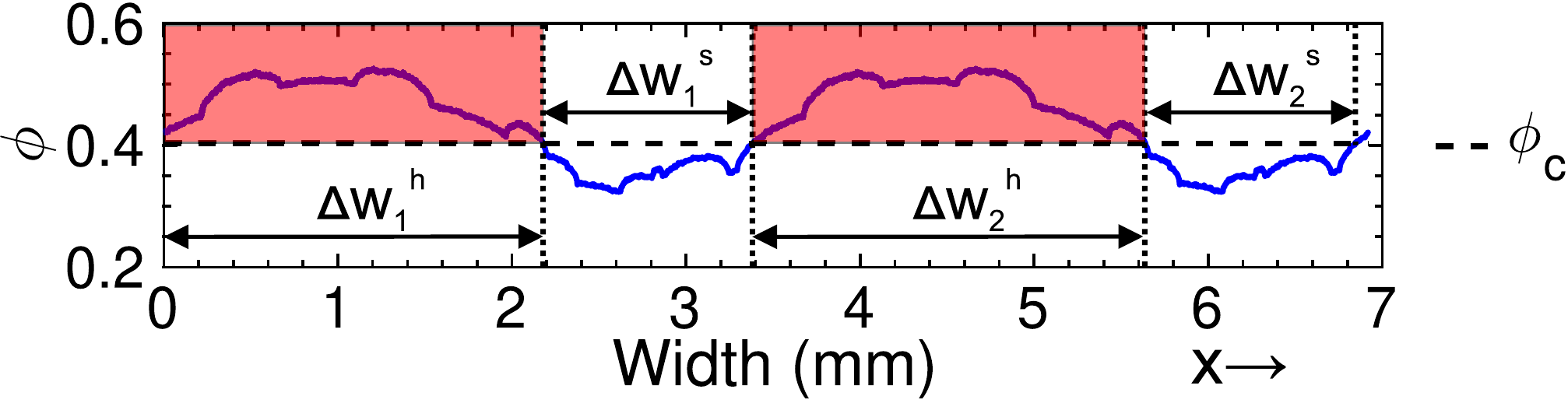}
	\centering
	\caption{Schematic diagram showing the discretization of the laminate into hard (shaded region) and soft phases based on the choice of threshold, $\phi_c$ for $\phi_{\rm min}<\phi_c<\phi_{\rm max}$.\label{img:f_illus}}
\end{figure}

To estimate the elastic behavior of hard and soft springs, we proceed as follows. The elastic behavior of hard and soft springs depends on the matrix, the reinforcements, on the choice of the critical threshold $\phi_c$, and the spatial distribution of $\phi$ (x-fiber) and $\phi$ (y-fiber). The choice of $ \phi_c $, other than being constrained to be between $ \phi_{\rm max} $ and $ \phi_{\rm min} $ of y-fiber, is arbitrary. However, the macroscopic elastic behavior of the composite cannot depend on the choice of $ \phi_c $. For this purpose, elastic moduli of hard regions ($ E_{h} $), represented by hard springs, and soft regions ($ E_{s} $), represented by soft springs, of the laminate at different $ \phi_{c} $ are calculated using the rule of mixtures. {A given choice of $ \phi_c $ decides the fraction $f$ of the laminate which will be designated as hard. The fraction $f$ of the laminate for a chosen $\phi_{c}$ is evaluated as
\be
f = \dfrac{\sum \Delta w_i^h}{\sum \Delta w_i^h + \sum \Delta w_i^s},
\label{eqn:f_eval}
\ee
where $\Delta w_i^h$ and $\Delta w_i^s$ denote the hard and soft phases respectively as defined in Fig.~\ref{img:f_illus}. To ensure effective elastic behavior of the laminate to be independent of the choice of $\phi_c$,} the elastic modulus of the laminate, $E_{\rm laminate}$, is then given by the rule of mixtures to be
\begin{equation}
E_{\rm laminate} = E_{h}f+E_{s}(1-f),
\label{eqn:rom_eff}
\end{equation}
We assume that a quadratic dependence well captures the dependence of $E_h$ and $E_s$ on the fraction $f$:
\bea
E_{h}(f)&=&a_{0}+a_{1}f+a_{2}f^2,
\label{eqn:eh}\\
E_{s}(f)&=&b_{0}+b_{1}f+b_{2}f^2.
\label{eqn:es}
\eea
The six constants $a_i$, $b_i$ can be determined as follows. Eq.~(\ref{eqn:rom_eff}) now expresses $E_{\rm laminate}$ as a polynomial of degree three in $f$. Since $E_{\rm laminate}$ cannot depend on the choice of $\phi_c$ (and hence $f$), each of the polynomial coefficients, other than the constant term, must be zero. This gives $b_0=E_{\rm laminate}$, $a_0=b_0-b_1$, $a_{1}=b_{1}-b_{2}$, $a_{2}=b_{2}$. Thus, only two constants need to be determined. It is easy to check that the solution is consistent with the constraint that when $f=0$, i.e. $\phi_c=\phi_{\rm max}$, then $E_{s}=E_{\rm laminate}$. Likewise for the constraint when $f=1$, i.e. $\phi_{c}=\phi_{\rm min}$, then $E_{h}=E_{\rm laminate}$.

Two more conditions obeyed by the constraint may be obtained by examining the limits $f \to 0$, and $f \to 1$ more closely. As the volume fraction $\phi$ of the y-fibers vary in space, the local elastic modulus of the laminate, using the rule of mixtures, varies between its maximum and minimum values as
\bea
E_{\rm max}&=&\phi_{\rm max} E_{\rm fibers}+(1-\phi_{\rm max})E_{\rm matrix},
\label{eqn:rom_max}\\
E_{\rm min}&=&\phi_{\rm min} E_{\rm fibers}+(1-\phi_{\rm min}) E_{\rm matrix}.
\label{eqn:rom_min}
\eea
When $\phi_c \to \phi_{\rm min}$, i.e. $f \to 1$, the material represented by soft regions corresponds to the laminate part that has fibers at $\phi_{\rm min}$. This implies that
\begin{equation}
E_{s}(f=1) = E_{\rm min}.
\label{eqn:es_eq}
\end{equation}
Applying the same argument to the limit $\phi_c \to \phi_{\rm max}$, i.e. $f \to 0$, we obtain
\begin{equation}
E_{h}(f=0) = E_{\rm max}.
\label{eqn:eh_eq}
\end{equation}
Eqs.~(\ref{eqn:es_eq}) and (\ref{eqn:eh_eq}) provide us with the two extra conditions that allow us to determine the constants $a_i$, $b_i$. Thus, $E_h$ and $E_s$ are determined in terms of $E_{\rm fibers}$, $E_{\rm matrix}$, and $E_{\rm laminate}$. 

We assume that both hard and soft regions have the same Poisson's ratio $\nu$. This, in turn, allows us to determine the values of the spring constants $k$ and $c$ for both hard and soft springs using Eqs.~(\ref{eqn:e_relation}) and (\ref{eqn:nu_relation}). The numerical values for the elastic constants are taken to be the same as reported for a typical plain weave laminate (for example, see Ref.~\cite{dhatreyi-2014}). These are tabulated in Table~\ref{table:rve_prop}.
\begin{table}
	\centering
	\caption{ Effective material properties for a typical plain weave laminate (taken from Ref.~\cite{dhatreyi-2014}).}\label{table:rve_prop}
	\begin{ruledtabular}
		\begin{tabular} {m{3.5cm} m{3.5cm}}
			Properties & Values \\
			$ E_{\rm matrix} $ (GPa) & 3.12 \\
			$ E_{\rm fibers}$ (GPa) & 136.5 \\
			$ E_{\rm laminate} $ (GPa) & 46.7 \\
			$ \nu $ (laminate) & 0.0654 
		\end{tabular}
	\end{ruledtabular}
\end{table}

\begin{figure}\centering
	\includegraphics[width=\columnwidth]{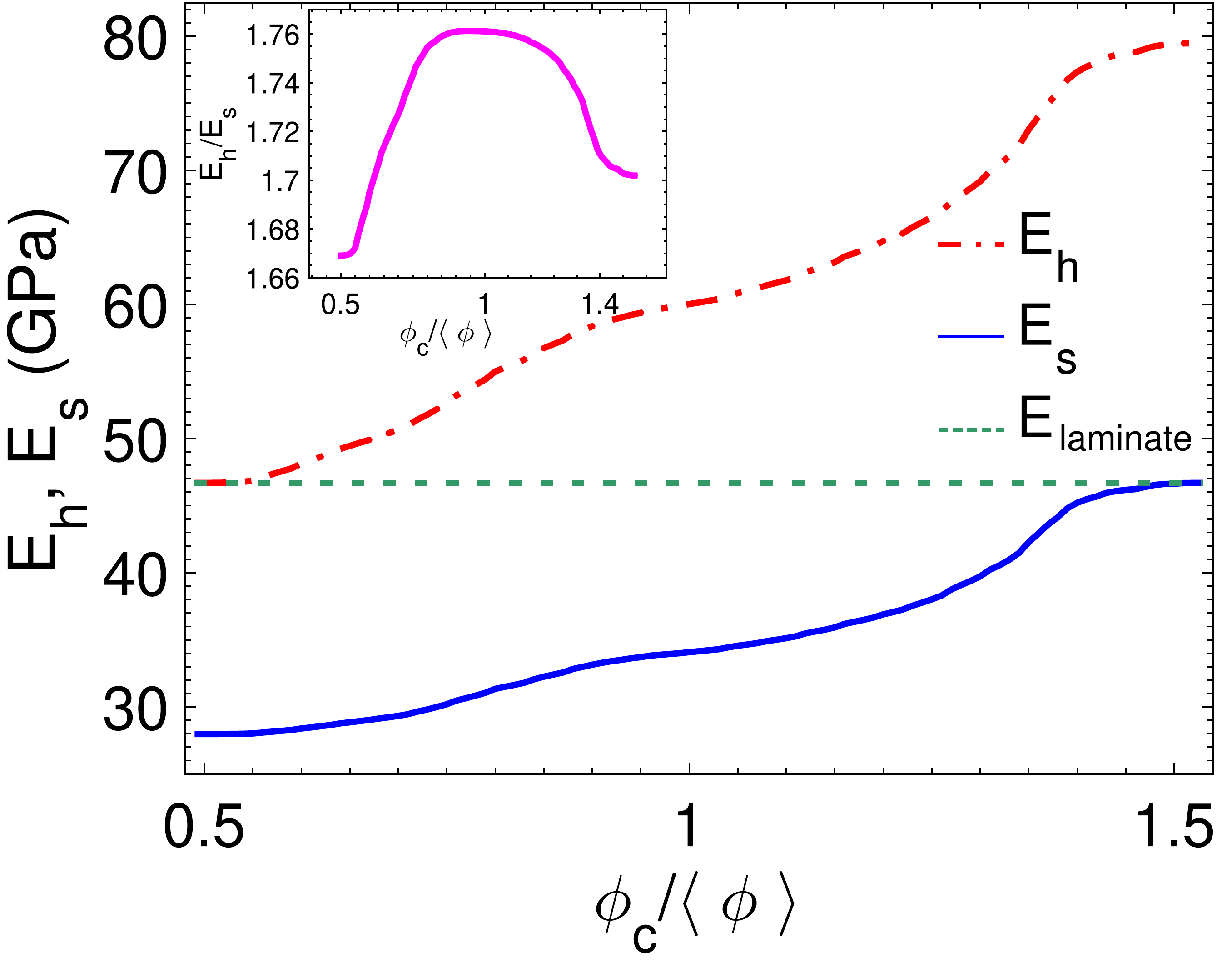}
	\caption{Elastic moduli of hard region, $E_h$, and soft region, $E_s$, as a function of the threshold $\phi_c$ [see Eqs.~(\ref{eqn:eh}) and (\ref{eqn:es})]. $E_h$ and $E_s$ satisfy the constraint that $E_{\rm laminate}$ is constant for any choice of $\phi_c$. Inset: The ratio $E_h/E_s$ has a maximum at $\phi_c=\langle \phi \rangle$ (of y-fiber). \label{img:eh_es_var_random}}
\end{figure}

The solution from the above quadratic interpolation of $ E_{h} $ and $ E_{s} $ at different $ \phi_c $ for representative laminate is plotted in Fig.~\ref{img:eh_es_var_random}. That this scheme works well and reproduces the elastic response correctly can be seen from the macroscopic stress-strain curves shown later [see Fig.~\ref{img:fd_comp_micro}].
\begin{figure}
	\includegraphics[width=0.5\columnwidth]{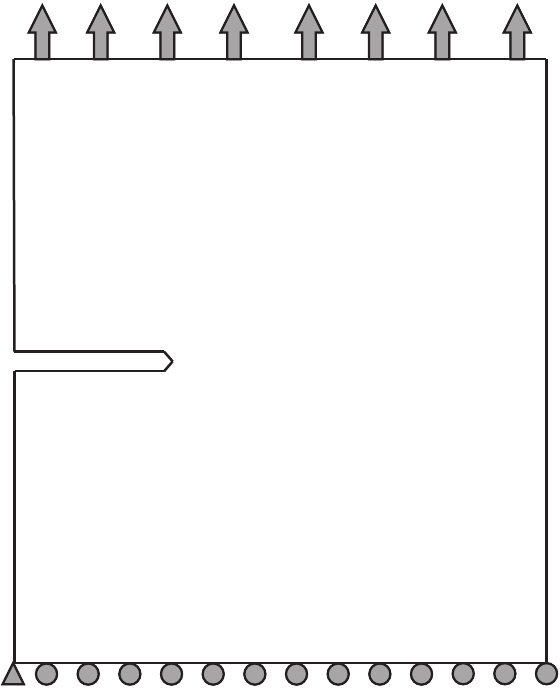}
	\centering
	\caption{A schematic diagram of the loading condition. The left bottom corner is pinned while the other lattice points on the bottom corner can only displace in the horizontal direction. \label{img:loading_condition}}
\end{figure}

\subsection{\label{subsec:md}Simulation procedure}

The lattice points have two translational degrees of freedom. To simulate mode I fracture, lattice points at the top edge of the domain was given upward displacement while those at the bottom edge was restricted to translate only in the x-direction (see Fig.~\ref{img:loading_condition}). In the simulation, the total macroscopic strain of 0.1 was applied incrementally in 400 steps. In each step, the top row of lattice particles were displaced as per the macroscopic strain. The resulting deformation of the spring network was evaluated iteratively. In the iterative procedure, the equation of motion
\be 
\overrightarrow{a_i} = -\nabla_{\overrightarrow{r_i}} \Phi - \gamma \frac{d{\overrightarrow{r_i}}}{dt},
\label{eqn:motion_eqn}
\ee
where $\overrightarrow{a_i}$ is the acceleration of lattice site $i$, {$\gamma = 0.25$,} is the damping coefficient which drives the system to equilibrium and the mass has been set to $1$, is integrated using the 
velocity-verlet algorithm:
\bea
	\overrightarrow{r_i}(t+\Delta t) &=& \overrightarrow{r_i}(t)(2-\gamma \Delta t)-\overrightarrow{r_i}(t-\Delta t)
	(1-\gamma \Delta t)\nonumber \\
	&&+\overrightarrow{a_i}^{\rm bond}(\Delta t)^2,
	\label{eqn:verlet_algo}
	\eea
for each incremental time step, {$ \Delta t \approx 10^{-2}\gamma^{-1}$. We note that strains are applied quasistatically. The dissipiation term in Eq.~(\ref{eqn:motion_eqn}) is to ensure relaxation to equilibrium. The time step in these numerical integration is taken to be much smaller than $1/\gamma$ to ensure time step independence.}

Critical strain and stress approach are used for failure analysis, i.e., spring breaks if strain/stress in the respective spring becomes greater than the critical strain/stress value. If a spring breaks, the system is again brought into equilibrium until no further spring breakage occurs. To speed up computation, we implement a parallelized version of the algorithm.

\section{\label{sec:results}Results}

\subsection{\label{sec:elastic}Elastic properties}

\begin{figure}\centering
\subfloat[\label{img:stress_homogeneous_1}]{
\includegraphics[width=0.48\columnwidth]{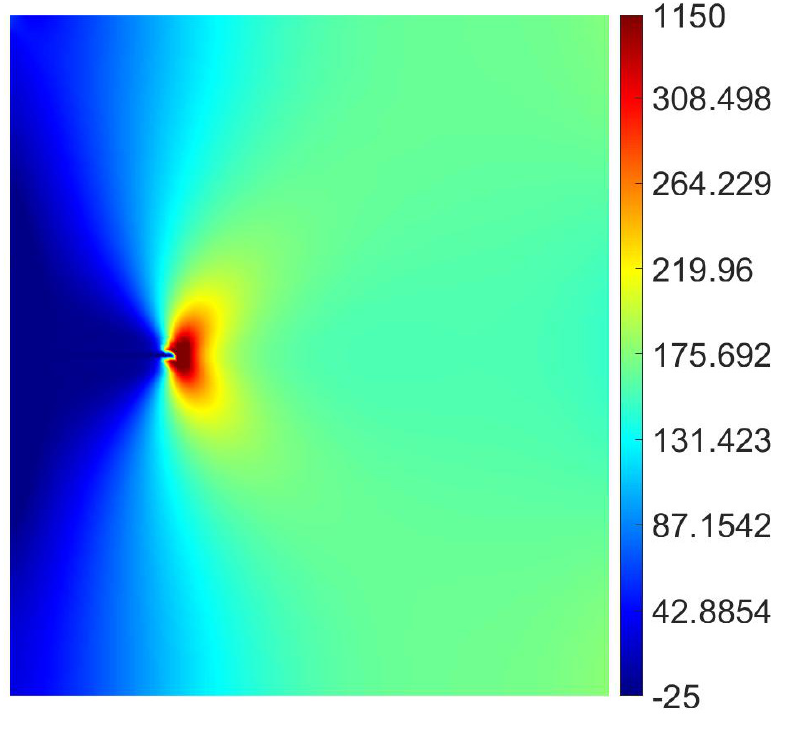}}
\hspace{0cm}
\subfloat[\label{img:random_stress_70}]{
\includegraphics[width=0.48\columnwidth]{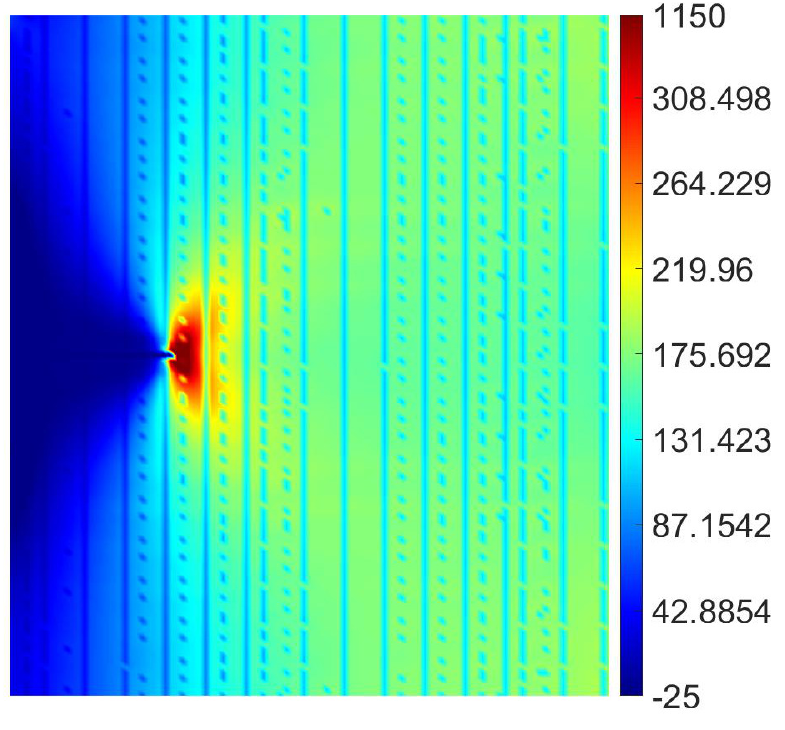}}
\vspace{0cm}
\subfloat[\label{img:random_stress_100}]{
\includegraphics[width=0.48\columnwidth]{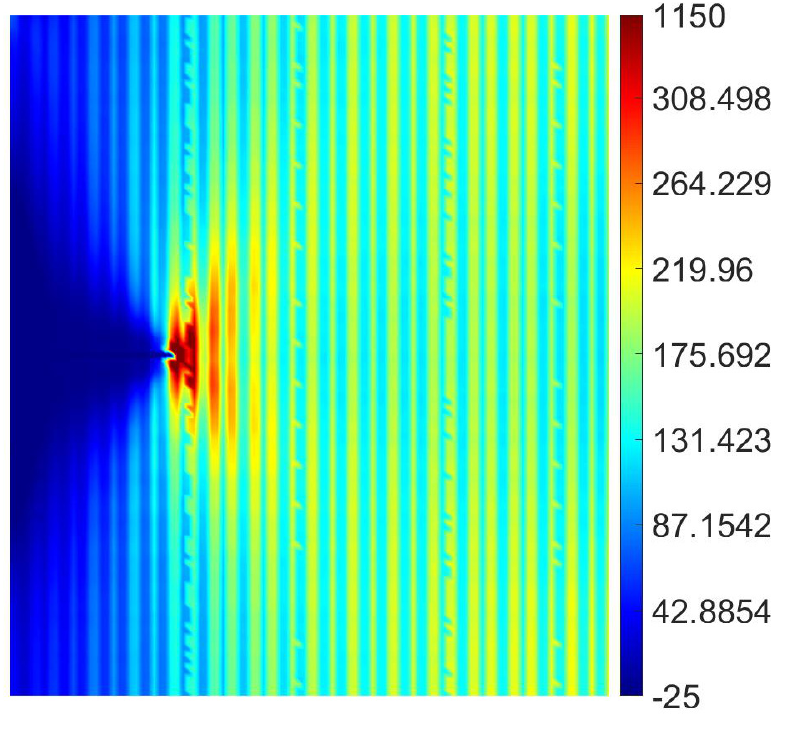}}
\hspace{0cm}
\subfloat[\label{img:random_stress_136}]{
\includegraphics[width=0.48\columnwidth]{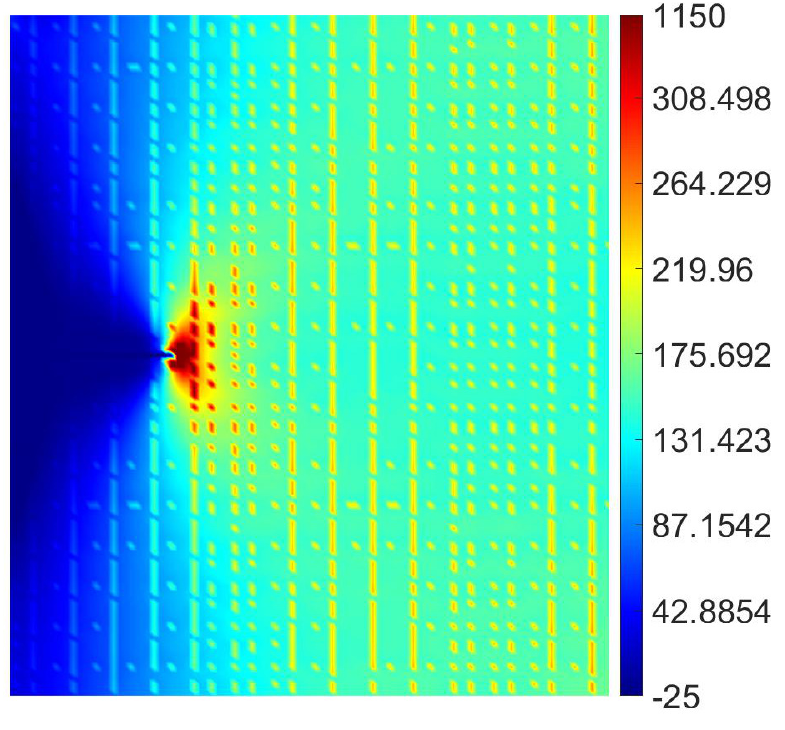}}
\caption{The spatial distribution of the stress $\sigma_{yy}$ for a laminate whose layers are offset from each other in the $xy$-plane by a random amount. The macroscopic strain is the same in all the panels which correspond to different cut-off thresholds $\phi_c$. (a) Homogeneous model. (b) $\phi_c = 0.7 \langle \phi \rangle$. (c) $\phi_c = 1.0 \langle \phi \rangle$. (d) $\phi_c = 1.36 \langle \phi \rangle$. The data are for a single realization. \label{img:stress_distribution_1}}
\end{figure}

\begin{figure}
	\includegraphics[width=7cm]{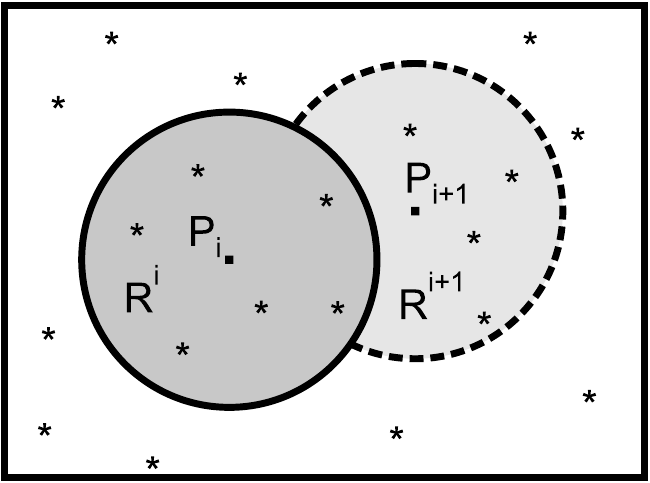}
	\centering
	\caption{A schematic diagram illustrating the moving least square method approximation to determine displacement of $P_i$ based on the weighted least square fit of the displacements of the lattice sites (shown by crosses) in a circular region $R^i$ encircling $P_i$. Another example is shown for point $P_{i+1}$. \label{img:mls}}
\end{figure}

We first study the effects of the choice of the cut-off threshold, $\phi_c$, and heterogeneity arising from the meso-structure of the laminate on the elastic stress distribution in the presence of a pre-existing crack. To study the effects of the choice of $\phi_c$, we characterize the spatial stress distribution in a representative laminate, in which the offsets in the $x$- and $y$-directions between layers are randomly chosen, for three choices of $\phi_c$, namely $\phi_c/\langle \phi \rangle = 0.70, 1.0, 1.36$. The contours of the stress $\sigma_{yy}$ in the $y$-direction (when the crack is in the $x$-direction) are shown in Fig.~\ref{img:stress_distribution_1} for a fixed macroscopic strain. These contours were obtained from the discrete displacement field using two-dimensional linear interpolation based on the moving least square method, as discussed in Ref.~\cite{liu-lsqm}. In this method, the value of displacement at an arbitrary location, $P_i$ (shown in Fig.~\ref{img:mls}) is computed from the weighted least square fit of the displacements of the lattice sites contained in a circular region $R^i$ encircling $P_i$. The point $P_i$ is then moved over the entire surface. In Fig.~\ref{img:stress_homogeneous_1}, the stress distribution is shown for the homogeneous case. This refers to the case when the Young's modulus is equal to the effective modulus of the laminate, and is constant throughout the domain such that the system has discrete translational symmetry. The stress pattern seen resembles the standard pattern that is observed in a homogeneous system with a crack, for example, see Ref.~\cite{chona-1982}. When heterogeneity is introduced [see Fig.~\ref{img:random_stress_70}-\ref{img:random_stress_136}], the stresses no longer have a smooth transition from highly intensified stresses near the crack tip to far-field lower stresses. Rather, the stress contours exhibit discontinuous behavior. Also, the highly stressed zones close to the crack tip, seen in the homogeneous case, become delocalized when heterogeneity is introduced. When the stress patterns are compared with the hard and soft bond distributions, as shown in Fig.~\ref{img:laminates_random}, it is evident that the hard and soft regions are distinguishable as they develop significantly different stresses.

\begin{figure}
	\includegraphics[width=\columnwidth]{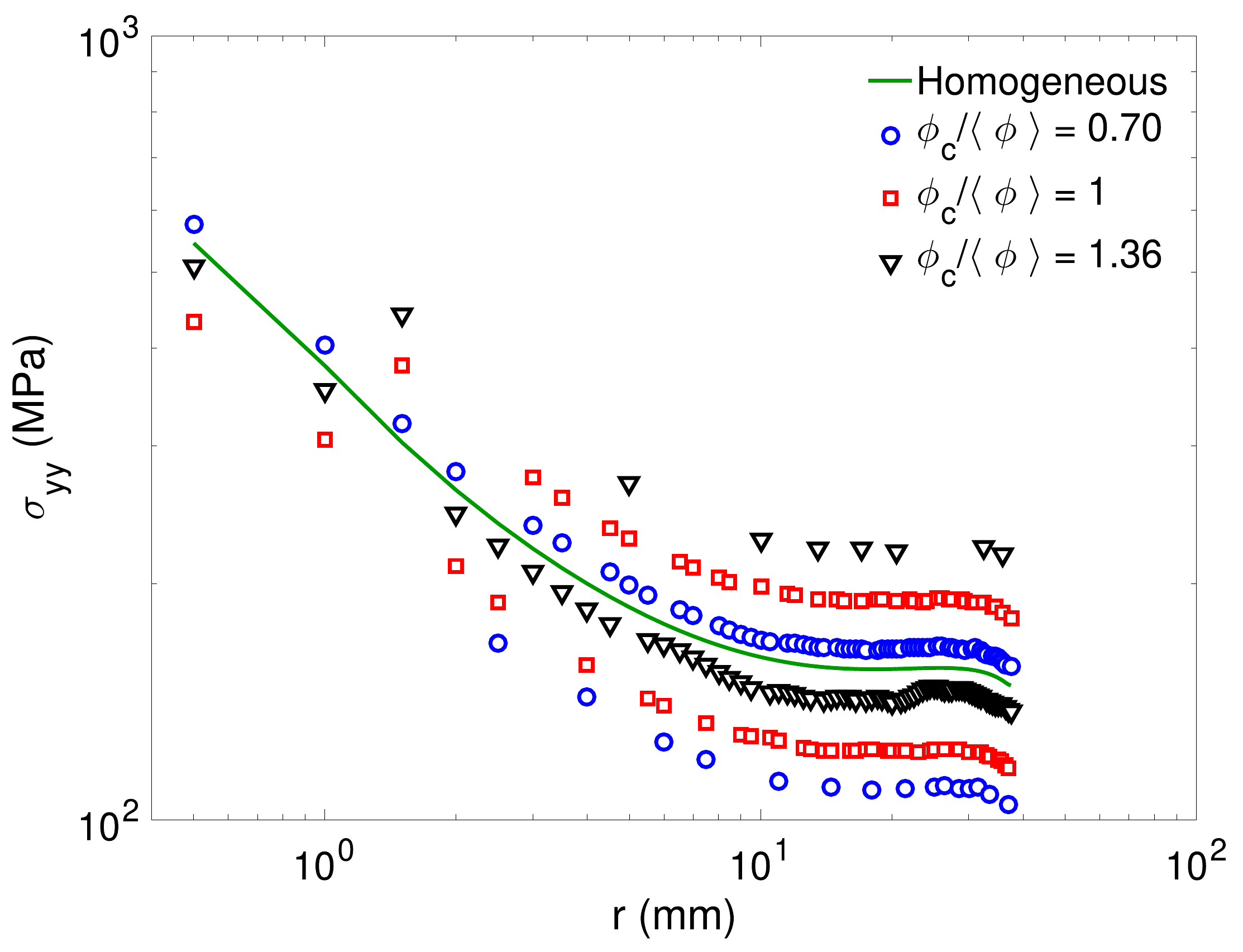}
	\centering
	\caption{The variation of the stress $\sigma_{yy}$ with distance ahead of the crack tip ($r$) for different choices of the cut-off threshold. The data are for a laminate whose offsets between layers are randomly chosen. The results for the homogeneous model are shown for comparison. \label{img:stress_ahead_crack}}
\end{figure}

The sharpest contrast between neighboring regions nearer to crack tip is observed when $ \phi_c $ is chosen to be $\langle \phi \rangle$. Increasing or decreasing $ \phi_c $ away from the mean reduces the contrast and the contours are closer to the homogeneous continuous patterns. These differences are quantified in terms of the stress normal to the plane of the initial crack ahead of the crack tip, as seen in Fig.~\ref{img:stress_ahead_crack}. Ahead of the crack tip, in the homogeneous solution, the opening stress decays rapidly in a smooth manner. In the heterogeneous combinations, the maximum deviations from the homogeneous solution are seen when $ \phi_c = \langle \phi \rangle $, which is a consequence of $ E_h/E_s $ being the highest for a given $ E_{\rm eff} $. As the threshold is increased, even though the hard phase develops significantly higher stresses, there are fewer locations that belong to the hard phase and the stresses in the softer phase are, thus, closer to the homogeneous solution. The same effect is also seen for low threshold as very few locations now belong to the softer phase. In the absence of any compelling reason to choose otherwise, as well as to enhance the effects of heterogeneity, we will consider $\phi_c=\langle \phi \rangle$ in the remainder of the paper.

\begin{figure}
	\centering
	\subfloat[\label{img:zero_offset_illus}]{
		\includegraphics[width=0.48\columnwidth]{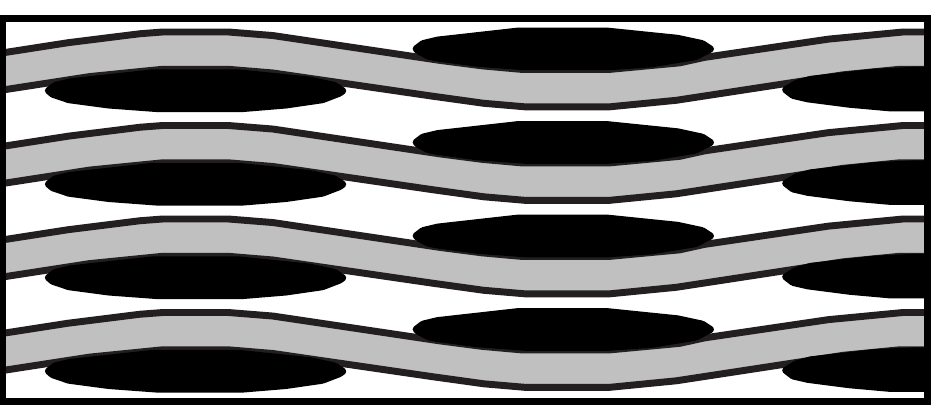}}
	\hspace{0cm}
	\subfloat[\label{img:staggered_illus}]{
		\includegraphics[width=0.48\columnwidth]{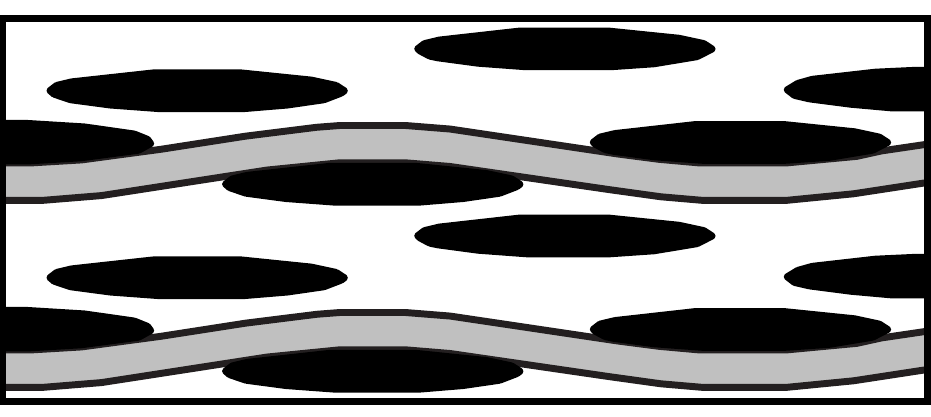}}
	\caption{Cross-sectional view of the two extreme configurations of the laminate. The panels correspond to meso-structures with (a) zero offset between layers and (b) staggered offset between layers.\label{img:offset_illus}}
\end{figure}

\begin{figure}
\centering
\subfloat[\label{img:stress_homogeneous_2}]{
\includegraphics[width=0.48\columnwidth]{stress_homogeneous.pdf}}
\hspace{0cm}
\subfloat[\label{img:zero_shift_stress_100}]{
\includegraphics[width=0.48\columnwidth]{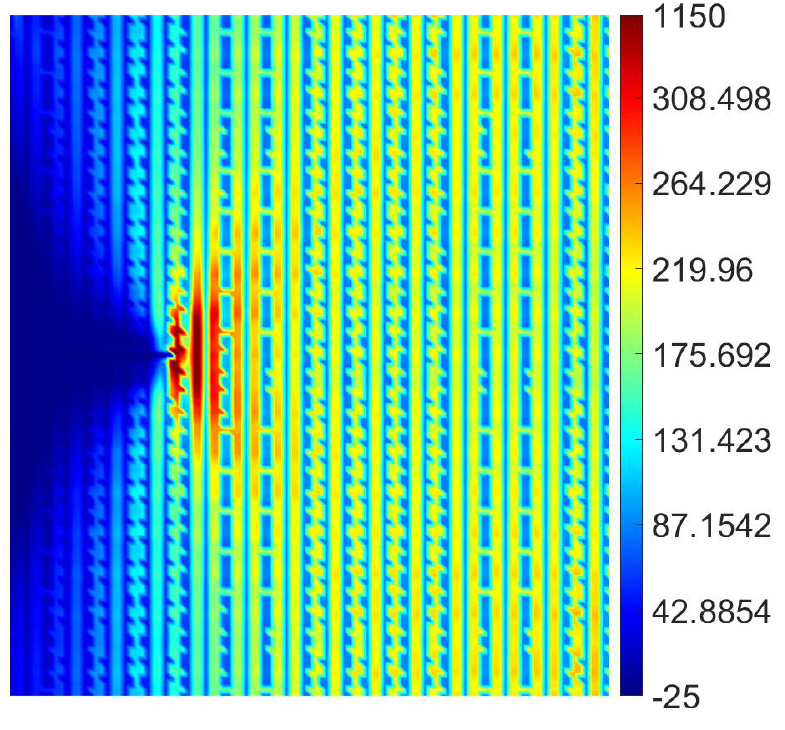}}
\vspace{0cm}
\subfloat[\label{img:staggered_stress_100}]{
\includegraphics[width=0.48\columnwidth]{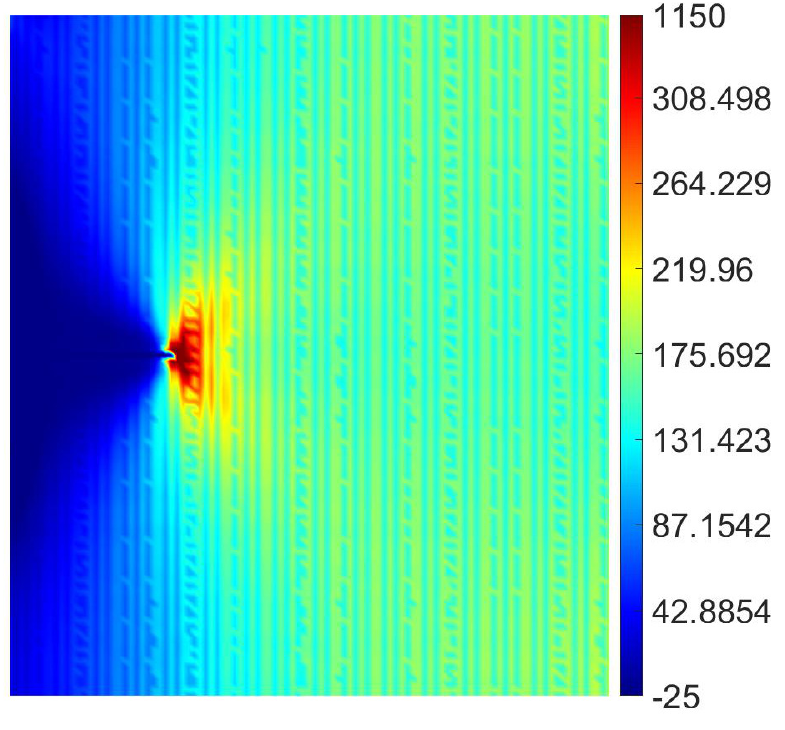}}
\caption{The spatial distribution of the stress $\sigma_{yy}$ for fixed cut-off $\phi_c= \langle \phi \rangle$ and different laminate configurations. The macroscopic strain is the same in all the panels which correspond to (a) homogeneous model (b) zero offset meso-structure and(c) staggered offset meso-structure. The data are for a single realization.
\label{img:stress_distribution_2}}
\end{figure}

For the same effective elastic modulus, the relative offset between the layers can also have a strong effect on the spatial distribution of the volume fractions of $x$- and $y$-fibers as well as its maximum and minimum limits. To establish the role of relative offset in comparison to the homogeneous material, {we consider two extreme configurations which we call as zero offset and staggered offset configurations. In the zero offset configuration, the fibers are perfectly aligned across the thickness, as shown in Fig.~\ref{img:zero_offset_illus}, such that, along the length of the laminate, the fiber volume fraction varies from maxima to zero (similar to a lamina). In the staggered offset configuration, each adjacent layer has an offset of exactly half the wavelength of the undulation in both $x$- and $y$-directions, as shown in Fig.~\ref{img:staggered_illus}, resulting in fiber volume fraction having double the frequency of the undulations}. Contours of the opening stress field $\sigma_{yy}$ are presented in Fig.~\ref{img:stress_distribution_2} for different types of laminate configurations in comparison to that for a homogeneous material. In the zero-offset configuration shown in Fig.~\ref{img:zero_shift_stress_100}, meso-structure patterns dominate the stress distribution, accompanied by significant delocalization of the high stresses near the crack tip. However, in the staggered configuration, in Fig.~\ref{img:staggered_stress_100}, even though the stress distribution has patterns of the heterogeneity, contour shapes are much closer to the homogeneous model. Higher disorder in the elastic field, as seen in the case of the zero-offset configuration contributes to its observed increase in strength and toughness, as we will see later.

\begin{figure}
	\includegraphics[width=\columnwidth]{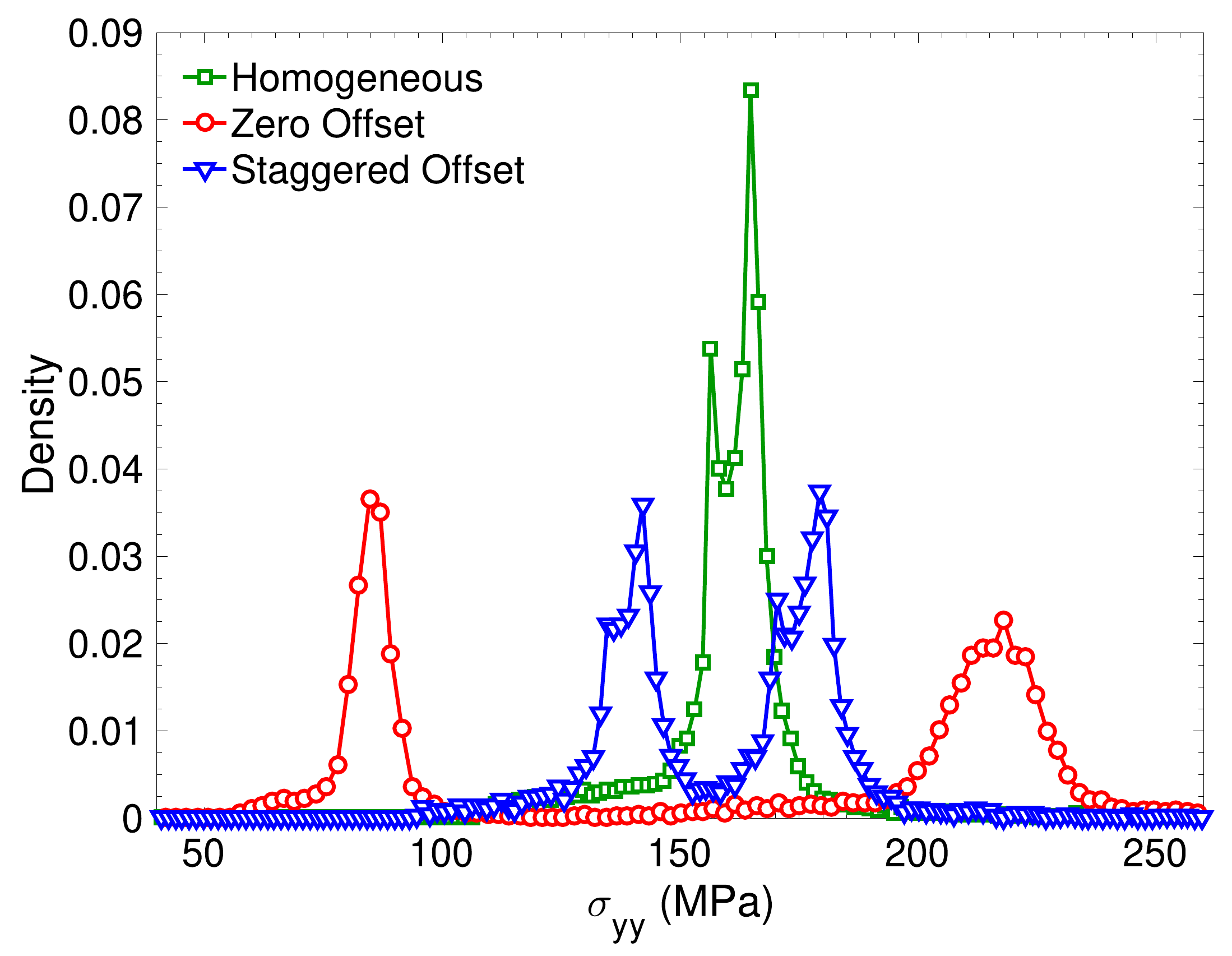}
	\centering
	\caption{Distribution of stress $ \sigma_{yy} $ of lattice points for homogeneous system and system{\tiny {\tiny }} when heterogeneity is included. All the data correspond to the same macroscopic strain for single realization. \label{img:force_distribution}}
\end{figure}

The distribution of $ \sigma_{yy} $, for the different configurations considered, is presented in Fig.~\ref{img:force_distribution}. The single peak observed in the homogeneous case corresponds to the lattice points in the region away from the crack tip. On the introduction of heterogeneity, two peaks form as the strains are compatible in the neighborhood but two different material behavior result in differences in the $ \sigma_{yy} $. In the zero offset configuration, the ratio between the elastic moduli of the two phases is largest, and thus the peaks are furthest away from each other, leading to higher stresses for a larger number of lattice points, whereas, in staggered offset, peaks are much closer. For random-offset configuration, we find that the peaks are in between (not plotted for the sake of clarity).

\subsection{\label{sec:fracture}Fracture properties}

\begin{figure} 
\centering
\subfloat[\label{img:stress_fl_crit}]{
\includegraphics[width=0.48\columnwidth]{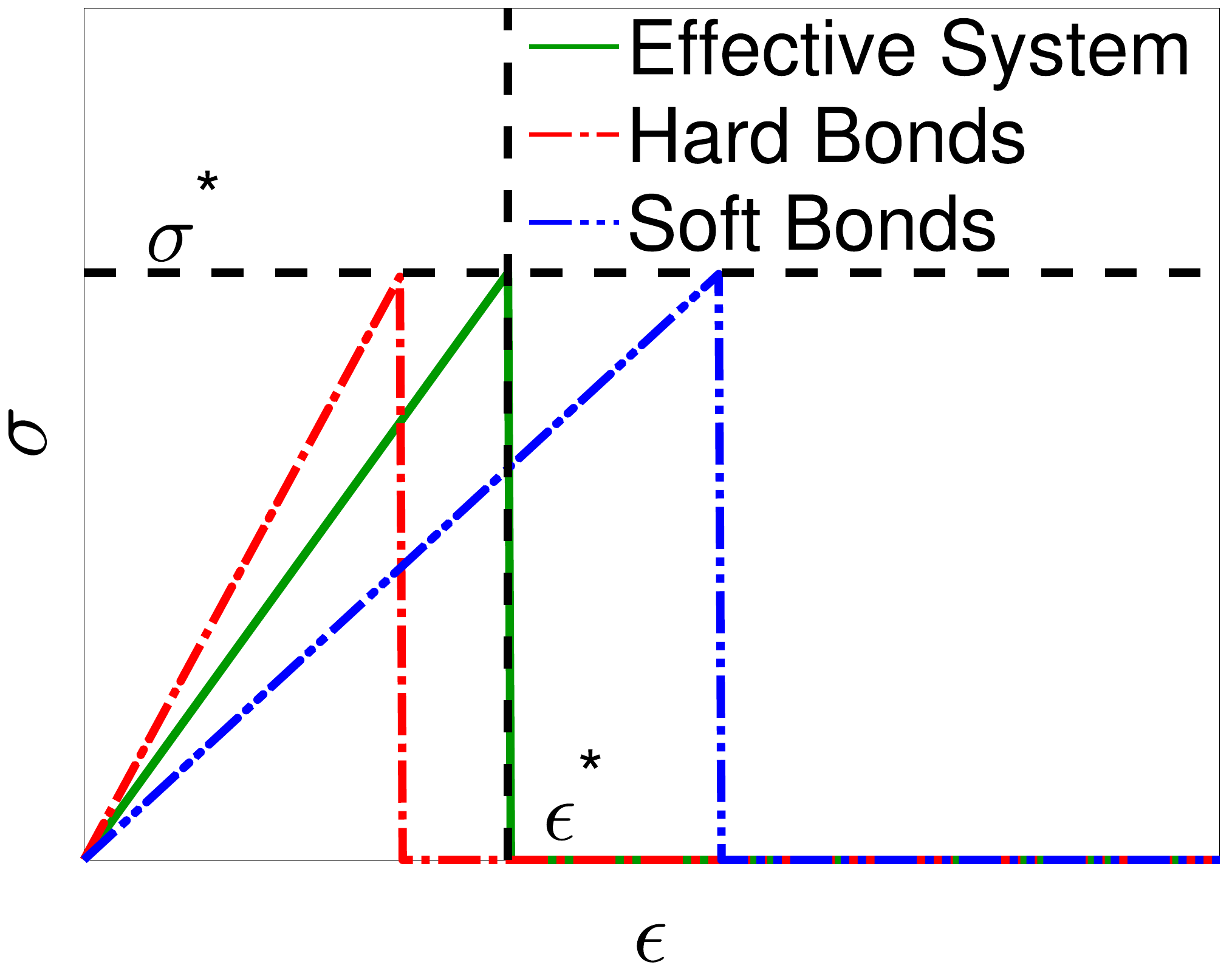}}
\hspace{0cm}
\subfloat[\label{img:strain_fl_crit}]{
\includegraphics[width=0.48\columnwidth]{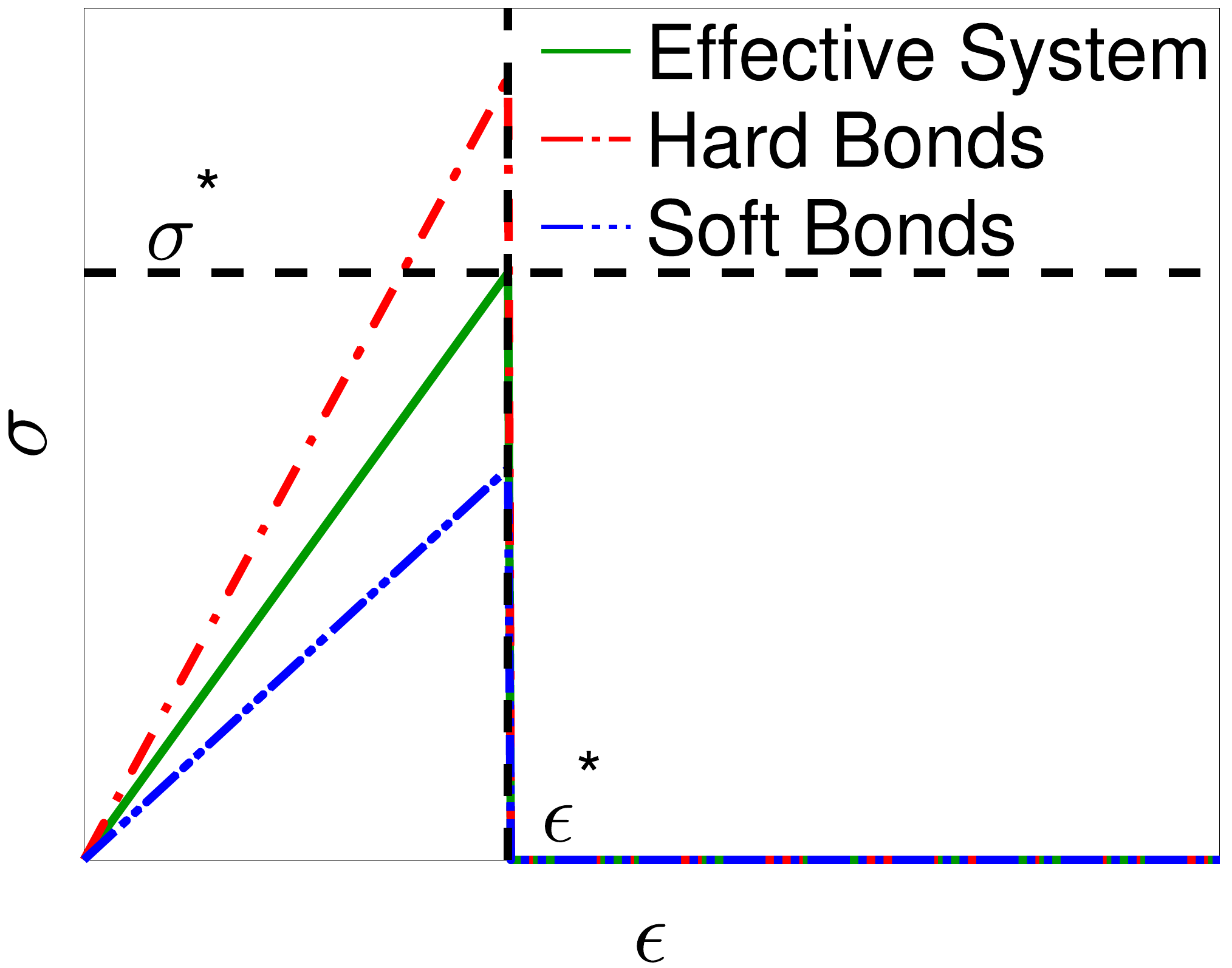}}
\caption{The stress-strain diagram for springs depicting their failure criteria based on (a) common stress threshold and (b) common strain threshold.\label{img:fail_mode}}
\end{figure}

We now study the effect of elastic heterogeneity, inherent in the laminates, on the fracture behavior. In the existing studies of fracture behavior of heterogeneous materials using network models, two main simplifications have been adopted in relation to the failure criteria for inter-particle interaction: a common threshold failure strain~\cite{dimas-2014a} or a common failure threshold stress~\cite{urabe-2010}, as shown in Fig.~\ref{img:fail_mode}. Since both elastic and failure properties of the springs affect the macroscopic response significantly, in the present work, we employ both the criteria to establish a comprehensive understanding of the role of spatial patterns in heterogeneity on its fracture process. In a common stress threshold criteria, we choose failure criteria for the bonds to be such that for all springs, the stress threshold, $ \sigma^*$, is taken to be the same, as shown in Fig.~\ref{img:stress_fl_crit}. This results in softer bonds requiring the highest energy for failure while harder bonds require the least, and the bonds of the effective homogeneous medium require energy in between. For a common strain threshold criteria, as the failure strain threshold is taken to be the same ($=\epsilon^*$) for all phases, softer bonds require, thus, the least energy for failure compared to the harder bonds, as evident in Fig.~\ref{img:strain_fl_crit}. We do simulations for laminates with zero and staggered offsets, and for comparison, we also simulate fracture response of an effective homogeneous laminate.

\begin{figure} \centering
\subfloat[\label{img:fd_comp_micro_stress}]{
\includegraphics[width=0.48\columnwidth]{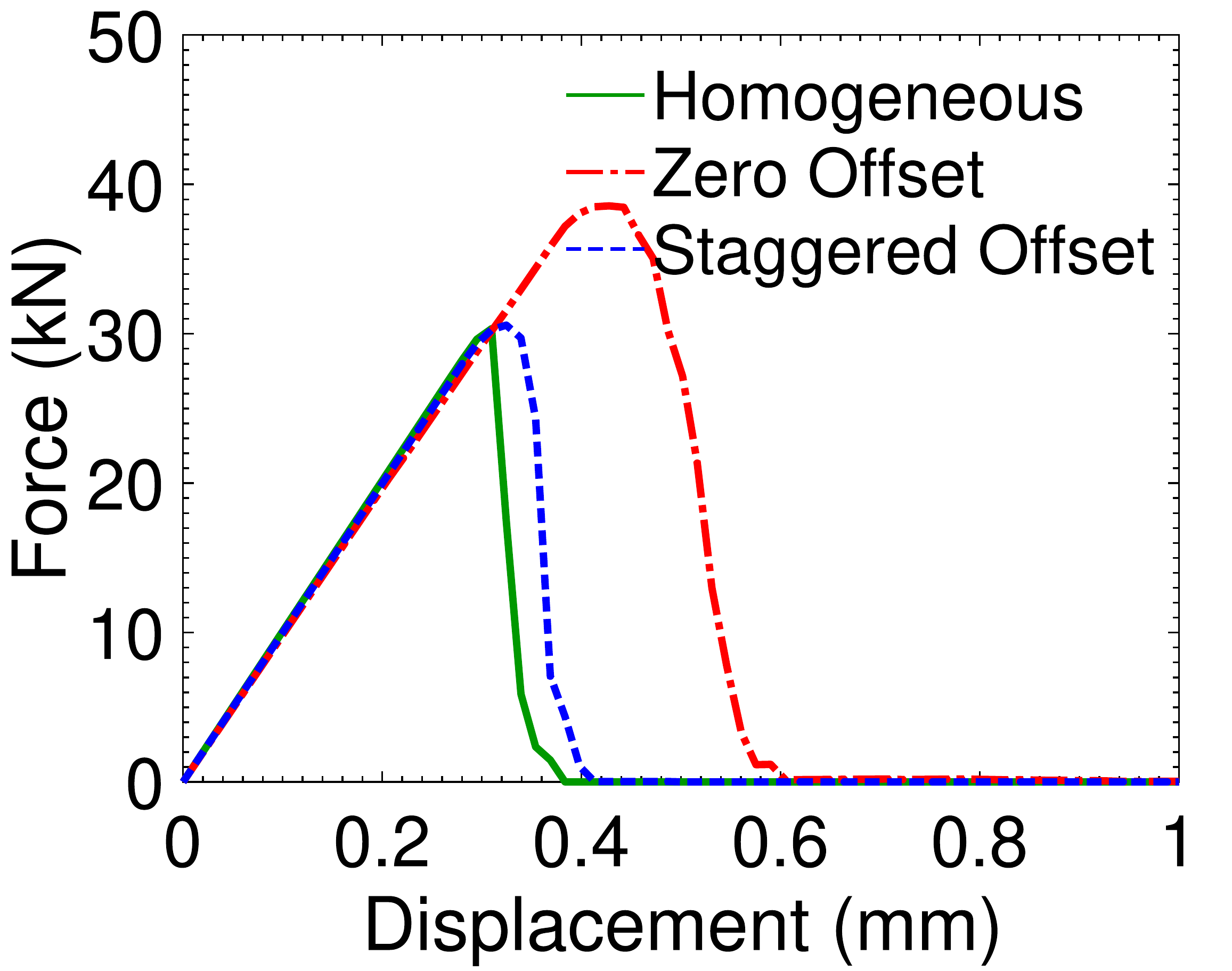}}
\hspace{0cm}
\subfloat[\label{img:fd_comp_micro_strain}]{
\includegraphics[width=0.48\columnwidth]{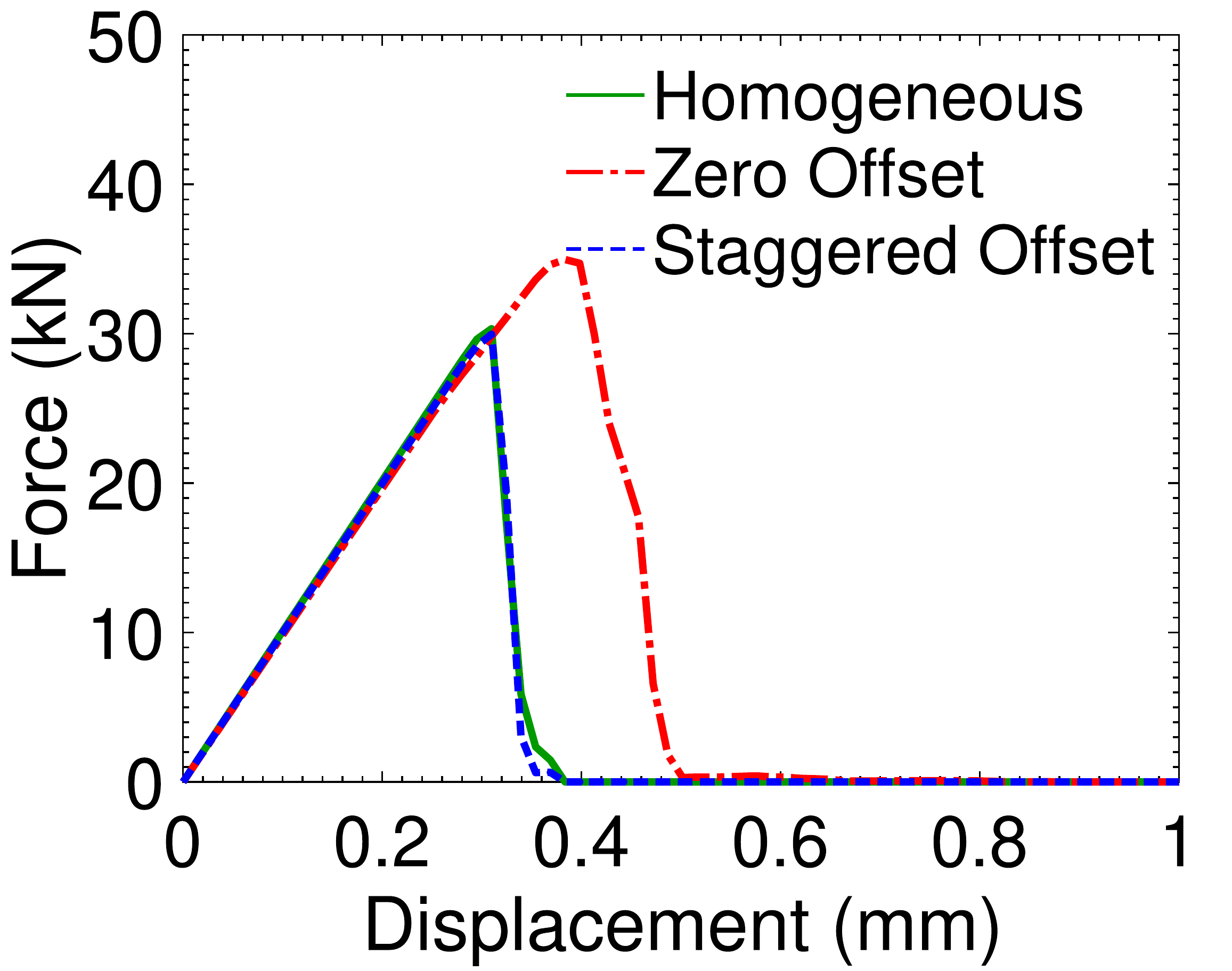}}
\caption{The force-displacement curve of laminates with different configuration for a crack of length $ a_0 = 0.28 w $. The disorder in the failure threshold corresponds to standard deviation $ 5\% $ of the mean. The panels correspond to data for (a) common stress threshold and (b) common strain threshold. \label{img:fd_comp_micro}}
\end{figure}

In addition to the elastic heterogeneity, we also account for the inherent disorder in the fracture properties, arising from microscopic variations and defects, by assigning the failure threshold strain for each spring independently from a Gaussian distribution with mean at $\sigma^*/E$ or $\epsilon^*$ as shown in Fig.~\ref{img:fail_mode} and a standard deviation taken to be within the range of zero to 10\% in all the simulations of fracture.
To show that the comparative responses of the three laminate meso-structures differ from each other, depending on the type of failure criteria chosen, we take an initial crack of length $ a_0 = 0.28 w$ and perform 25 realizations each for 3 different meso-structures: effective homogeneous, zero offset and staggered offset. We assume the disorder in the failure strain threshold to have a standard deviation of $ 5\% $ of the mean. We choose the failure strain threshold for the effective homogeneous laminate arbitrarily to be $ \epsilon^{*}= 0.025 $. From the responses based on a common failure stress threshold shown in Fig.~\ref{img:fd_comp_micro_stress}, the staggered configuation has a response very similar to the homogeneous case, while the configuration with zero offset exhibits significantly enhanced strength. The area under the response curve, a measure of the material's toughness, is also highest for the zero offset configuration.

\begin{figure} \centering
\subfloat[\label{img:bond_count_zs}]{
\includegraphics[width=0.48\columnwidth]{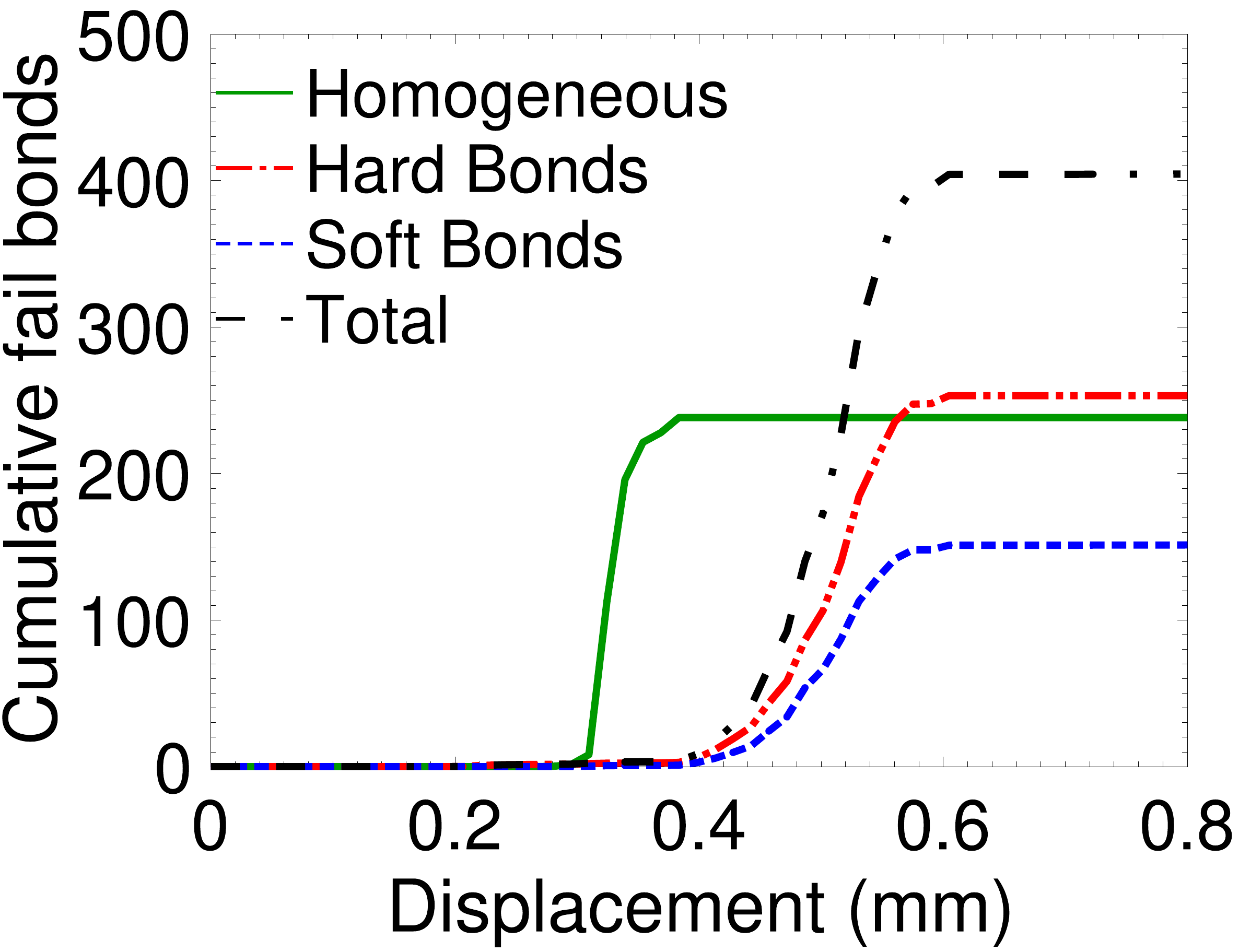}}
\hspace{0cm}
\subfloat[\label{img:bond_count_stg}]{
\includegraphics[width=0.48\columnwidth]{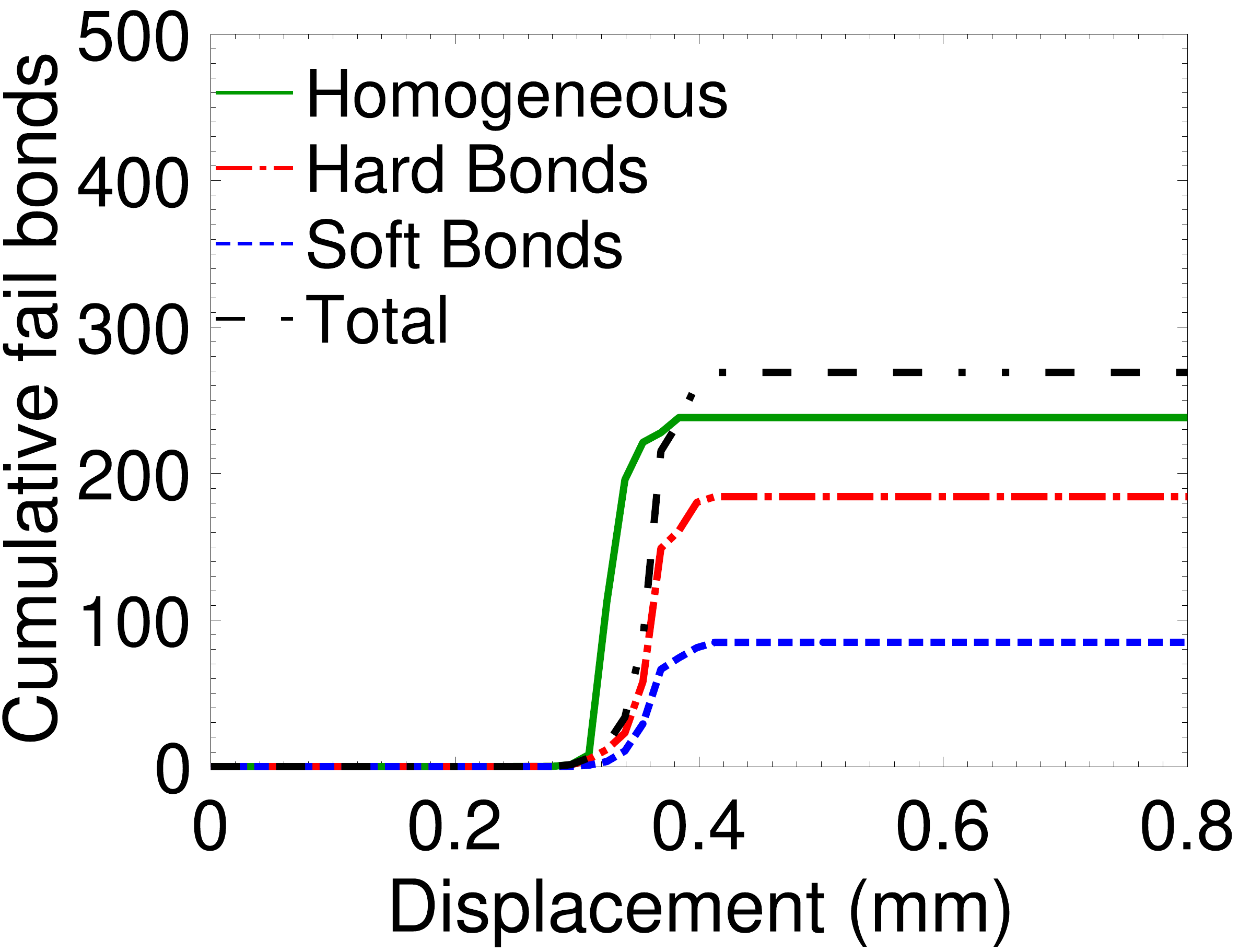}}
\vspace{0cm}
\subfloat[\label{img:bond_count_zs_strain}]{
\includegraphics[width=0.48\columnwidth]{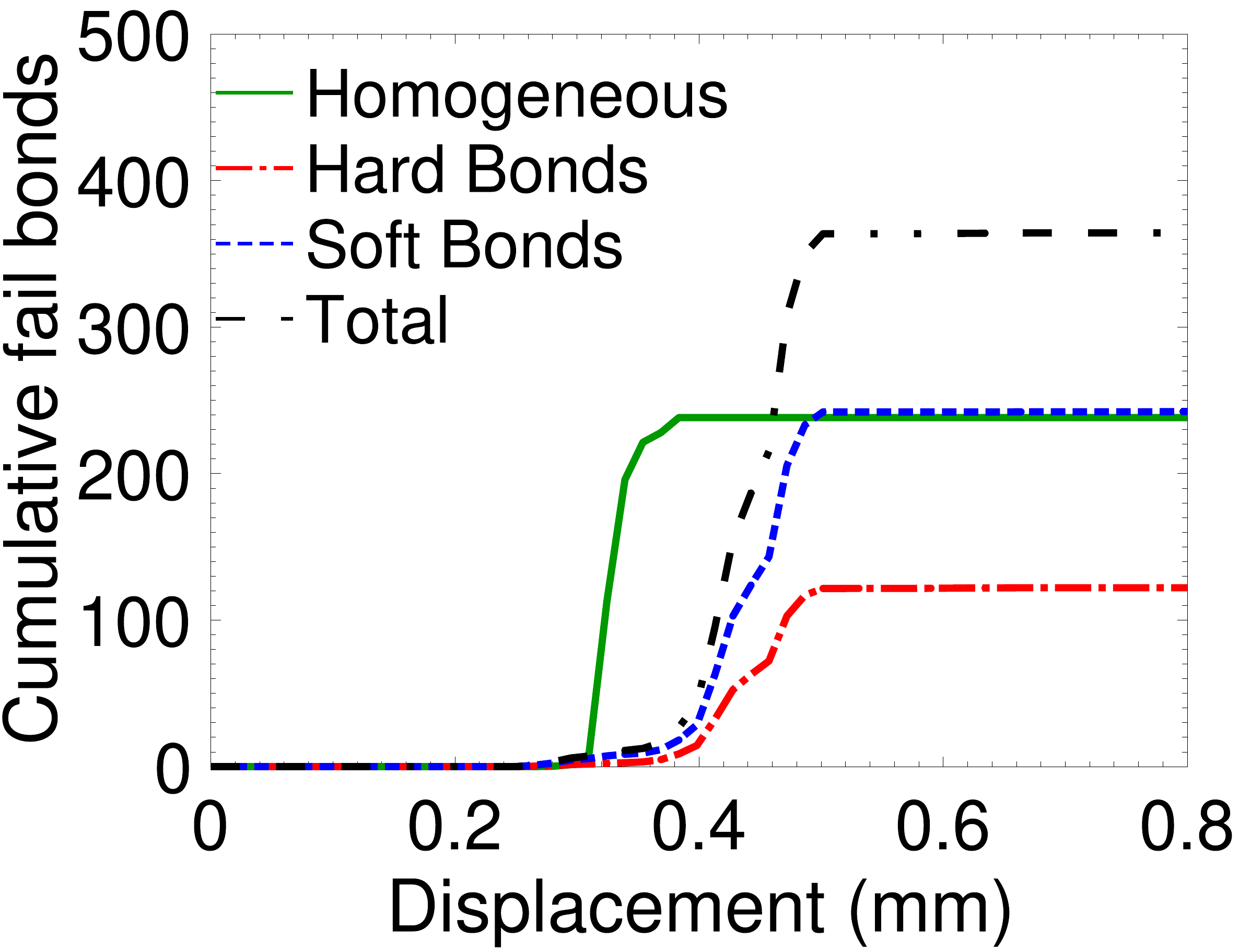}}
\hspace{0cm}
\subfloat[\label{img:bond_count_stg_strain}]{
\includegraphics[width=0.48\columnwidth]{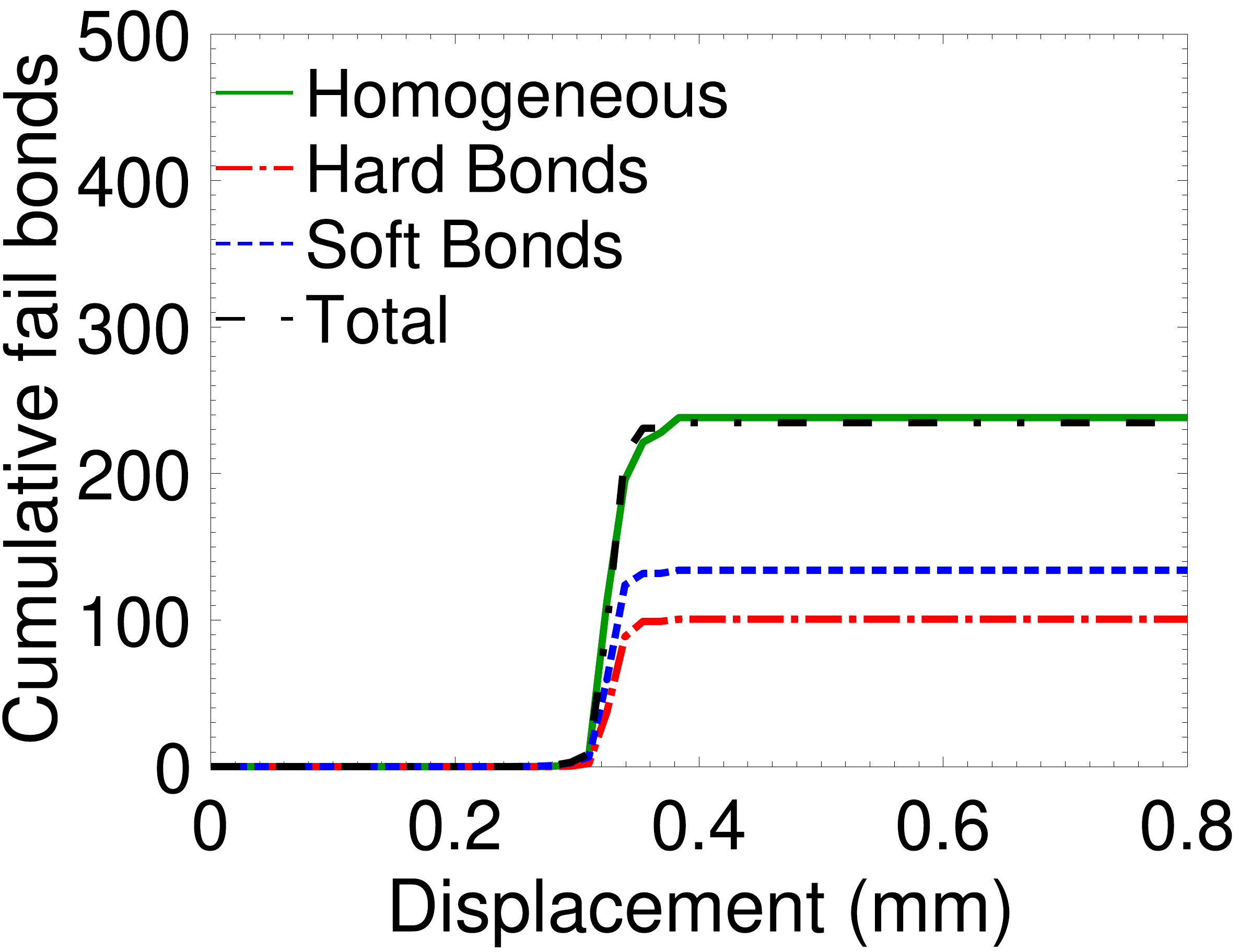}}
\caption{The cumulative number of broken or failed bonds with increasing displacement. The panels (a) and (c) are for the zero-offset meso-structure while the panels (b) and (d) are for the staggered offset meso-structure. The panels (a) and (b) correspond to a common stress threshold failure criteria while the panels (c) and (d) correspond to a common constant strain threshold criteria.\label{img:bond_count}}
\end{figure}

Even though the meso-structure with zero-offset exhibits toughening in both the common stress and strain threshold criteria, there are differences in dissipative mechanisms when considering the relative contribution of hard bonds and soft bonds towards the total energy that is dissipated during crack growth. The number of extensional springs failing during crack growth is primarily driven by the energy costs involved. In the common stress threshold criterion, since the soft bonds are tougher, fewer soft bonds break compared to hard bonds. This is true for both zero-shift as well as staggered configuration, as is evident from Fig.~\ref{img:bond_count_zs} and~\ref{img:bond_count_stg}. Similarly, the common strain threshold criterion implies the harder bonds are tougher resulting in fewer hard bonds breaking compared to soft bonds, as can be seen for both meso-structures in Fig.~\ref{img:bond_count_zs_strain} and~\ref{img:bond_count_stg_strain}, respectively. However, for both the criteria, the total number of failed springs are higher for the zero-shift configuration than the effective homogeneous counterpart, as seen in Fig.~\ref{img:bond_count_zs} and~\ref{img:bond_count_zs_strain}, resulting in its enhanced toughness. 

\begin{figure}\centering
\subfloat[\label{img:mesh_fail_zero_shift}]{
\includegraphics[width=0.96\columnwidth]{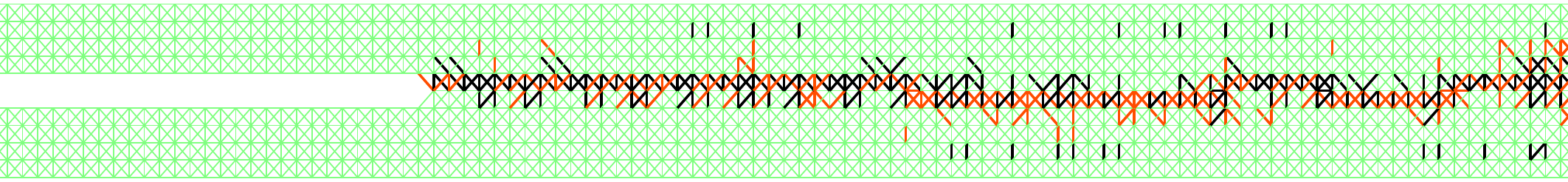}}
\vspace{0cm}
\subfloat[\label{img:mesh_fail_staggered}]{
\includegraphics[width=0.96\columnwidth]{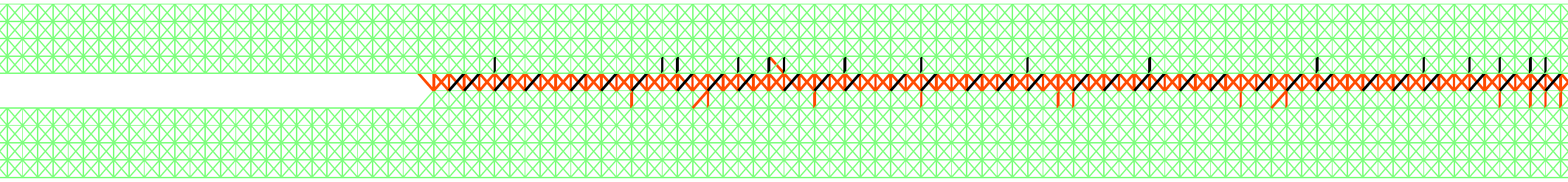}}
\vspace{0cm}
\subfloat[\label{img:mesh_fail_zero_shift_strain}]{
\includegraphics[width=0.96\columnwidth]{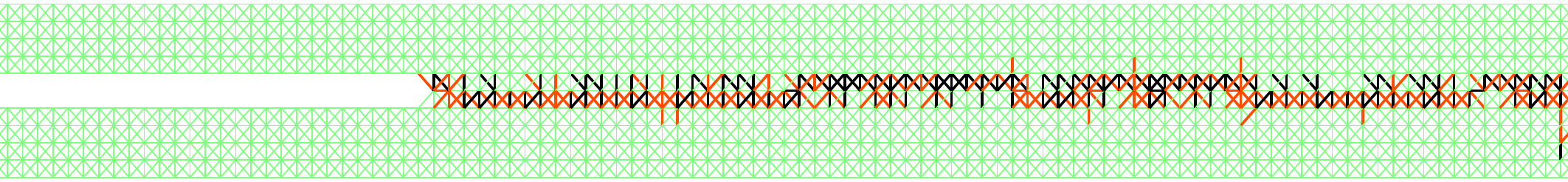}}
\vspace{0cm}
\subfloat[\label{img:mesh_fail_staggered_strain}]{
\includegraphics[width=0.96\columnwidth]{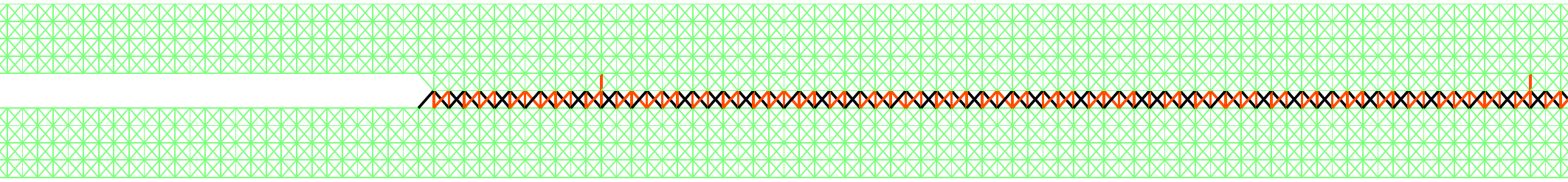}}
\caption{The final geometry of the crack for different meso-structures and failure criteria. The {orange [grey], black [black] and light green [light grey]} colors represent broken hard, soft, and unbroken bonds respectively. {The panels (a) and (c) are for the zero-offset meso-structure while the panels (b) and (d) are for the staggered offset meso-structure. The panels (a) and (b) correspond to a common stress threshold failure criterion while the panels (c) and (d) correspond to a common constant strain threshold criterion.}\label{img:crack_path}}
\end{figure}

The spatial location of the broken bonds is shown in Fig.~\ref{img:crack_path} for representative realizations of zero-offset and staggered offset meso-structures for both common stress as well as strain threshold failure criteria. The broken hard, soft, and unbroken bonds are shown in {orange [grey], black [black] and light green [light grey]} colors respectively. The crack path is most tortuous for zero-offset configuration with a common stress failure threshold, as can be seen in Fig.~\ref{img:mesh_fail_zero_shift}. This is because the crack preferably avoids breaking soft bonds as they cost larger energy. It is also observed that a few hard bonds that are not directly connected to the crack path also fail. Thus, the damage is spread over a larger area, presumably resulting in higher toughness. The crack path for zero-offset configuration for a common strain failure criterion has a comparatively less tortuous path, though as expected, there is a clear preference for breakage of softer bonds (see Fig.~\ref{img:mesh_fail_zero_shift_strain}). Also, all the broken bonds are connected to the main crack. On the other hand, the staggered configuration, for both choices of failure criteria, exhibits a very localized path of the propagating crack along the initial crack plane, as shown in Figs.~\ref{img:mesh_fail_staggered} and \ref{img:mesh_fail_staggered_strain}. This is because $E_h/E_s$ is closer to unity and the differentiation between soft and hard bonds is minimal. 

\begin{figure} \centering
\subfloat[\label{img:force_cl_29_sd_all_stress}]{
\includegraphics[width=0.48\columnwidth]{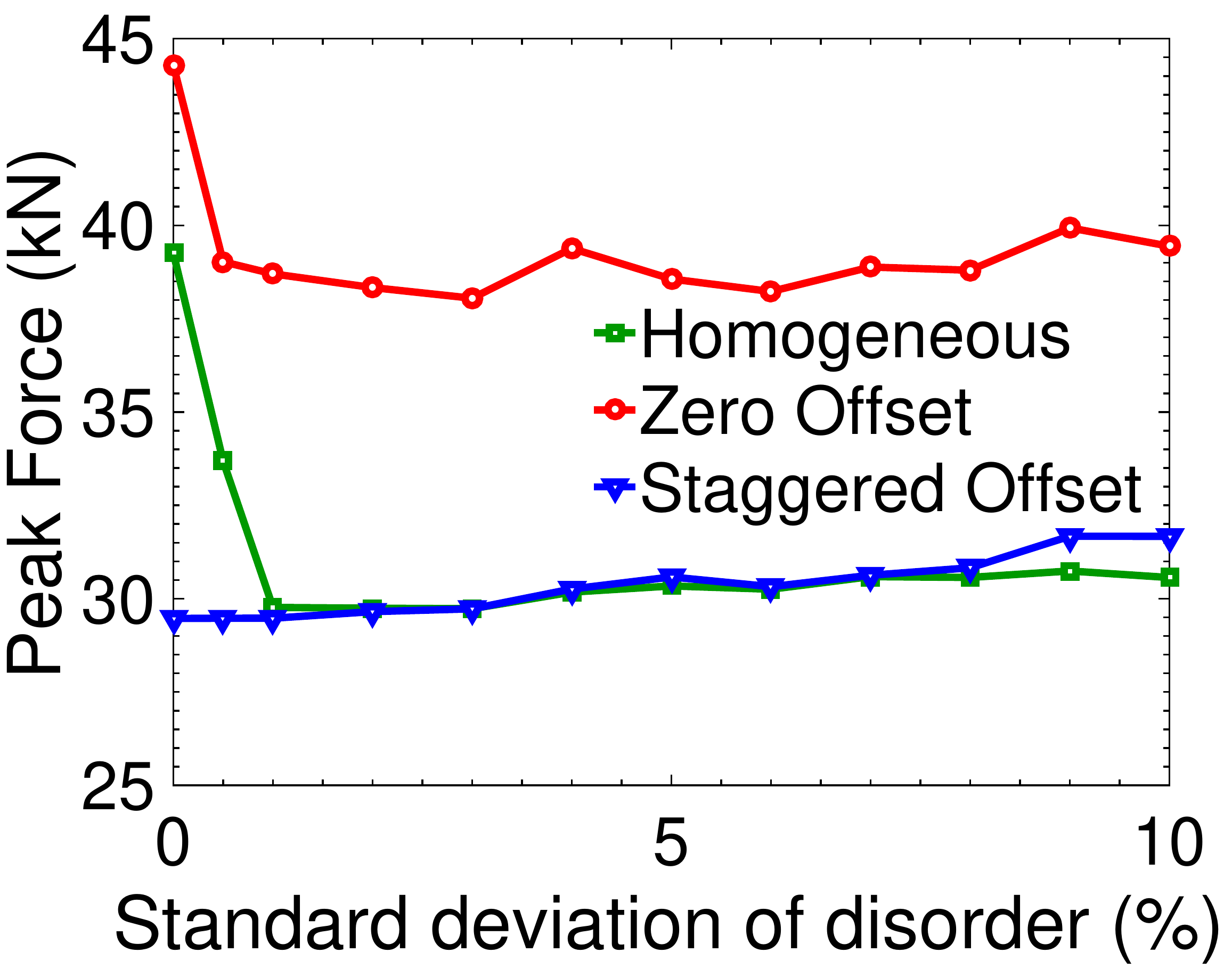}}
\hspace{0cm}
\subfloat[\label{img:toughness_cl_29_sd_all_stress}]{
\includegraphics[width=0.48\columnwidth]{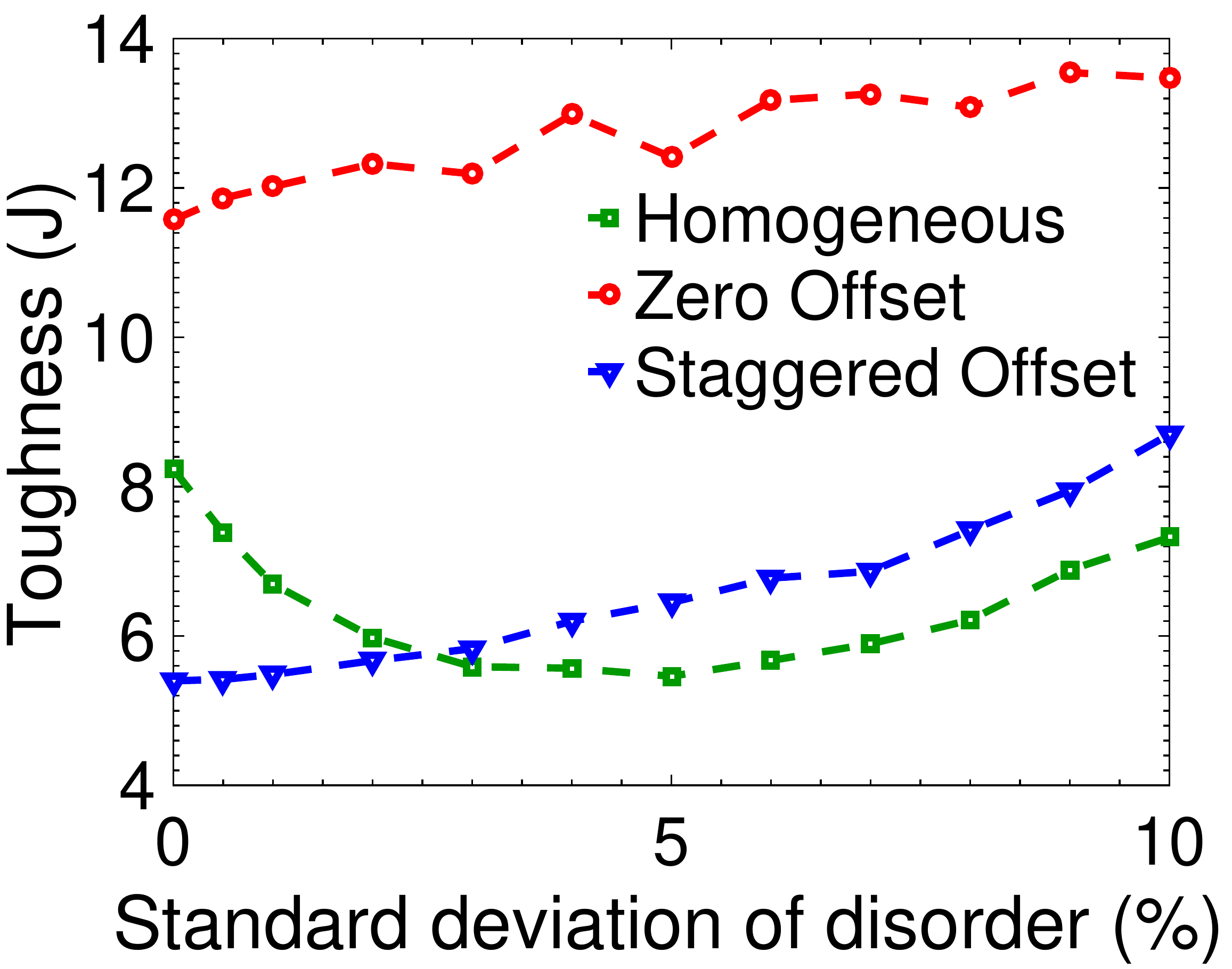}}
\vspace{0cm}
\subfloat[\label{img:force_cl_29_sd_all_strain}]{
\includegraphics[width=0.48\columnwidth]{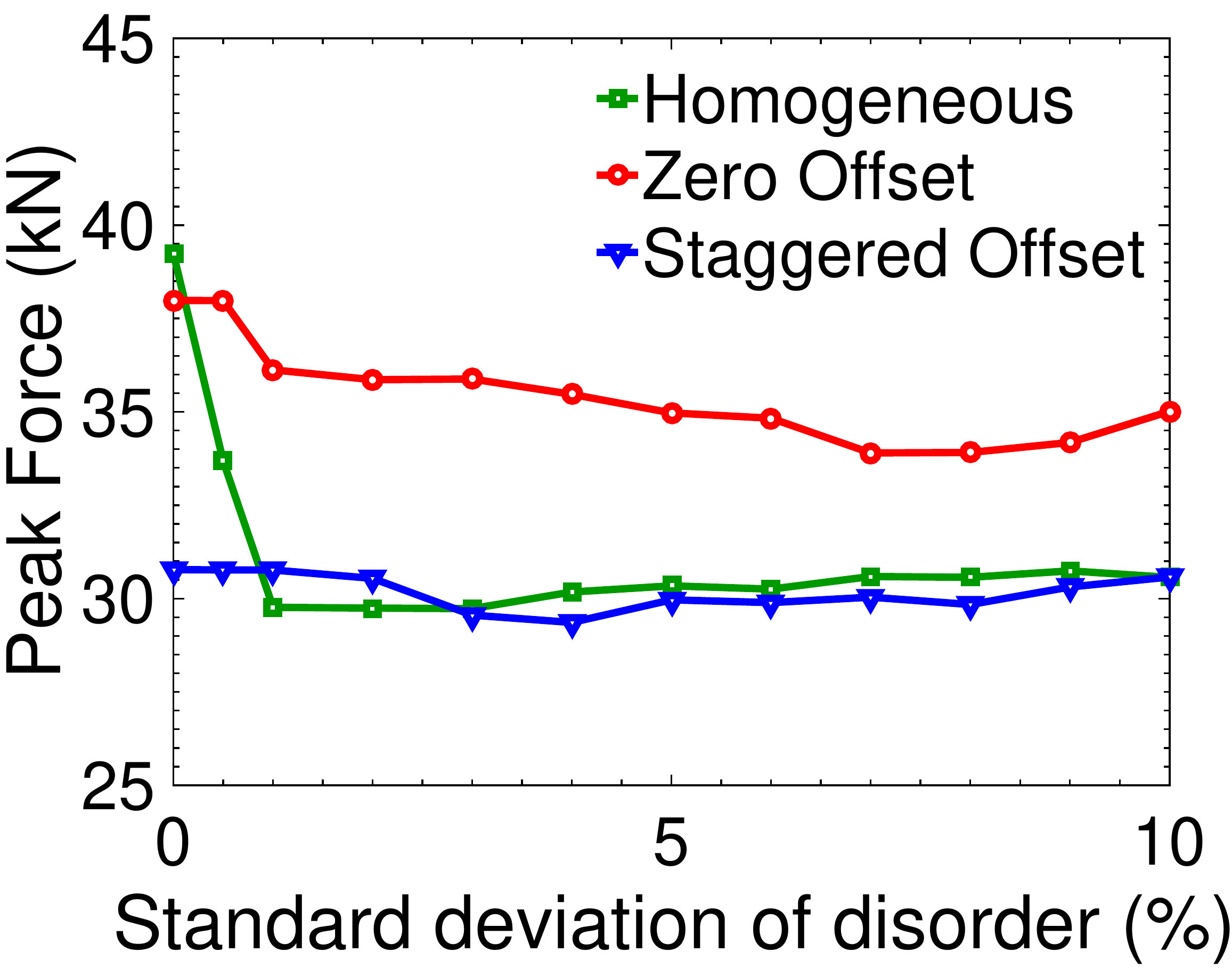}}
\hspace{0cm}
\subfloat[\label{img:toughness_cl_29_sd_all_strain}]{
\includegraphics[width=0.48\columnwidth]{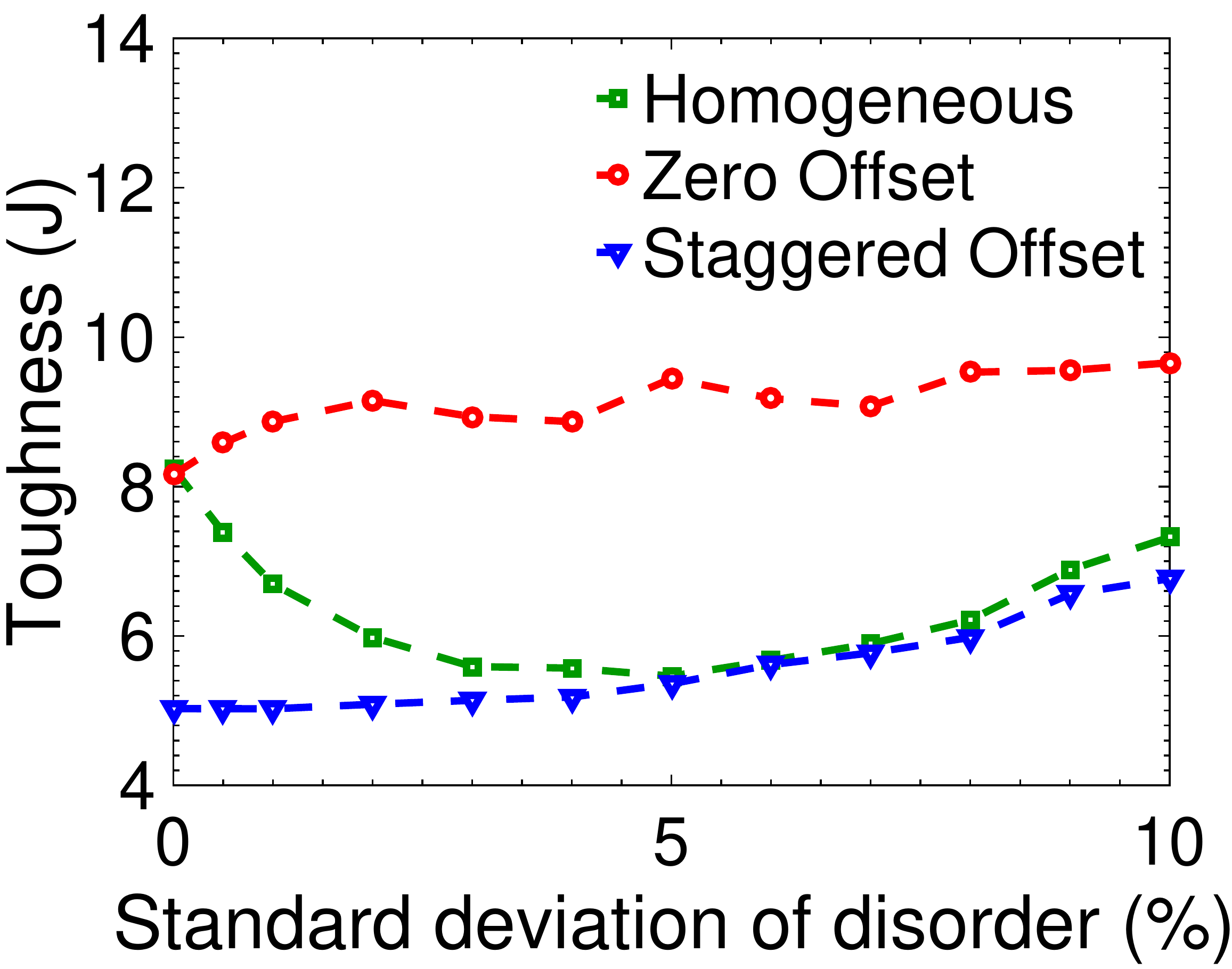}}
\caption{Effect of disorder in failure criteria on the peak force and fracture toughness of laminate. The different panels correspond to (a), (b): common stress threshold criterion and (c), (d): common strain threshold criterion. The data are for a crack of length $ a_{0}= 0.28 w$, and have been averaged over 25 realizations (50 for homogeneous).\label{img:force_toughness_cl_29_sd_all}}
\end{figure}

To gain an insight into the role of disorder in failure threshold on fracture for different laminate configurations with their inherent elastic heterogeneity, we simulate the fracture response for a range of disorder by varying the standard deviation of the Gaussian distribution for failure threshold from $0\%$ - $10\%$. When the hard and soft phase have a common stress threshold, the zero-offset configuration develops the highest strength as well as toughness for the entire range of disorder considered, as seen in Fig.~\ref{img:force_cl_29_sd_all_stress}-\ref{img:toughness_cl_29_sd_all_strain}. With slightest increase in disorder, strength of the elastically homogeneous solid drops sharply and is nearly a constant thereafter. The effect of increasing disorder has a marginal effect on the strength of staggered configuration and, except for very low disorder, the failure threshold the strength is very close to the effective elastically homogeneous laminate.

\begin{figure} 
\centering
\subfloat[\label{img:force_cl_all_sd_05_stress}]{
\includegraphics[width=0.96\columnwidth]{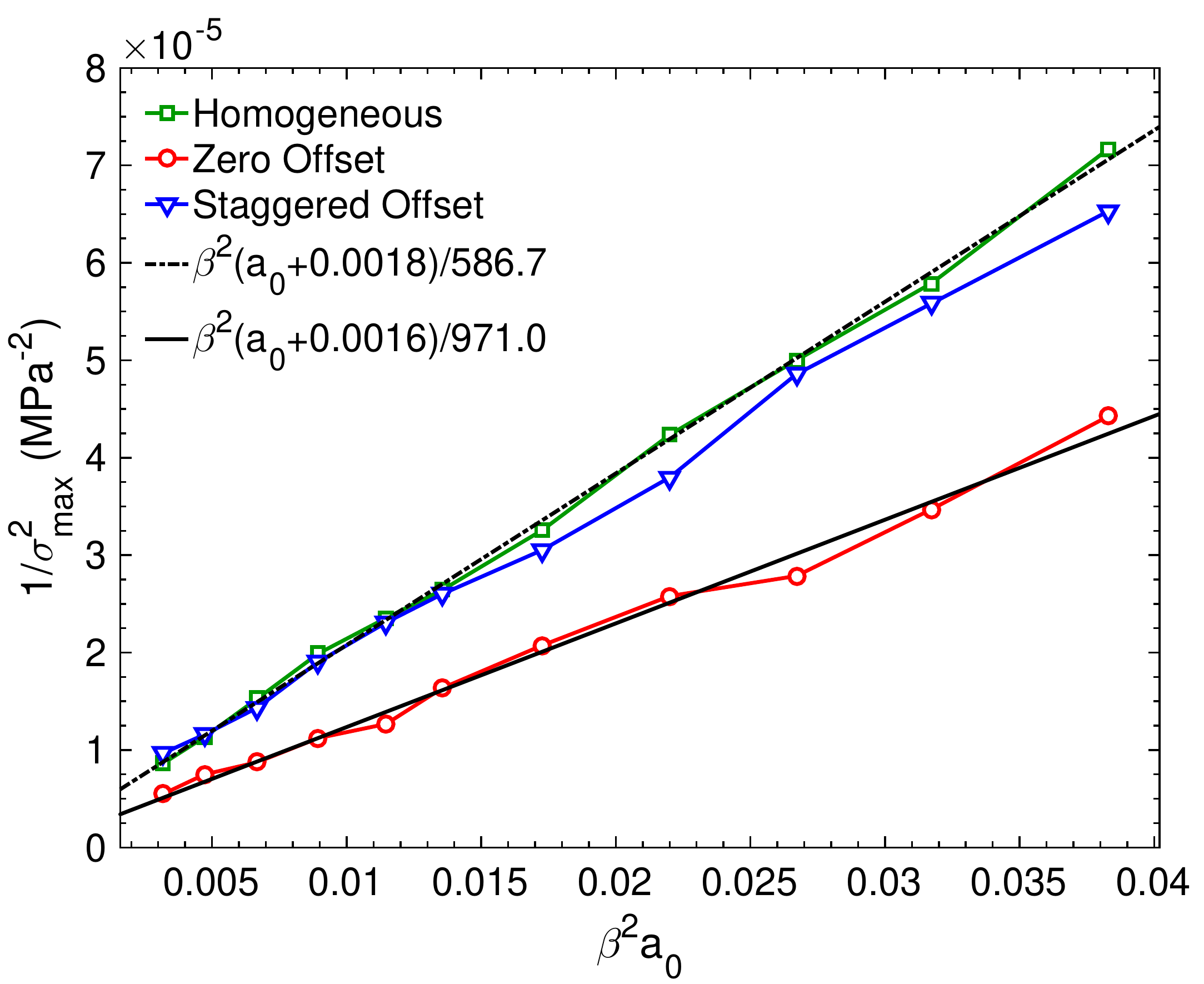}}
\vspace{0cm}
\subfloat[\label{img:force_cl_all_sd_05_strain}]{
\includegraphics[width=0.96\columnwidth]{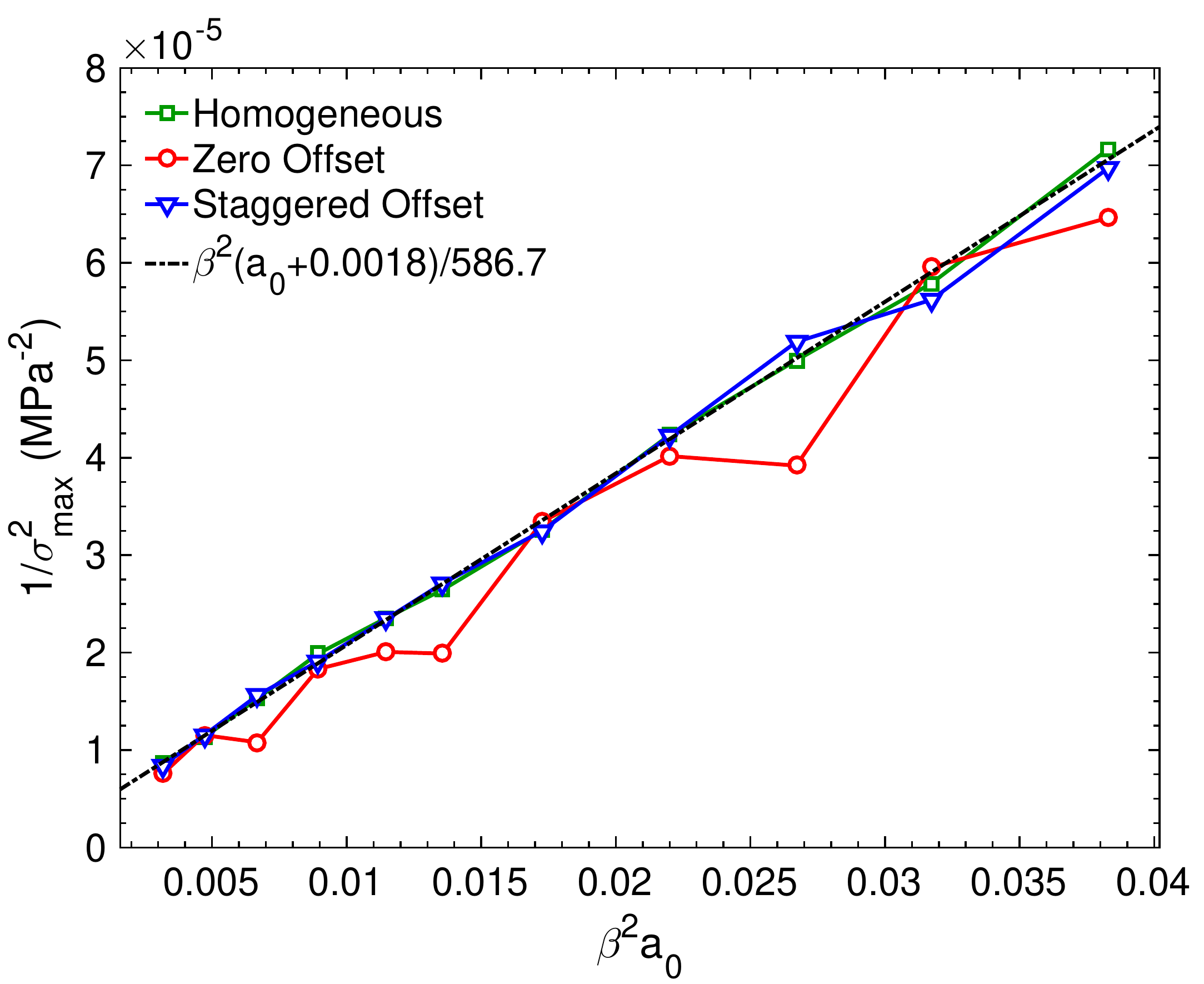}}
\caption{The variation of the failure stress $\sigma_{\rm max}$ with crack length $a_0$. The results are for laminates with staggered offset and zero offset, as well as the homogeneous model, with $ 5\% $ standard deviation in the failure threshold. The panels correspond to failure thresholds based on (a) common stress criterion and (b) common strain criterion. Each data point has been averaged over 25 realizations (50 for homogeneous).\label{img:frc_cl_all_sd_05}}
\end{figure}

We now study the effect of elastic heterogeneity and the failure threshold criteria on the critical stress intensity factor or fracture toughness $K_{Ic}$. As per the linear elastic fracture theory, the failure stress $\sigma_{\rm max}$ depends on the crack length $a_0$ as
 \be
 \frac{1}{\sigma^2_{\rm max}} = \frac{\beta^2 a_0}{K^2_{Ic}},
 \label{eqn:lefm}
 \ee
where the geometric correction factor $\beta$ for a given height to width ratio of the specimen geometry is known~\cite{hammond-2016}. {In presence of a fracture process zone, it has been proposed that Eq.~(\ref{eqn:lefm}) is modified to
\be
\frac{1}{\sigma^2_{\rm max}} = \frac{\beta^2 (a_0+\xi)}{K^2_{Ic}},
\label{eqn:lefm_mod}
\ee
where $\xi$ is another additional length length scale~\cite{bazant-1984, bazant-1996a, bazant-2004}. In presence of strong disorder, it has been argued that the crack length $a_0$ in Eq.~(\ref{eqn:lefm_mod}) is dependent also on the disorder~\cite{alava-2008, papanikolaou-2019}. In our simulation, the disorder is small, with damage limited to near crack, and we will therefore compare our data with Eq.~(\ref{eqn:lefm_mod}).} To confirm the linear relation between $1/\sigma^2_{\rm max}$ and $a_0$, as well as to measure $K_{Ic}$, we simulate the fracture response at $5\%$ standard deviation in failure strength for a range of initial crack lengths, $a_0$, as shown in Fig.~\ref{img:frc_cl_all_sd_05}. It is evident that, in our simulations, $1/\sigma^2_{\rm max}$ is linearly proportional to $a_0$ for different elastic heterogeneities as well as different failure threshold criteria. Interestingly, for the constant stress failure threshold [see Fig.~\ref{img:force_cl_all_sd_05_stress}], we find that $K_{Ic}$ is larger for zero-offset laminate than the homogeneous or staggered laminate, thus increasing fracture toughness. For a common strain threshold failure criterion [see Fig.~\ref{img:force_cl_all_sd_05_strain}], we find that on an average $K_{Ic}$ is independent of the elastic heterogeneity. However, on a more local scale, there are deviations from linearity for the zero-offset laminate, and $K_{Ic}$ depends on the initial location of the tip of the crack front.

\begin{figure}
	\centering
	\subfloat[\label{img:stress_avalanche}]{
		\includegraphics[width=0.96\columnwidth]{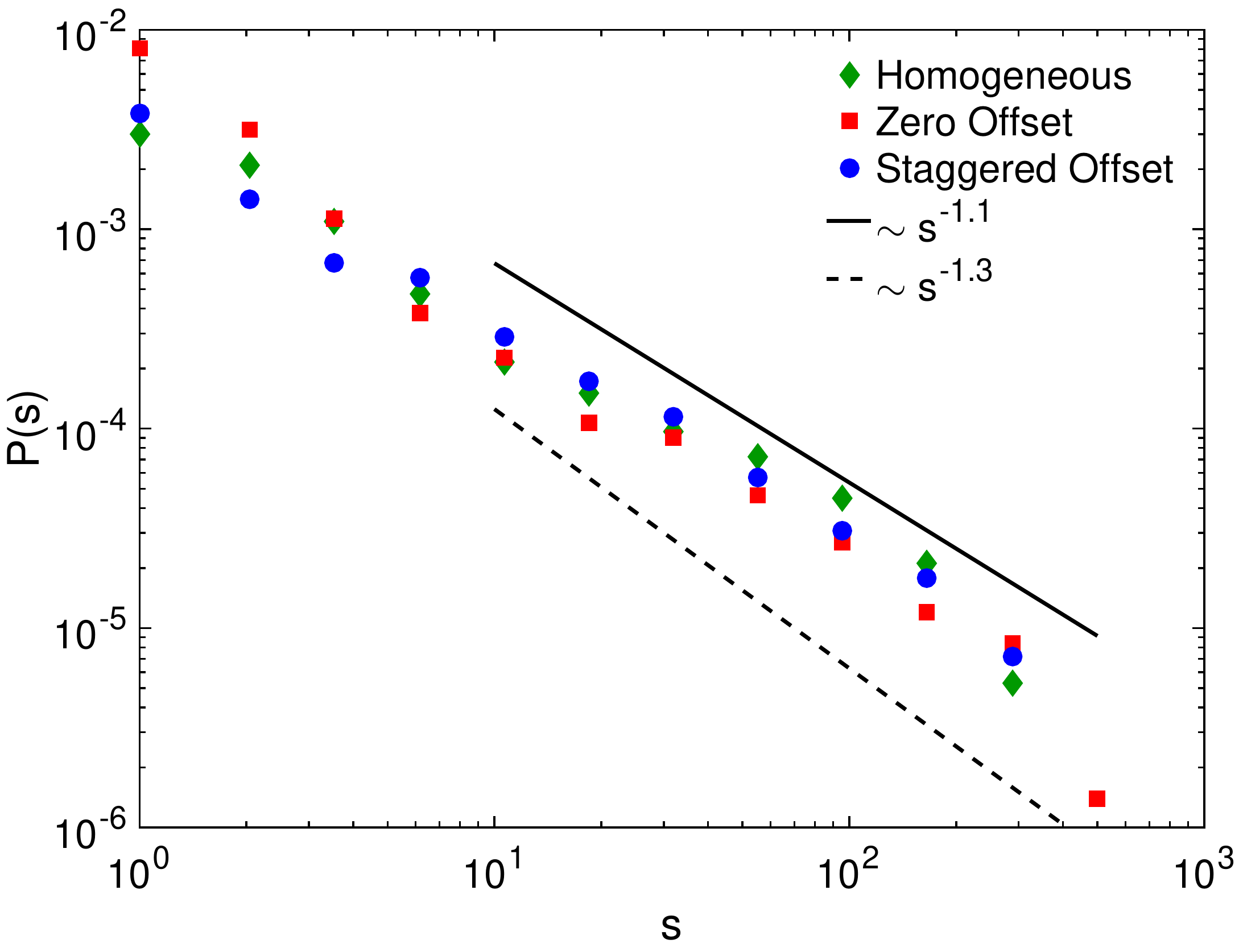}}
	\vspace{0cm}
	\subfloat[\label{img:strain_avalanche}]{
		\includegraphics[width=0.96\columnwidth]{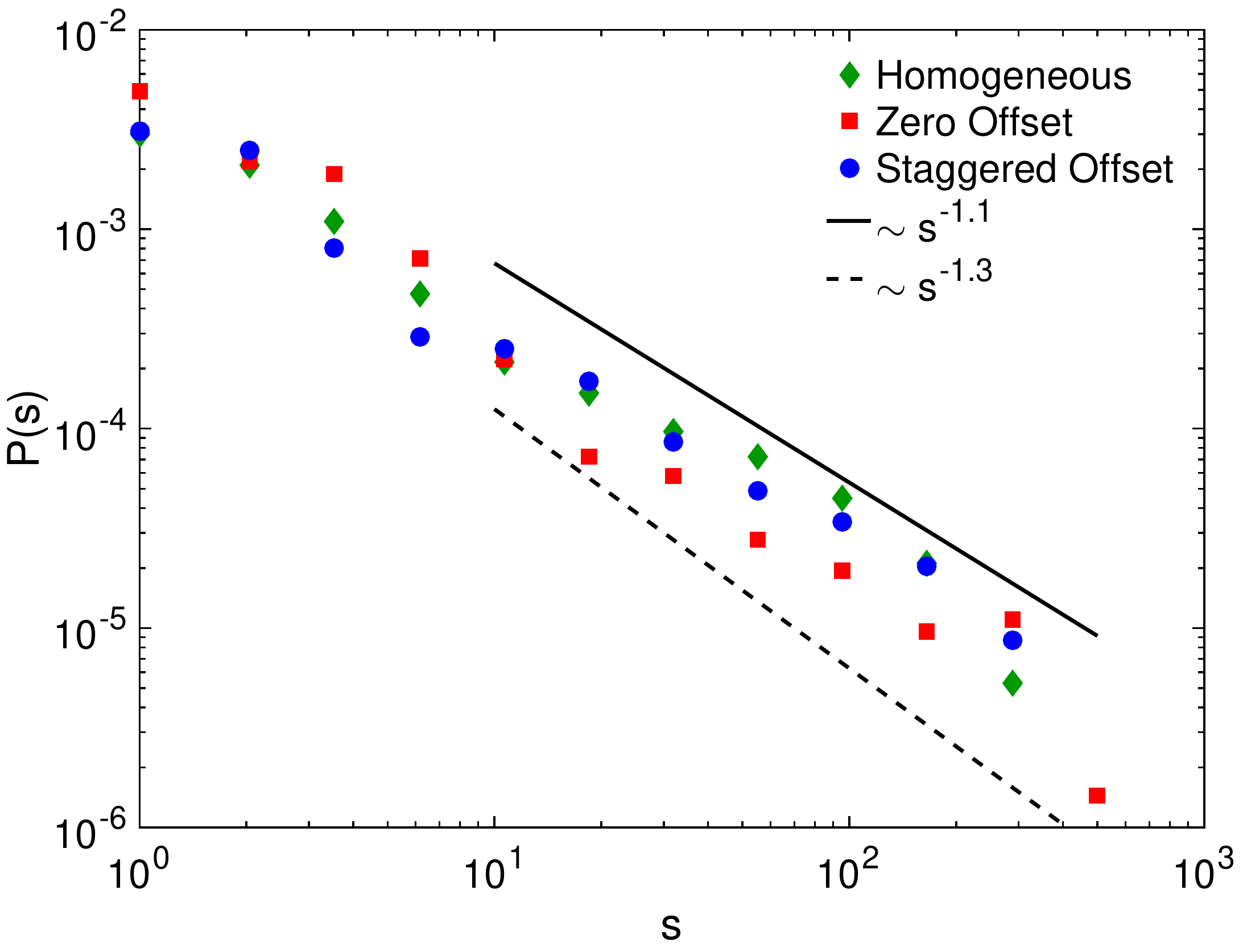}}
	\caption{{The variation of the avalanche size distribution, $P(s)$, with avalanche size, $s$, for different laminate configurations, as well as for the homogeneous model. The panels represent data for (a) a common stress failure thresholds, and (b) a common strain failure thresholds. The data are for $10\%$ standard deviation in failure threshold and averaged over 250 realizations.\label{img:avalanche}}}
\end{figure}
Another commonly used probe to characterize the fracture process is statistics of acoustic emissions during fracture, or equivalently avalanche statistics in simulations. An avalanche is defined as the number of bonds that are broken in one increment of applied strain. Let $P(s)$ denote the probability density function for avalanches of size $s$. The distribution is usually insensitive to the nature of the material undergoing fracture and follows a power law distribution $P(s) \sim s^{-\alpha}$ for large $s$. The avalanche size distribution, for $10\%$ standard deviation in failure strength, is shown in Fig.~\ref{img:avalanche} for the different laminate configurations for both choices of the critical thresholds and compared to that for the homogeneous model. {For intermediate $s$, we see that the data for the different meso-structure follow power law distributions for both common stress theshold (see Fig.~\ref{img:stress_avalanche}) as well as common strain threshold (see Fig.~\ref{img:strain_avalanche}). In this range of s, there is no distinguishable effect of meso-structure on the exponent $\alpha$. However, the range of the power law is limited, and a more definite statement on the universality requires simulation of much larger system sizes. We conclude that exponent $\alpha$ is atmost weakly dependent on the meso-structure as well as the choice of threshold.} Note that the power law distribution is quite different from the avalanche distributions obtained from RSNM simulations, in the absence of a pre-existing notch or crack, which have exponents close to 2.5~\cite{alava-2006}. However, close to breakdown, it is known from exact solution of the fiber bundle model as well as simulations of the random fuse model that the exponent for critical avalanches could be very different from the exponents of avalanche distribution away from breakdown~\cite{pradhan-2006}. For instance, in the fiber bundle model, the exponent changes from $5/2$ to $3/2$ near breakdown{, while in a network of electric fuses the exponent decreases from 3 to 2 near the catastrophic failure. The corresponding exponent for critical avalanches in RSNM, while not known, is expected to decrease from 2.5~\cite{pradhan-2005, pradhan-2006}. In our simulations, where there is a pre-existing crack in the system as well as when the standard deviation of the failure threshold distributions is very small, the system can be thought of as one close to breakdown and the avalanche distribution is presumably closer to near failure condition, and thus in analogy with other models it is expected to have an exponent significantly lower than 2.5 (between 1.1 and 1.3), as shown in Fig.~\ref{img:avalanche}.

We note that the avalanche distribution is not stationary, i.e., it is not independent of the value of the strain. This is clear since there are no avalanches for small strains. For the choice of disorder, the integrated $P(s)$ is dominated by avalanche near breakdown.}

\section{\label{sec:summary}Conclusion}

Fracture processes in composite materials strongly depend not only on the properties of the constituents but also equally on their architecture. In this paper, we develop a methodology for constructing a discrete element model for composites, particularly a plain weave laminate, which incorporates the inherent spatial patterns of the meso-structure. Though the macroscopic elastic behavior does not depend on the spatial patterning of its constituents, the spatial distribution of the stresses for a composite is distinctly different from that of an elastically equivalent homogeneous system as the high stresses near the crack tip are delocalized significantly.

For studying fracture behavior, two failure threshold criteria were used: a common stress threshold and a common strain threshold. Interestingly, the enhancement in strength and fracture toughness is observed for all initial crack lengths when the failure criteria have a common stress threshold. The enhancement is most pronounced for the composite laminate with zero offset. The dominant mechanism is observed to be the breakage of bonds that are energetically less dissipative. Consequently, when energetically more dissipative bonds are encountered in the path of a growing crack, it tends to deflect from its plane, resulting in a more tortuous path and a larger number of total bonds broken. In contrast, for a common strain threshold, while there is no enhancement in toughness on an average, occasional enhancement in toughness is observed depending on the initial position of the crack tip.

We also studied the effect of spatial patterning on the avalanche size distribution. {Within the simulation data, the avalanche distribution is independent of the elastic heterogeneity as well as the failure criteria. However, the value of the avalanche exponent is much smaller than the value reported for RSNM. In the presence of a crack as well as very little disorder, we argued that the avalanche resemble critical avalanche close to breakdown. In such case, it is known for other models like fiber bundle models and random fuse model that the critical avalanche exponent could be significantly lower than off-critical exponents. Determining this exponent more accurately for RSNM is a promising area for future study.}

The simulations provide valuable insights into the role of elastic heterogeneity in a two-phase network in the enhancement of toughness for the two specific scenarios of comparative failure thresholds. In particular, a common stress threshold criterion in which the harder phase is weaker and the softer phase is tougher is a common comparative material failure response. However, for the realistic approximation of actual experimental data, the imposition of a constraint such as a common failure threshold would be inapplicable. Equivalent statistical signatures of two-phase behavior, as observed in the count of broken bonds, if found in experimental data such as acoustic emission activity, can provide vital inputs in establishing a failure criterion for the two phases.

\bibliographystyle{apsrev4-2}

\begin{thebibliography}{47}%
\makeatletter
\providecommand \@ifxundefined [1]{%
 \@ifx{#1\undefined}
}%
\providecommand \@ifnum [1]{%
 \ifnum #1\expandafter \@firstoftwo
 \else \expandafter \@secondoftwo
 \fi
}%
\providecommand \@ifx [1]{%
 \ifx #1\expandafter \@firstoftwo
 \else \expandafter \@secondoftwo
 \fi
}%
\providecommand \natexlab [1]{#1}%
\providecommand \enquote  [1]{``#1''}%
\providecommand \bibnamefont  [1]{#1}%
\providecommand \bibfnamefont [1]{#1}%
\providecommand \citenamefont [1]{#1}%
\providecommand \href@noop [0]{\@secondoftwo}%
\providecommand \href [0]{\begingroup \@sanitize@url \@href}%
\providecommand \@href[1]{\@@startlink{#1}\@@href}%
\providecommand \@@href[1]{\endgroup#1\@@endlink}%
\providecommand \@sanitize@url [0]{\catcode `\\12\catcode `\$12\catcode
  `\&12\catcode `\#12\catcode `\^12\catcode `\_12\catcode `\%12\relax}%
\providecommand \@@startlink[1]{}%
\providecommand \@@endlink[0]{}%
\providecommand \url  [0]{\begingroup\@sanitize@url \@url }%
\providecommand \@url [1]{\endgroup\@href {#1}{\urlprefix }}%
\providecommand \urlprefix  [0]{URL }%
\providecommand \Eprint [0]{\href }%
\providecommand \doibase [0]{https://doi.org/}%
\providecommand \selectlanguage [0]{\@gobble}%
\providecommand \bibinfo  [0]{\@secondoftwo}%
\providecommand \bibfield  [0]{\@secondoftwo}%
\providecommand \translation [1]{[#1]}%
\providecommand \BibitemOpen [0]{}%
\providecommand \bibitemStop [0]{}%
\providecommand \bibitemNoStop [0]{.\EOS\space}%
\providecommand \EOS [0]{\spacefactor3000\relax}%
\providecommand \BibitemShut  [1]{\csname bibitem#1\endcsname}%
\let\auto@bib@innerbib\@empty
\bibitem [{\citenamefont {Gao}\ \emph {et~al.}(2003)\citenamefont {Gao},
  \citenamefont {Ji}, \citenamefont {J{\"a}ger}, \citenamefont {Arzt},\ and\
  \citenamefont {Fratzl}}]{gao-2003}%
  \BibitemOpen
  \bibfield  {author} {\bibinfo {author} {\bibfnamefont {H.}~\bibnamefont
  {Gao}}, \bibinfo {author} {\bibfnamefont {B.}~\bibnamefont {Ji}}, \bibinfo
  {author} {\bibfnamefont {I.~L.}\ \bibnamefont {J{\"a}ger}}, \bibinfo {author}
  {\bibfnamefont {E.}~\bibnamefont {Arzt}},\ and\ \bibinfo {author}
  {\bibfnamefont {P.}~\bibnamefont {Fratzl}},\ }\href
  {https://doi.org/10.1073/pnas.0631609100} {\bibfield  {journal} {\bibinfo
  {journal} {Proceedings of the National Academy of Sciences}\ }\textbf
  {\bibinfo {volume} {100}},\ \bibinfo {pages} {5597} (\bibinfo {year}
  {2003})}\BibitemShut {NoStop}%
\bibitem [{\citenamefont {Gupta}\ \emph {et~al.}(2009)\citenamefont {Gupta},
  \citenamefont {Krauss}, \citenamefont {Seto}, \citenamefont {Wagermaier},
  \citenamefont {Kerschnitzki}, \citenamefont {Benecke}, \citenamefont
  {Zaslansky}, \citenamefont {Boesecke}, \citenamefont {Funari}, \citenamefont
  {Kirchner},\ and\ \citenamefont {Fratzl}}]{gupta-2009}%
  \BibitemOpen
  \bibfield  {author} {\bibinfo {author} {\bibfnamefont {H.}~\bibnamefont
  {Gupta}}, \bibinfo {author} {\bibfnamefont {S.}~\bibnamefont {Krauss}},
  \bibinfo {author} {\bibfnamefont {J.}~\bibnamefont {Seto}}, \bibinfo {author}
  {\bibfnamefont {W.}~\bibnamefont {Wagermaier}}, \bibinfo {author}
  {\bibfnamefont {M.}~\bibnamefont {Kerschnitzki}}, \bibinfo {author}
  {\bibfnamefont {G.}~\bibnamefont {Benecke}}, \bibinfo {author} {\bibfnamefont
  {P.}~\bibnamefont {Zaslansky}}, \bibinfo {author} {\bibfnamefont
  {P.}~\bibnamefont {Boesecke}}, \bibinfo {author} {\bibfnamefont
  {S.}~\bibnamefont {Funari}}, \bibinfo {author} {\bibfnamefont
  {H.}~\bibnamefont {Kirchner}},\ and\ \bibinfo {author} {\bibfnamefont
  {P.}~\bibnamefont {Fratzl}},\ }\href
  {https://doi.org/https://doi.org/10.1016/j.bone.2009.01.084} {\bibfield
  {journal} {\bibinfo  {journal} {Bone}\ }\textbf {\bibinfo {volume} {44}},\
  \bibinfo {pages} {S33} (\bibinfo {year} {2009})}\BibitemShut {NoStop}%
\bibitem [{\citenamefont {Sen}\ and\ \citenamefont {Buehler}(2011)}]{sen-2011}%
  \BibitemOpen
  \bibfield  {author} {\bibinfo {author} {\bibfnamefont {D.}~\bibnamefont
  {Sen}}\ and\ \bibinfo {author} {\bibfnamefont {M.~J.}\ \bibnamefont
  {Buehler}},\ }\href {https://doi.org/10.1038/srep00035} {\bibfield  {journal}
  {\bibinfo  {journal} {Scientific Reports}\ }\textbf {\bibinfo {volume} {1}},\
  \bibinfo {pages} {35} (\bibinfo {year} {2011})}\BibitemShut {NoStop}%
\bibitem [{\citenamefont {Piggott}(1994)}]{piggott-1994}%
  \BibitemOpen
  \bibfield  {author} {\bibinfo {author} {\bibfnamefont {M.~R.}\ \bibnamefont
  {Piggott}},\ }\href {https://doi.org/10.1177/002199839402800701} {\bibfield
  {journal} {\bibinfo  {journal} {Journal of Composite Materials}\ }\textbf
  {\bibinfo {volume} {28}},\ \bibinfo {pages} {588} (\bibinfo {year}
  {1994})}\BibitemShut {NoStop}%
\bibitem [{\citenamefont {Puck}\ and\ \citenamefont
  {Schürmann}(2002)}]{puck-2002}%
  \BibitemOpen
  \bibfield  {author} {\bibinfo {author} {\bibfnamefont {A.}~\bibnamefont
  {Puck}}\ and\ \bibinfo {author} {\bibfnamefont {H.}~\bibnamefont
  {Schürmann}},\ }\href
  {https://doi.org/https://doi.org/10.1016/S0266-3538(01)00208-1} {\bibfield
  {journal} {\bibinfo  {journal} {Composites Science and Technology}\ }\textbf
  {\bibinfo {volume} {62}},\ \bibinfo {pages} {1633} (\bibinfo {year}
  {2002})}\BibitemShut {NoStop}%
\bibitem [{\citenamefont {Mishnaevsky}\ and\ \citenamefont
  {Brøndsted}(2009)}]{mishnaevsky-2009}%
  \BibitemOpen
  \bibfield  {author} {\bibinfo {author} {\bibfnamefont {L.}~\bibnamefont
  {Mishnaevsky}}\ and\ \bibinfo {author} {\bibfnamefont {P.}~\bibnamefont
  {Brøndsted}},\ }\href
  {https://doi.org/https://doi.org/10.1016/j.commatsci.2008.09.004} {\bibfield
  {journal} {\bibinfo  {journal} {Computational Materials Science}\ }\textbf
  {\bibinfo {volume} {44}},\ \bibinfo {pages} {1351} (\bibinfo {year}
  {2009})}\BibitemShut {NoStop}%
\bibitem [{\citenamefont {Daniel}\ \emph {et~al.}(2009)\citenamefont {Daniel},
  \citenamefont {Luo}, \citenamefont {Schubel},\ and\ \citenamefont
  {Werner}}]{daniel-2009}%
  \BibitemOpen
  \bibfield  {author} {\bibinfo {author} {\bibfnamefont {I.~M.}\ \bibnamefont
  {Daniel}}, \bibinfo {author} {\bibfnamefont {J.-J.}\ \bibnamefont {Luo}},
  \bibinfo {author} {\bibfnamefont {P.~M.}\ \bibnamefont {Schubel}},\ and\
  \bibinfo {author} {\bibfnamefont {B.~T.}\ \bibnamefont {Werner}},\ }\href
  {https://doi.org/https://doi.org/10.1016/j.compscitech.2008.04.016}
  {\bibfield  {journal} {\bibinfo  {journal} {Composites Science and
  Technology}\ }\textbf {\bibinfo {volume} {69}},\ \bibinfo {pages} {764}
  (\bibinfo {year} {2009})}\BibitemShut {NoStop}%
\bibitem [{\citenamefont {Wicks}\ \emph {et~al.}(2014)\citenamefont {Wicks},
  \citenamefont {Wang}, \citenamefont {Williams},\ and\ \citenamefont
  {Wardle}}]{wicks-2014}%
  \BibitemOpen
  \bibfield  {author} {\bibinfo {author} {\bibfnamefont {S.~S.}\ \bibnamefont
  {Wicks}}, \bibinfo {author} {\bibfnamefont {W.}~\bibnamefont {Wang}},
  \bibinfo {author} {\bibfnamefont {M.~R.}\ \bibnamefont {Williams}},\ and\
  \bibinfo {author} {\bibfnamefont {B.~L.}\ \bibnamefont {Wardle}},\ }\href
  {https://doi.org/https://doi.org/10.1016/j.compscitech.2014.06.003}
  {\bibfield  {journal} {\bibinfo  {journal} {Composites Science and
  Technology}\ }\textbf {\bibinfo {volume} {100}},\ \bibinfo {pages} {128}
  (\bibinfo {year} {2014})}\BibitemShut {NoStop}%
\bibitem [{\citenamefont {Das}\ \emph {et~al.}(2018)\citenamefont {Das},
  \citenamefont {Kandan}, \citenamefont {Kazemahvazi}, \citenamefont {Wadley},\
  and\ \citenamefont {Deshpande}}]{das-2018}%
  \BibitemOpen
  \bibfield  {author} {\bibinfo {author} {\bibfnamefont {S.}~\bibnamefont
  {Das}}, \bibinfo {author} {\bibfnamefont {K.}~\bibnamefont {Kandan}},
  \bibinfo {author} {\bibfnamefont {S.}~\bibnamefont {Kazemahvazi}}, \bibinfo
  {author} {\bibfnamefont {H.}~\bibnamefont {Wadley}},\ and\ \bibinfo {author}
  {\bibfnamefont {V.}~\bibnamefont {Deshpande}},\ }\href
  {https://doi.org/https://doi.org/10.1016/j.ijsolstr.2017.12.011} {\bibfield
  {journal} {\bibinfo  {journal} {International Journal of Solids and
  Structures}\ }\textbf {\bibinfo {volume} {136-137}},\ \bibinfo {pages} {137}
  (\bibinfo {year} {2018})}\BibitemShut {NoStop}%
\bibitem [{\citenamefont {Tan}\ \emph {et~al.}(2000)\citenamefont {Tan},
  \citenamefont {Tong}, \citenamefont {Steven},\ and\ \citenamefont
  {Ishikawa}}]{tan-2000}%
  \BibitemOpen
  \bibfield  {author} {\bibinfo {author} {\bibfnamefont {P.}~\bibnamefont
  {Tan}}, \bibinfo {author} {\bibfnamefont {L.}~\bibnamefont {Tong}}, \bibinfo
  {author} {\bibfnamefont {G.}~\bibnamefont {Steven}},\ and\ \bibinfo {author}
  {\bibfnamefont {T.}~\bibnamefont {Ishikawa}},\ }\href
  {https://doi.org/https://doi.org/10.1016/S1359-835X(99)00070-6} {\bibfield
  {journal} {\bibinfo  {journal} {Composites Part A: Applied Science and
  Manufacturing}\ }\textbf {\bibinfo {volume} {31}},\ \bibinfo {pages} {259}
  (\bibinfo {year} {2000})}\BibitemShut {NoStop}%
\bibitem [{\citenamefont {Carol}\ \emph
  {et~al.}(2001{\natexlab{a}})\citenamefont {Carol}, \citenamefont {Rizzi},\
  and\ \citenamefont {Willam}}]{carol-2001a}%
  \BibitemOpen
  \bibfield  {author} {\bibinfo {author} {\bibfnamefont {I.}~\bibnamefont
  {Carol}}, \bibinfo {author} {\bibfnamefont {E.}~\bibnamefont {Rizzi}},\ and\
  \bibinfo {author} {\bibfnamefont {K.}~\bibnamefont {Willam}},\ }\href
  {https://doi.org/https://doi.org/10.1016/S0020-7683(00)00030-5} {\bibfield
  {journal} {\bibinfo  {journal} {International Journal of Solids and
  Structures}\ }\textbf {\bibinfo {volume} {38}},\ \bibinfo {pages} {491}
  (\bibinfo {year} {2001}{\natexlab{a}})}\BibitemShut {NoStop}%
\bibitem [{\citenamefont {Carol}\ \emph
  {et~al.}(2001{\natexlab{b}})\citenamefont {Carol}, \citenamefont {Rizzi},\
  and\ \citenamefont {Willam}}]{carol-2001b}%
  \BibitemOpen
  \bibfield  {author} {\bibinfo {author} {\bibfnamefont {I.}~\bibnamefont
  {Carol}}, \bibinfo {author} {\bibfnamefont {E.}~\bibnamefont {Rizzi}},\ and\
  \bibinfo {author} {\bibfnamefont {K.}~\bibnamefont {Willam}},\ }\href
  {https://doi.org/https://doi.org/10.1016/S0020-7683(00)00031-7} {\bibfield
  {journal} {\bibinfo  {journal} {International Journal of Solids and
  Structures}\ }\textbf {\bibinfo {volume} {38}},\ \bibinfo {pages} {519}
  (\bibinfo {year} {2001}{\natexlab{b}})}\BibitemShut {NoStop}%
\bibitem [{\citenamefont {Luccioni}\ and\ \citenamefont
  {Oller}(2003)}]{luccioni-2003}%
  \BibitemOpen
  \bibfield  {author} {\bibinfo {author} {\bibfnamefont {B.}~\bibnamefont
  {Luccioni}}\ and\ \bibinfo {author} {\bibfnamefont {S.}~\bibnamefont
  {Oller}},\ }\href
  {https://doi.org/https://doi.org/10.1016/S0045-7825(02)00577-7} {\bibfield
  {journal} {\bibinfo  {journal} {Computer Methods in Applied Mechanics and
  Engineering}\ }\textbf {\bibinfo {volume} {192}},\ \bibinfo {pages} {1119}
  (\bibinfo {year} {2003})}\BibitemShut {NoStop}%
\bibitem [{\citenamefont {Ansar}\ \emph {et~al.}(2011)\citenamefont {Ansar},
  \citenamefont {Xinwei},\ and\ \citenamefont {Chouwei}}]{ansar-2011}%
  \BibitemOpen
  \bibfield  {author} {\bibinfo {author} {\bibfnamefont {M.}~\bibnamefont
  {Ansar}}, \bibinfo {author} {\bibfnamefont {W.}~\bibnamefont {Xinwei}},\ and\
  \bibinfo {author} {\bibfnamefont {Z.}~\bibnamefont {Chouwei}},\ }\href
  {https://doi.org/https://doi.org/10.1016/j.compstruct.2011.03.010} {\bibfield
   {journal} {\bibinfo  {journal} {Composite Structures}\ }\textbf {\bibinfo
  {volume} {93}},\ \bibinfo {pages} {1947} (\bibinfo {year}
  {2011})}\BibitemShut {NoStop}%
\bibitem [{\citenamefont {Zhang}\ \emph {et~al.}(2012)\citenamefont {Zhang},
  \citenamefont {Li}, \citenamefont {Kadkhodaei},\ and\ \citenamefont
  {Gao}}]{zhang-2012}%
  \BibitemOpen
  \bibfield  {author} {\bibinfo {author} {\bibfnamefont {T.}~\bibnamefont
  {Zhang}}, \bibinfo {author} {\bibfnamefont {X.}~\bibnamefont {Li}}, \bibinfo
  {author} {\bibfnamefont {S.}~\bibnamefont {Kadkhodaei}},\ and\ \bibinfo
  {author} {\bibfnamefont {H.}~\bibnamefont {Gao}},\ }\href
  {https://doi.org/10.1021/nl301908b} {\bibfield  {journal} {\bibinfo
  {journal} {Nano Letters}\ }\textbf {\bibinfo {volume} {12}},\ \bibinfo
  {pages} {4605} (\bibinfo {year} {2012})}\BibitemShut {NoStop}%
\bibitem [{\citenamefont {Camanho}\ \emph {et~al.}(2013)\citenamefont
  {Camanho}, \citenamefont {Bessa}, \citenamefont {Catalanotti}, \citenamefont
  {Vogler},\ and\ \citenamefont {Rolfes}}]{camanho-2013}%
  \BibitemOpen
  \bibfield  {author} {\bibinfo {author} {\bibfnamefont {P.}~\bibnamefont
  {Camanho}}, \bibinfo {author} {\bibfnamefont {M.}~\bibnamefont {Bessa}},
  \bibinfo {author} {\bibfnamefont {G.}~\bibnamefont {Catalanotti}}, \bibinfo
  {author} {\bibfnamefont {M.}~\bibnamefont {Vogler}},\ and\ \bibinfo {author}
  {\bibfnamefont {R.}~\bibnamefont {Rolfes}},\ }\href
  {https://doi.org/https://doi.org/10.1016/j.mechmat.2012.12.001} {\bibfield
  {journal} {\bibinfo  {journal} {Mechanics of Materials}\ }\textbf {\bibinfo
  {volume} {59}},\ \bibinfo {pages} {36} (\bibinfo {year} {2013})}\BibitemShut
  {NoStop}%
\bibitem [{\citenamefont {Vogler}\ \emph {et~al.}(2013)\citenamefont {Vogler},
  \citenamefont {Rolfes},\ and\ \citenamefont {Camanho}}]{volger-2013}%
  \BibitemOpen
  \bibfield  {author} {\bibinfo {author} {\bibfnamefont {M.}~\bibnamefont
  {Vogler}}, \bibinfo {author} {\bibfnamefont {R.}~\bibnamefont {Rolfes}},\
  and\ \bibinfo {author} {\bibfnamefont {P.}~\bibnamefont {Camanho}},\ }\href
  {https://doi.org/https://doi.org/10.1016/j.mechmat.2012.12.002} {\bibfield
  {journal} {\bibinfo  {journal} {Mechanics of Materials}\ }\textbf {\bibinfo
  {volume} {59}},\ \bibinfo {pages} {50} (\bibinfo {year} {2013})}\BibitemShut
  {NoStop}%
\bibitem [{\citenamefont {Tan}\ and\ \citenamefont {Falzon}(2016)}]{tan-2016}%
  \BibitemOpen
  \bibfield  {author} {\bibinfo {author} {\bibfnamefont {W.}~\bibnamefont
  {Tan}}\ and\ \bibinfo {author} {\bibfnamefont {B.~G.}\ \bibnamefont
  {Falzon}},\ }\href
  {https://doi.org/https://doi.org/10.1016/j.compscitech.2016.02.008}
  {\bibfield  {journal} {\bibinfo  {journal} {Composites Science and
  Technology}\ }\textbf {\bibinfo {volume} {126}},\ \bibinfo {pages} {60}
  (\bibinfo {year} {2016})}\BibitemShut {NoStop}%
\bibitem [{\citenamefont {Rosti}\ \emph {et~al.}(2009)\citenamefont {Rosti},
  \citenamefont {Illa}, \citenamefont {Koivisto},\ and\ \citenamefont
  {Alava}}]{rosti-2009}%
  \BibitemOpen
  \bibfield  {author} {\bibinfo {author} {\bibfnamefont {J.}~\bibnamefont
  {Rosti}}, \bibinfo {author} {\bibfnamefont {X.}~\bibnamefont {Illa}},
  \bibinfo {author} {\bibfnamefont {J.}~\bibnamefont {Koivisto}},\ and\
  \bibinfo {author} {\bibfnamefont {M.~J.}\ \bibnamefont {Alava}},\ }\href
  {https://doi.org/10.1088/0022-3727/42/21/214013} {\bibfield  {journal}
  {\bibinfo  {journal} {Journal of Physics D: Applied Physics}\ }\textbf
  {\bibinfo {volume} {42}},\ \bibinfo {pages} {214013} (\bibinfo {year}
  {2009})}\BibitemShut {NoStop}%
\bibitem [{\citenamefont {Bar\'o}\ \emph {et~al.}(2013)\citenamefont {Bar\'o},
  \citenamefont {Corral}, \citenamefont {Illa}, \citenamefont {Planes},
  \citenamefont {Salje}, \citenamefont {Schranz}, \citenamefont {Soto-Parra},\
  and\ \citenamefont {Vives}}]{baro-2013}%
  \BibitemOpen
  \bibfield  {author} {\bibinfo {author} {\bibfnamefont {J.}~\bibnamefont
  {Bar\'o}}, \bibinfo {author} {\bibfnamefont {A.}~\bibnamefont {Corral}},
  \bibinfo {author} {\bibfnamefont {X.}~\bibnamefont {Illa}}, \bibinfo {author}
  {\bibfnamefont {A.}~\bibnamefont {Planes}}, \bibinfo {author} {\bibfnamefont
  {E.~K.~H.}\ \bibnamefont {Salje}}, \bibinfo {author} {\bibfnamefont
  {W.}~\bibnamefont {Schranz}}, \bibinfo {author} {\bibfnamefont {D.~E.}\
  \bibnamefont {Soto-Parra}},\ and\ \bibinfo {author} {\bibfnamefont
  {E.}~\bibnamefont {Vives}},\ }\href
  {https://doi.org/10.1103/PhysRevLett.110.088702} {\bibfield  {journal}
  {\bibinfo  {journal} {Phys. Rev. Lett.}\ }\textbf {\bibinfo {volume} {110}},\
  \bibinfo {pages} {088702} (\bibinfo {year} {2013})}\BibitemShut {NoStop}%
\bibitem [{\citenamefont {Curtin}\ and\ \citenamefont
  {Scher}(1990)}]{curtin-1990}%
  \BibitemOpen
  \bibfield  {author} {\bibinfo {author} {\bibfnamefont {W.~A.}\ \bibnamefont
  {Curtin}}\ and\ \bibinfo {author} {\bibfnamefont {H.}~\bibnamefont {Scher}},\
  }\href {https://doi.org/10.1557/JMR.1990.0535} {\bibfield  {journal}
  {\bibinfo  {journal} {Journal of Materials Research}\ }\textbf {\bibinfo
  {volume} {5}},\ \bibinfo {pages} {535} (\bibinfo {year} {1990})}\BibitemShut
  {NoStop}%
\bibitem [{\citenamefont {Alava}\ \emph {et~al.}(2006)\citenamefont {Alava},
  \citenamefont {Nukala},\ and\ \citenamefont {Zapperi}}]{alava-2006}%
  \BibitemOpen
  \bibfield  {author} {\bibinfo {author} {\bibfnamefont {M.~J.}\ \bibnamefont
  {Alava}}, \bibinfo {author} {\bibfnamefont {P.~K.}\ \bibnamefont {Nukala}},\
  and\ \bibinfo {author} {\bibfnamefont {S.}~\bibnamefont {Zapperi}},\ }\href
  {https://doi.org/10.1080/00018730300741518} {\bibfield  {journal} {\bibinfo
  {journal} {Advances in Physics}\ }\textbf {\bibinfo {volume} {55}},\ \bibinfo
  {pages} {349} (\bibinfo {year} {2006})}\BibitemShut {NoStop}%
\bibitem [{\citenamefont {Pan}\ \emph {et~al.}(2018)\citenamefont {Pan},
  \citenamefont {Ma}, \citenamefont {Wang},\ and\ \citenamefont
  {Chen}}]{pan-2018}%
  \BibitemOpen
  \bibfield  {author} {\bibinfo {author} {\bibfnamefont {Z.}~\bibnamefont
  {Pan}}, \bibinfo {author} {\bibfnamefont {R.}~\bibnamefont {Ma}}, \bibinfo
  {author} {\bibfnamefont {D.}~\bibnamefont {Wang}},\ and\ \bibinfo {author}
  {\bibfnamefont {A.}~\bibnamefont {Chen}},\ }\href
  {https://doi.org/https://doi.org/10.1016/j.engfracmech.2017.12.037}
  {\bibfield  {journal} {\bibinfo  {journal} {Engineering Fracture Mechanics}\
  }\textbf {\bibinfo {volume} {190}},\ \bibinfo {pages} {382} (\bibinfo {year}
  {2018})}\BibitemShut {NoStop}%
\bibitem [{\citenamefont {Ray}(2006)}]{ray-2006}%
  \BibitemOpen
  \bibfield  {author} {\bibinfo {author} {\bibfnamefont {P.}~\bibnamefont
  {Ray}},\ }\href
  {https://doi.org/https://doi.org/10.1016/j.commatsci.2005.12.012} {\bibfield
  {journal} {\bibinfo  {journal} {Computational Materials Science}\ }\textbf
  {\bibinfo {volume} {37}},\ \bibinfo {pages} {141} (\bibinfo {year}
  {2006})}\BibitemShut {NoStop}%
\bibitem [{\citenamefont {Pradhan}\ \emph {et~al.}(2006)\citenamefont
  {Pradhan}, \citenamefont {Hansen},\ and\ \citenamefont
  {Hemmer}}]{pradhan-2006}%
  \BibitemOpen
  \bibfield  {author} {\bibinfo {author} {\bibfnamefont {S.}~\bibnamefont
  {Pradhan}}, \bibinfo {author} {\bibfnamefont {A.}~\bibnamefont {Hansen}},\
  and\ \bibinfo {author} {\bibfnamefont {P.~C.}\ \bibnamefont {Hemmer}},\
  }\href {https://doi.org/10.1103/PhysRevE.74.016122} {\bibfield  {journal}
  {\bibinfo  {journal} {Phys. Rev. E}\ }\textbf {\bibinfo {volume} {74}},\
  \bibinfo {pages} {016122} (\bibinfo {year} {2006})}\BibitemShut {NoStop}%
\bibitem [{\citenamefont {Karihaloo}\ \emph {et~al.}(2003)\citenamefont
  {Karihaloo}, \citenamefont {Shao},\ and\ \citenamefont
  {Xiao}}]{karihaloo-2003}%
  \BibitemOpen
  \bibfield  {author} {\bibinfo {author} {\bibfnamefont {B.}~\bibnamefont
  {Karihaloo}}, \bibinfo {author} {\bibfnamefont {P.}~\bibnamefont {Shao}},\
  and\ \bibinfo {author} {\bibfnamefont {Q.}~\bibnamefont {Xiao}},\ }\href
  {https://doi.org/https://doi.org/10.1016/S0013-7944(03)00004-3} {\bibfield
  {journal} {\bibinfo  {journal} {Engineering Fracture Mechanics}\ }\textbf
  {\bibinfo {volume} {70}},\ \bibinfo {pages} {2385} (\bibinfo {year}
  {2003})}\BibitemShut {NoStop}%
\bibitem [{\citenamefont {Berton}\ and\ \citenamefont
  {Bolander}(2006)}]{berton-2006}%
  \BibitemOpen
  \bibfield  {author} {\bibinfo {author} {\bibfnamefont {S.}~\bibnamefont
  {Berton}}\ and\ \bibinfo {author} {\bibfnamefont {J.~E.}\ \bibnamefont
  {Bolander}},\ }\href
  {https://doi.org/https://doi.org/10.1016/j.cma.2005.04.020} {\bibfield
  {journal} {\bibinfo  {journal} {Computer Methods in Applied Mechanics and
  Engineering}\ }\textbf {\bibinfo {volume} {195}},\ \bibinfo {pages} {7172}
  (\bibinfo {year} {2006})}\BibitemShut {NoStop}%
\bibitem [{\citenamefont {Grassl}\ and\ \citenamefont
  {Bažant}(2009)}]{grassl-2009}%
  \BibitemOpen
  \bibfield  {author} {\bibinfo {author} {\bibfnamefont {P.}~\bibnamefont
  {Grassl}}\ and\ \bibinfo {author} {\bibfnamefont {Z.~P.}\ \bibnamefont
  {Bažant}},\ }\href {https://doi.org/10.1061/(ASCE)0733-9399(2009)135:2(85)}
  {\bibfield  {journal} {\bibinfo  {journal} {Journal of Engineering
  Mechanics}\ }\textbf {\bibinfo {volume} {135}},\ \bibinfo {pages} {85}
  (\bibinfo {year} {2009})}\BibitemShut {NoStop}%
\bibitem [{\citenamefont {Mayya}\ \emph {et~al.}(2016)\citenamefont {Mayya},
  \citenamefont {Praveen}, \citenamefont {Banerjee},\ and\ \citenamefont
  {Rajesh}}]{ashwij-2016}%
  \BibitemOpen
  \bibfield  {author} {\bibinfo {author} {\bibfnamefont {A.}~\bibnamefont
  {Mayya}}, \bibinfo {author} {\bibfnamefont {P.}~\bibnamefont {Praveen}},
  \bibinfo {author} {\bibfnamefont {A.}~\bibnamefont {Banerjee}},\ and\
  \bibinfo {author} {\bibfnamefont {R.}~\bibnamefont {Rajesh}},\ }\href
  {https://royalsocietypublishing.org/doi/full/10.1098/rsif.2016.0809}
  {\bibfield  {journal} {\bibinfo  {journal} {Journal of the Royal Society
  Interface}\ }\textbf {\bibinfo {volume} {13}},\ \bibinfo {pages} {20160809}
  (\bibinfo {year} {2016})}\BibitemShut {NoStop}%
\bibitem [{\citenamefont {Mayya}\ \emph {et~al.}(2017)\citenamefont {Mayya},
  \citenamefont {Banerjee},\ and\ \citenamefont {Rajesh}}]{ashwij-2017}%
  \BibitemOpen
  \bibfield  {author} {\bibinfo {author} {\bibfnamefont {A.}~\bibnamefont
  {Mayya}}, \bibinfo {author} {\bibfnamefont {A.}~\bibnamefont {Banerjee}},\
  and\ \bibinfo {author} {\bibfnamefont {R.}~\bibnamefont {Rajesh}},\ }\href
  {https://doi.org/10.1103/PhysRevE.96.053001} {\bibfield  {journal} {\bibinfo
  {journal} {Phys. Rev. E}\ }\textbf {\bibinfo {volume} {96}},\ \bibinfo
  {pages} {053001} (\bibinfo {year} {2017})}\BibitemShut {NoStop}%
\bibitem [{\citenamefont {Mayya}\ \emph {et~al.}(2018)\citenamefont {Mayya},
  \citenamefont {Banerjee},\ and\ \citenamefont {Rajesh}}]{ashwij-2018}%
  \BibitemOpen
  \bibfield  {author} {\bibinfo {author} {\bibfnamefont {A.}~\bibnamefont
  {Mayya}}, \bibinfo {author} {\bibfnamefont {A.}~\bibnamefont {Banerjee}},\
  and\ \bibinfo {author} {\bibfnamefont {R.}~\bibnamefont {Rajesh}},\ }\href
  {https://doi.org/https://doi.org/10.1016/j.jmbbm.2018.04.013} {\bibfield
  {journal} {\bibinfo  {journal} {Journal of the Mechanical Behavior of
  Biomedical Materials}\ }\textbf {\bibinfo {volume} {83}},\ \bibinfo {pages}
  {108} (\bibinfo {year} {2018})}\BibitemShut {NoStop}%
\bibitem [{\citenamefont {Urabe}\ and\ \citenamefont
  {Takesue}(2010)}]{urabe-2010}%
  \BibitemOpen
  \bibfield  {author} {\bibinfo {author} {\bibfnamefont {C.}~\bibnamefont
  {Urabe}}\ and\ \bibinfo {author} {\bibfnamefont {S.}~\bibnamefont
  {Takesue}},\ }\href {https://doi.org/10.1103/PhysRevE.82.016106} {\bibfield
  {journal} {\bibinfo  {journal} {Phys. Rev. E}\ }\textbf {\bibinfo {volume}
  {82}},\ \bibinfo {pages} {016106} (\bibinfo {year} {2010})}\BibitemShut
  {NoStop}%
\bibitem [{\citenamefont {Boyina}\ \emph {et~al.}(2015)\citenamefont {Boyina},
  \citenamefont {Kirubakaran}, \citenamefont {Banerjee},\ and\ \citenamefont
  {Velmurugan}}]{dhatreyi-2015}%
  \BibitemOpen
  \bibfield  {author} {\bibinfo {author} {\bibfnamefont {D.}~\bibnamefont
  {Boyina}}, \bibinfo {author} {\bibfnamefont {T.}~\bibnamefont {Kirubakaran}},
  \bibinfo {author} {\bibfnamefont {A.}~\bibnamefont {Banerjee}},\ and\
  \bibinfo {author} {\bibfnamefont {R.}~\bibnamefont {Velmurugan}},\ }\href
  {https://doi.org/https://doi.org/10.1016/j.mechmat.2015.07.013} {\bibfield
  {journal} {\bibinfo  {journal} {Mechanics of Materials}\ }\textbf {\bibinfo
  {volume} {91}},\ \bibinfo {pages} {64} (\bibinfo {year} {2015})}\BibitemShut
  {NoStop}%
\bibitem [{\citenamefont {Dimas}\ \emph {et~al.}(2014)\citenamefont {Dimas},
  \citenamefont {Giesa},\ and\ \citenamefont {Buehler}}]{dimas-2014a}%
  \BibitemOpen
  \bibfield  {author} {\bibinfo {author} {\bibfnamefont {L.~S.}\ \bibnamefont
  {Dimas}}, \bibinfo {author} {\bibfnamefont {T.}~\bibnamefont {Giesa}},\ and\
  \bibinfo {author} {\bibfnamefont {M.~J.}\ \bibnamefont {Buehler}},\ }\href
  {https://doi.org/https://doi.org/10.1016/j.jmps.2013.07.006} {\bibfield
  {journal} {\bibinfo  {journal} {Journal of the Mechanics and Physics of
  Solids}\ }\textbf {\bibinfo {volume} {63}},\ \bibinfo {pages} {481} (\bibinfo
  {year} {2014})}\BibitemShut {NoStop}%
\bibitem [{\citenamefont {Dimas}\ \emph
  {et~al.}(2015{\natexlab{a}})\citenamefont {Dimas}, \citenamefont {Veneziano},
  \citenamefont {Giesa},\ and\ \citenamefont {Buehler}}]{dimas-2015a}%
  \BibitemOpen
  \bibfield  {author} {\bibinfo {author} {\bibfnamefont {L.~S.}\ \bibnamefont
  {Dimas}}, \bibinfo {author} {\bibfnamefont {D.}~\bibnamefont {Veneziano}},
  \bibinfo {author} {\bibfnamefont {T.}~\bibnamefont {Giesa}},\ and\ \bibinfo
  {author} {\bibfnamefont {M.~J.}\ \bibnamefont {Buehler}},\ }\href
  {https://doi.org/10.1115/1.4028783} {\bibfield  {journal} {\bibinfo
  {journal} {Journal of Applied Mechanics}\ }\textbf {\bibinfo {volume} {82}},\
  \bibinfo {pages} {011003} (\bibinfo {year} {2015}{\natexlab{a}})}\BibitemShut
  {NoStop}%
\bibitem [{\citenamefont {Dimas}\ \emph
  {et~al.}(2015{\natexlab{b}})\citenamefont {Dimas}, \citenamefont {Veneziano},
  \citenamefont {Giesa},\ and\ \citenamefont {Buehler}}]{dimas-2015b}%
  \BibitemOpen
  \bibfield  {author} {\bibinfo {author} {\bibfnamefont {L.~S.}\ \bibnamefont
  {Dimas}}, \bibinfo {author} {\bibfnamefont {D.}~\bibnamefont {Veneziano}},
  \bibinfo {author} {\bibfnamefont {T.}~\bibnamefont {Giesa}},\ and\ \bibinfo
  {author} {\bibfnamefont {M.~J.}\ \bibnamefont {Buehler}},\ }\href
  {https://doi.org/https://doi.org/10.1016/j.jmps.2015.06.016} {\bibfield
  {journal} {\bibinfo  {journal} {Journal of the Mechanics and Physics of
  Solids}\ }\textbf {\bibinfo {volume} {84}},\ \bibinfo {pages} {116} (\bibinfo
  {year} {2015}{\natexlab{b}})}\BibitemShut {NoStop}%
\bibitem [{\citenamefont {Monette}\ and\ \citenamefont
  {Anderson}(1994)}]{monette-1994}%
  \BibitemOpen
  \bibfield  {author} {\bibinfo {author} {\bibfnamefont {L.}~\bibnamefont
  {Monette}}\ and\ \bibinfo {author} {\bibfnamefont {M.~P.}\ \bibnamefont
  {Anderson}},\ }\href {http://stacks.iop.org/0965-0393/2/i=1/a=004} {\bibfield
   {journal} {\bibinfo  {journal} {Modelling and Simulation in Materials
  Science and Engineering}\ }\textbf {\bibinfo {volume} {2}},\ \bibinfo {pages}
  {53} (\bibinfo {year} {1994})}\BibitemShut {NoStop}%
\bibitem [{\citenamefont {Boyina}\ \emph {et~al.}(2014)\citenamefont {Boyina},
  \citenamefont {Banerjee},\ and\ \citenamefont {Velmurugan}}]{dhatreyi-2014}%
  \BibitemOpen
  \bibfield  {author} {\bibinfo {author} {\bibfnamefont {D.}~\bibnamefont
  {Boyina}}, \bibinfo {author} {\bibfnamefont {A.}~\bibnamefont {Banerjee}},\
  and\ \bibinfo {author} {\bibfnamefont {R.}~\bibnamefont {Velmurugan}},\
  }\href {https://doi.org/https://doi.org/10.1016/j.compositesb.2013.12.052}
  {\bibfield  {journal} {\bibinfo  {journal} {Composites Part B: Engineering}\
  }\textbf {\bibinfo {volume} {60}},\ \bibinfo {pages} {21} (\bibinfo {year}
  {2014})}\BibitemShut {NoStop}%
\bibitem [{\citenamefont {Liu}\ and\ \citenamefont {Long}(2015)}]{liu-lsqm}%
  \BibitemOpen
  \bibfield  {author} {\bibinfo {author} {\bibfnamefont {W.}~\bibnamefont
  {Liu}}\ and\ \bibinfo {author} {\bibfnamefont {R.}~\bibnamefont {Long}},\
  }\href {https://doi.org/10.1115/1.4031763} {\bibfield  {journal} {\bibinfo
  {journal} {Journal of Applied Mechanics}\ }\textbf {\bibinfo {volume} {83}},\
  \bibinfo {pages} {011006} (\bibinfo {year} {2015})}\BibitemShut {NoStop}%
\bibitem [{\citenamefont {Chona}\ \emph {et~al.}(1982)\citenamefont {Chona},
  \citenamefont {Irwin},\ and\ \citenamefont {Shukla}}]{chona-1982}%
  \BibitemOpen
  \bibfield  {author} {\bibinfo {author} {\bibfnamefont {R.}~\bibnamefont
  {Chona}}, \bibinfo {author} {\bibfnamefont {G.~R.}\ \bibnamefont {Irwin}},\
  and\ \bibinfo {author} {\bibfnamefont {A.}~\bibnamefont {Shukla}},\ }\href
  {https://doi.org/10.1243/03093247V172079} {\bibfield  {journal} {\bibinfo
  {journal} {The Journal of Strain Analysis for Engineering Design}\ }\textbf
  {\bibinfo {volume} {17}},\ \bibinfo {pages} {79} (\bibinfo {year}
  {1982})}\BibitemShut {NoStop}%
\bibitem [{\citenamefont {Hammond}\ and\ \citenamefont
  {Fawaz}(2016)}]{hammond-2016}%
  \BibitemOpen
  \bibfield  {author} {\bibinfo {author} {\bibfnamefont {M.~J.}\ \bibnamefont
  {Hammond}}\ and\ \bibinfo {author} {\bibfnamefont {S.~A.}\ \bibnamefont
  {Fawaz}},\ }\href
  {https://doi.org/https://doi.org/10.1016/j.engfracmech.2015.12.022}
  {\bibfield  {journal} {\bibinfo  {journal} {Engineering Fracture Mechanics}\
  }\textbf {\bibinfo {volume} {153}},\ \bibinfo {pages} {25} (\bibinfo {year}
  {2016})}\BibitemShut {NoStop}%
\bibitem [{\citenamefont {Ba{\v{z}}ant}(1984)}]{bazant-1984}%
  \BibitemOpen
  \bibfield  {author} {\bibinfo {author} {\bibfnamefont {Z.~P.}\ \bibnamefont
  {Ba{\v{z}}ant}},\ }\href
  {https://doi.org/10.1061/(ASCE)0733-9399(1984)110:4(518)} {\bibfield
  {journal} {\bibinfo  {journal} {Journal of Engineering Mechanics}\ }\textbf
  {\bibinfo {volume} {110}},\ \bibinfo {pages} {518} (\bibinfo {year}
  {1984})}\BibitemShut {NoStop}%
\bibitem [{\citenamefont {Ba{\v z}ant}(1996)}]{bazant-1996a}%
  \BibitemOpen
  \bibfield  {author} {\bibinfo {author} {\bibfnamefont {Z.~P.}\ \bibnamefont
  {Ba{\v z}ant}},\ }\href
  {https://doi.org/https://doi.org/10.1016/S1065-7355(96)90081-4} {\bibfield
  {journal} {\bibinfo  {journal} {Advanced Cement Based Materials}\ }\textbf
  {\bibinfo {volume} {4}},\ \bibinfo {pages} {128 } (\bibinfo {year}
  {1996})}\BibitemShut {NoStop}%
\bibitem [{\citenamefont {Ba{\v z}ant}(2004)}]{bazant-2004}%
  \BibitemOpen
  \bibfield  {author} {\bibinfo {author} {\bibfnamefont {Z.~P.}\ \bibnamefont
  {Ba{\v z}ant}},\ }\href {https://doi.org/10.1073/pnas.0404096101} {\bibfield
  {journal} {\bibinfo  {journal} {Proceedings of the National Academy of
  Sciences}\ }\textbf {\bibinfo {volume} {101}},\ \bibinfo {pages} {13400}
  (\bibinfo {year} {2004})}\BibitemShut {NoStop}%
\bibitem [{\citenamefont {Alava}\ \emph {et~al.}(2008)\citenamefont {Alava},
  \citenamefont {Nukala},\ and\ \citenamefont {Zapperi}}]{alava-2008}%
  \BibitemOpen
  \bibfield  {author} {\bibinfo {author} {\bibfnamefont {M.~J.}\ \bibnamefont
  {Alava}}, \bibinfo {author} {\bibfnamefont {P.~K. V.~V.}\ \bibnamefont
  {Nukala}},\ and\ \bibinfo {author} {\bibfnamefont {S.}~\bibnamefont
  {Zapperi}},\ }\href {https://doi.org/10.1103/PhysRevLett.100.055502}
  {\bibfield  {journal} {\bibinfo  {journal} {Phys. Rev. Lett.}\ }\textbf
  {\bibinfo {volume} {100}},\ \bibinfo {pages} {055502} (\bibinfo {year}
  {2008})}\BibitemShut {NoStop}%
\bibitem [{\citenamefont {Papanikolaou}\ \emph {et~al.}(2019)\citenamefont
  {Papanikolaou}, \citenamefont {Shanthraj}, \citenamefont {Thibault},
  \citenamefont {Woodward},\ and\ \citenamefont {Roters}}]{papanikolaou-2019}%
  \BibitemOpen
  \bibfield  {author} {\bibinfo {author} {\bibfnamefont {S.}~\bibnamefont
  {Papanikolaou}}, \bibinfo {author} {\bibfnamefont {P.}~\bibnamefont
  {Shanthraj}}, \bibinfo {author} {\bibfnamefont {J.}~\bibnamefont {Thibault}},
  \bibinfo {author} {\bibfnamefont {C.}~\bibnamefont {Woodward}},\ and\
  \bibinfo {author} {\bibfnamefont {F.}~\bibnamefont {Roters}},\ }\href
  {https://doi.org/10.1186/s41313-019-0017-0} {\bibfield  {journal} {\bibinfo
  {journal} {Materials Theory}\ }\textbf {\bibinfo {volume} {3}},\ \bibinfo
  {pages} {5} (\bibinfo {year} {2019})}\BibitemShut {NoStop}%
\bibitem [{\citenamefont {Pradhan}\ \emph {et~al.}(2005)\citenamefont
  {Pradhan}, \citenamefont {Hansen},\ and\ \citenamefont
  {Hemmer}}]{pradhan-2005}%
  \BibitemOpen
  \bibfield  {author} {\bibinfo {author} {\bibfnamefont {S.}~\bibnamefont
  {Pradhan}}, \bibinfo {author} {\bibfnamefont {A.}~\bibnamefont {Hansen}},\
  and\ \bibinfo {author} {\bibfnamefont {P.~C.}\ \bibnamefont {Hemmer}},\
  }\href {https://doi.org/10.1103/PhysRevLett.95.125501} {\bibfield  {journal}
  {\bibinfo  {journal} {Phys. Rev. Lett.}\ }\textbf {\bibinfo {volume} {95}},\
  \bibinfo {pages} {125501} (\bibinfo {year} {2005})}\BibitemShut {NoStop}%
\end{thebibliography}
%

\end{document}